\documentclass[11pt, a4paper, onecolumn]{quantumarticle}
\pdfoutput=1
\usepackage[breaklinks=true, pdftex, 
  pdfborder={0 0 0}, colorlinks=true, linkcolor=blue, citecolor=blue, urlcolor=blue, 
  bookmarks=false, pdfpagelabels=false, pdftitle={atlas-ai-ml-detector-proposal}]{hyperref}
\usepackage{acronym}
\usepackage{fullpage}
\usepackage{ifthen}   
\usepackage{graphicx,graphics}
\usepackage{sidecap}
\usepackage[]{rotating}
\usepackage{setspace}
\usepackage{amssymb,amsmath,amsthm}
\usepackage{color}
\usepackage{subfigure,wrapfig}
\usepackage{multirow}
\usepackage{pdfpages}
\usepackage{longtable} 	
\usepackage{datetime}
\usepackage{xspace}
\usepackage{array,ragged2e}
\usepackage{physics}
\usepackage{enumitem} 
\usepackage[it,small]{caption}
\usepackage[top=1in,bottom=1in,left=1in,right=1in]{geometry} 
\usepackage[T1]{fontenc}
\usepackage{subfigure}
\usepackage{comment}

\usepackage{xcolor}
\usepackage{braket}
\usepackage{bm}
\usepackage{nicematrix}  

\usepackage{ifthen}
\usepackage{xkeyval}
\usepackage{calc}

\newcounter{mycomment}

\usepackage[numbers,sort&compress]{natbib} 
\usepackage[resetlabels]{multibib}  
\newcites{dsg}{DSG papers}
\usepackage{algorithmic}
\usepackage[noresetcount,lined,boxed]{algorithm2e} 
\usepackage{gensymb}
\usepackage{nicematrix}

\usepackage{amsfonts}
\usepackage{amsmath}
\usepackage{supertabular}
\usepackage{amsthm}
\usepackage{floatrow}
\usepackage{tikz}
\usepackage{float}
\usepackage{mathtools}

\usepackage{titlesec}
  \usepackage{etoolbox}
  \makeatletter
  \patchcmd{\ttlh@hang}{\parindent\z@}{\parindent\z@\leavevmode}{}{}
  \patchcmd{\ttlh@hang}{\noindent}{}{}{}
  \makeatother

\usepackage{mdframed}                 
\usepackage{colortbl}
\usepackage{xcolor}
\definecolor{highlight}{HTML}{cdeeff} 
\definecolor{highlight}{HTML}{DFEFF9}
\mdfdefinestyle{highlightbox}{leftmargin=0pt,rightmargin=0pt,%
innerleftmargin=3pt,innerrightmargin=2pt,%
innertopmargin=4pt,innerbottommargin=4pt,%
backgroundcolor=highlight,linecolor=white}
\definecolor{algm}{HTML}{D4EFDF} 
\mdfdefinestyle{highlightalg}{leftmargin=0pt,rightmargin=0pt,%
innerleftmargin=3pt,innerrightmargin=2pt,%
innertopmargin=4pt,innerbottommargin=4pt,%
backgroundcolor=algm,linecolor=white}
\definecolor{struct}{HTML}{F2D7D5} 
\mdfdefinestyle{highlightstruct}{leftmargin=0pt,rightmargin=0pt,%
innerleftmargin=3pt,innerrightmargin=2pt,%
innertopmargin=4pt,innerbottommargin=4pt,%
backgroundcolor=struct,linecolor=white}
\usepackage[section]{placeins}

\newcommand{\R}{\mathbb R}           

\newcommand{\bvec}{\left(\begin{array}{c}}
\newcommand{\evec}{\end{array}\right)}

\newcommand{\subsubsubsection}[1]{\paragraph{#1.}}

\newcommand{\call}[1]{ } 
 \newcommand{\team}[1]			
      {\textcolor{cyan}{#1}\marginpar{{\color{cyan}\textbf{Team}}}}

\newcommand{\ignore}[1]{}  

 \newcommand{\done}[1]{ } 

\newcommand{\pagebudget}[1]{ } 

\usepackage{fancyhdr}
\renewcommand{\sectionmark}[1]{\markright{#1}{}}
\newcommand{\bio}[3]{\newpage \phantomsection 
		 \pagestyle{fancy} \sectionmark{BIOGRAPHICAL SKETCHES \hfill #1}
   \IfFileExists{bio/bio-#3.tex}{\addcontentsline{toc}{subsection}{#1 -- #2} \input{bio/bio-#3}
    }{\IfFileExists{bio/#3.pdf}{\addcontentsline{toc}{subsection}{#1 -- #2}
         \includepdf[pages=-,offset=0 -65,pagecommand={\thispagestyle{fancy}}]{bio/#3}
      }{
	 \addcontentsline{toc}{subsection}{\textcolor{red}{#1 -- #2 MISSING}}
         \begin{center}
            \textbf{\color{red}\Large No file found for `#1'!\\
                    Tried bio/#3.tex and bio/#3.pdf} \\
         \end{center}}}}

\newcommand{\cnp}[3]{\newpage \phantomsection 
		 \pagestyle{fancy} \sectionmark{SUPPORT OF INVESTIGATORS \hfill #1}
   \IfFileExists{cnp/support#3.tex}{\addcontentsline{toc}{subsection}{#1 -- 
#2} \input{cnp/support#3.tex}
    }{\IfFileExists{cnp/#3CurrentPending.pdf}{ 
\addcontentsline{toc}{subsection}{#1 -- #2}
         \includepdf[pages=-,offset=0 -65,pagecommand={\thispagestyle{fancy}} 
]{cnp/#3CurrentPending.pdf}
      }{
	 \addcontentsline{toc}{subsection}{\textcolor{red}{#1 -- #2 MISSING!!!}}
         \begin{center}
            \textbf{\color{red}\Large No file found for `#1'!\\
                    Tried cnp/support#3.tex and cnp/#3CurrentPending.pdf} \\
         \end{center}}}}

\begin{document}

\title{A Survey of Quantum Computing for Finance}   

\author[1]{Dylan A. Herman$^\star$}
\author[2]{Cody Googin$^\star$}
\author[3]{Xiaoyuan Liu$^\star$}
\author[4]{Alexey Galda}
\author[3]{Ilya Safro}
\author[1]{Yue Sun}
\author[1]{Marco Pistoia}
\author[5]{Yuri Alexeev}
\affil[1]{JPMorgan Chase Bank, N.A., New York, NY, USA --- \{\href{mailto:dylan.a.herman@jpmorgan.com}{dylan.a.herman},\href{mailto:yue.sun@jpmorgan.com}{yue.sun},\href{mailto:marco.pistoia@jpmorgan.com}{marco.pistoia}\}@jpmorgan.com}
\affil[2]{University of Chicago, Chicago, IL, USA --- \href{mailto:codygoogin@uchicago.edu}{codygoogin@uchicago.edu}}
\affil[3]{University of Delaware, Newark, DE, USA --- \{\href{mailto:joeyxliu@udel.edu}{joeyxliu},\href{mailto:isafro@udel.edu}{isafro}\}@udel.edu}
\affil[4]{Menten AI, San Francisco, CA, USA --- \href{mailto:alexey.galda@menten.ai}{alexey.galda@menten.ai}}
\affil[5]{Argonne National Laboratory, Lemont, IL, USA --- \href{mailto:yuri@anl.gov}{yuri@anl.gov}}
\maketitle

\begin{abstract}
\noindent \textbf{Abstract.} Quantum computers are expected to surpass the computational capabilities of classical computers during this decade and have transformative impact on numerous industry sectors, particularly finance.  In fact, finance is estimated to be the first industry sector to benefit from quantum computing, not only in the medium and long terms, but even in the short term. This survey paper presents a comprehensive summary of the state of the art of quantum computing for financial applications, with particular emphasis on stochastic modeling, optimization, and machine learning, describing how these solutions, adapted to work on a quantum computer, can potentially help to solve financial problems, such as derivative pricing, risk modeling, portfolio optimization, natural language processing, and fraud detection, more efficiently and accurately. We also discuss  the feasibility of these algorithms on near-term quantum computers with various hardware implementations and demonstrate how they relate to a wide range of use cases in finance. We hope this article will not only serve as a reference for academic researchers and industry practitioners but also inspire new ideas for future research.
\end{abstract}

\vspace{80mm}

\noindent \footnotesize{$^\star$These authors contributed equally to this work.}

\pagestyle{plain}
\pagenumbering{roman}
\setcounter{page}{1}





\setcounter{secnumdepth}{3} 
\setcounter{tocdepth}{4}
\newcommand{\hide}[1]{}
\newpage \phantomsection
\pagestyle{plain}
 \renewcommand\contentsname{Contents}
 \subpdfbookmark{Table of Contents}{toc}
 \tableofcontents

 \newpage

\newpage
\pagestyle{fancy}


\pagenumbering{arabic}
\setcounter{page}{1}


\pagestyle{plain}
\section{Introduction  }
\label{sec:intro}
Quantum computation relies on a fundamentally different means of processing and storing information than  today's classical computers use.  The reason is that the information does not obey the laws of classical mechanics but those of quantum mechanics. Usually, quantum-mechanical effects  become apparent only at very small scales, when quantum systems are properly isolated from the surrounding environments. These conditions, however, make the realization of a quantum computer a  challenging task. Even so, according to a McKinsey \& Co. report~\cite{mckinsey_quantum}, finance is estimated to be the first industry sector to benefit from quantum computing (Section \ref{sec:financial_concepts}), largely because of the potential for many financial use cases to be formulated as problems that can be solved by quantum algorithms suitable for near-term quantum computers. This is important because current quantum computers are small-scale and noisy, yet the hope is that we can still find use cases for them. In addition, a variety of quantum algorithms will be more applicable when large-scale, robust quantum computers become available, which will significantly speed up computations used in finance.

The diversity in the potential hardware platforms or physical realizations of quantum computation is completely unlike any of the classical computational devices available today. There exist proposals based on vastly different physical systems: superconductors, trapped ions, neutral atoms, photons, and others (Section \ref{sec:currentquantumhardware}), yet no clear winner  has emerged. A large number of companies are competing to be the first to develop a quantum computer capable of running applications useful in a production environment.
In theory, any computation that runs on a quantum computer may also be executed on a classical computer---the benefit that quantum computing may bring is the potential reduction in \emph{time} and memory \emph{space} with which the computational tasks are performed \cite{nielsen2002quantum}, which in turn may lead to unprecedented scalability and accuracy of the computations. In addition, any classical algorithm can be modified (in a potentially nontrivial way) such that it can be executed on a universal quantum computer, but without  any speedup. Obtaining quantum speedup requires developing new algorithms that specifically take advantage of quantum-mechanical properties. Thus, classical and quantum computers will need to work together. Moreover, in order to solve a real-world problem, a classical computer should be able to efficiently insert  \emph{and} obtain the necessary data into and from a quantum computer.

This promised efficiency of quantum computers enables certain computations that are otherwise infeasible for current classical computers to complete in any reasonable amount of time (Section \ref{sec:quantumconcepts}).
 In general, however, the speedup for each task can vary greatly or may even be currently unknown (Section \ref{sec:quantum_algorithms}).  While  these speedups, if found,  can have a tremendous impact in practice, they are typically  difficult to obtain. And even if they are discovered, the underlying quantum hardware must be powerful enough to minimize errors without introducing an overhead that counteracts the algorithmic speedup. We do not know exactly when such robust hardware will exist. Thus, the goal of quantum computing research is to develop quantum algorithms (Section \ref{sec:quantum_algorithms}) that solve useful problems faster and to build robust hardware platforms to run them on. The industry needs to understand the problems that can best benefit from quantum computing and the extent of these benefits, in order to make full use of its revolutionary power when production-grade quantum devices are available.

To this end, we offer a comprehensive overview of the applicability of quantum computing to finance. Additionally, we provide insight into the nature of quantum computation, focusing particularly on specific financial problems for which quantum computing can provide potential speedups compared with classical computing. 
The sections in this article can be grouped into two parts. The \emph{first part} contains introductory material:  Section \ref{sec:financial_concepts} introduces computationally intensive financial problems that potentially benefit from quantum computing, whereas Sections \ref{sec:quantumconcepts} and \ref{sec:quantum_algorithms} introduce the core concepts of quantum computation and quantum algorithms.
The \emph{second part}---the main focus of the article---reviews research performed by the scientific community to develop quantum-enhanced versions of three major problem-solving techniques used in finance: stochastic modeling (Section \ref{sec:stochastic_modeling}), optimization (Section \ref{sec:optimization}), and machine learning (Section \ref{sec:ML}). The connections between financial problems and these problem-solving techniques are outlined in Table \ref{table:fintable}.  Section \ref{sec:qubit tech} covers experiments on current quantum hardware that have been performed by researchers to solve financial use cases.

\subsection*{Overview of Earlier Surveys}
Although the field of quantum computing is one of the most dynamic today and some scientific conclusions, approaches, and technologies become obsolete relatively quickly, we  reiterate the importance of getting information from different perspectives. Here we emphasize the difference between our survey and several existing surveys that have recently been published or appeared in open domain. One of the key differences is that in our survey we not only summarize  the recent achievements but also discuss the limitations of quantum devices and algorithmic approaches that are currently subject to massive abuse in the media.

In \cite{egger2020quantum}, Egger et al. focus on covering the hardware and algorithmic work done by IBM. We take a much broader view and survey the entire landscape of quantum technologies and their applicability in the finance domain. For example, we believe it is beneficial to also discuss the quantum annealing-based approaches and other gate-based devices. Also, with regard to optimization applications, we discuss a much wider variety of financial applications that make use of optimization and may take advantage of quantum hardware development. 

The article by Bouland et al. \cite{bouland2020prospects} focuses on works done by the QC Ware team with a central emphasis on   quantum Monte Carlo integration. Some more recent work has been done in that area and is included in our survey. Also Bouland et al. focus on portfolio optimization. While this is an  important financial problem that we also cover, we have tried to include other financial applications that  use optimization as well. Moreover, we tried to delve deeper into the various quantum machine learning approaches for generative modeling and neural networks.

The survey by Orus et al. \cite{orus2019quantum} does an excellent job at highlighting financial applications that make use of quantum annealers. However, we believe the quantitative finance and quantum computing communities would also benefit from hearing about other quantum optimization approaches and devices. We also tried to delve a little bit deeper into how quantum annealing works and discuss universal adiabatic computation. Given that the field of quantum computation is so dynamic, we believe the community can benefit from seeing more recent work.

A recent survey by Pistoia et al. \cite{pistoia2021quantum} covers a variety of quantum machine learning algorithms applicable to finance. In our review, we discuss a broader array of applications outside the realm of machine learning, such as financial applications that make use of stochastic modeling and optimization.

There are also several  problem-specific surveys such as the derivative pricing \cite{gomez2022survey}, supply chain finance \cite{griffin2021quantum} and high-frequency trading \cite{ganapathy2021quantum} that tackle different goals than that of our survey.

\section{Applicability of Quantum Computing to Finance} 
\label{sec:financial_concepts}
Numerous financial use cases require the ability to assess a wide range of potential outcomes. To this extent, financial institutions employ statistical models and algorithms to predict future outcomes. Such techniques are fairly effective but not infallible. In a world where huge amounts of data are generated daily, computers that can perform predictive computations accurately are becoming a predominant need. For this reason, several financial institutions are turning to quantum computing, given its promise to analyze vast amounts of data and compute results faster and more accurately than  any classical computer has ever been able to do. Financial institutions believe that once they are capable of leveraging quantum computing, they are likely to see important benefits.
This section introduces financial use cases that are projected to lend themselves to a quantum computing approach.
\subsection{Macroeconomic Challenges for Financial Institutions}
The financial services industry is expected to reach a market cap of \$28.53 trillion by 2025, with a compound annual growth rate of 6\% \cite{researchandmarkets}. While this industry is often considered incredibly stable and insulated from change, shifting government regulations, new technologies, and customer expectations present challenges that require the industry to adapt. We discuss here three key macroeconomic financial trends that can be impacted by quantum computing: keeping up with regulations, addressing customer expectations and big data requirements, and ensuring data security.

\subsubsection{Keeping Up with Regulations}
\label{sec:regulations}
The Third Basel Accord, commonly known as Basel III, is an international regulatory framework set up in response to the 2008 financial crisis. Basel III sets requirements on capital, risk coverage, leverage, risk management and supervision, market discipline, liquidity, and supervisory monitoring~\cite{basel}. These regulations require computing numerous risk metrics, such as value at risk (VaR) and conditional value at risk (CVaR). Because  of the stochastic nature of the parameters involved in these computations, closed-form analytical solutions are generally not available, and hence numerical methods  have to be employed. One of the most widely used numerical methods is Monte Carlo integration, since it is generic and scalable to high-dimensional problems~\cite{glasserman2004monte}. Quantum Monte Carlo integration (Section \ref{sec:simulation and pricing}) has been shown to offer up to a quadratic speedup over the performance of its  classical counterpart. Engaging in ways to determine risk metrics with higher levels of accuracy and efficiency will enable financial institutions to make better-informed loan-related decisions and financial projections. 

\subsubsection{Addressing Customer Expectations and Big-Data Requirements}
Customer personalization has become increasingly important in finance.  For example, financial technology (FinTech), in contrast to traditional finance, thrives on providing personalized services. A Goldman Sachs report has highlighted the necessity for financial institutions to
provide new big-data-driven services to align with changing consumer behavior~\cite{TheFutureofFinance}. According to a Salesforce 2020 publication \cite{Salesforce_trends}, 52\% of customers expect companies to always personalize offers, while only 27\% of customers feel that the financial services industry provides great service and support~\cite{Salesforce_trends}. Financial institutions must meet this expectation in order to continue to retain their customers. The current  classical techniques for analyzing data have large computational demands and require significant time to train algorithms, whereas quantum algorithms may be able to speed up these computationally intensive and time-consuming components. As financial institutions continue to generate data,  they must be able to employ that data in a functional way to improve their business strategy. Additionally, data organization can allow financial institutions to engage with customers' finances more specifically and effectively, supporting their customer service and keeping customers engaged despite other options such as FinTech. Much of this data analysis is done by dealing with large matrices, which can be computationally demanding for classical computers.

\subsubsection{Ensuring Data Security}
Ensuring data security is one of the most pressing concerns for the financial services industry, according to an article by McKinsey~\cite{mckinsey_data} discussing rising cybersecurity issues. An increase in digitization has occurred due to the rise of FinTech.  Additionally, as a result of the COVID-19 pandemic,  substantially more financial activity is performed online. According to the McKinsey article, 95\% of board committees discuss cyber and tech risks four or more times a year. Financial institutions must stay up to date with online security and remain vigilant against attackers. Quantum computing data modeling and machine learning techniques could allow financial institutions to identify potential risks with higher accuracy. Furthermore, even if one does not plan to use quantum computation, one must be knowledgeable about it  because  of its ability to break current public-key cryptographic standards \cite{shor1997, 2021factoring, nielsen2002quantum}, such as RSA and Diffie--Hellmann, including the common variants based on elliptic curves \cite{roetteler2017quantum}. This is due to quantum computing's ability to efficiently solve the Abelian hidden subgroup problem~\cite{childs2010quantum}. Although current quantum computers cannot achieve this task,  financial institutions need  to become ready by investigating how to utilize quantum-safe protocols \cite{bernstein2017post} and, potentially, quantum-enhanced cryptography \cite{gisin2002quantum}, each of which has its own potential usages in the layers of a network. While the situation is by no means as serious for cryptographically-secure hash functions or symmetric-key ciphers, it is important to be aware of the potential quantum-enhanced attacks \cite{amy2016estimating, hosoyamada2021quantum, jaques2020implementing, Bonnetain_encryption}. Here, however, we are going to focus on quantum algorithms that have direct applications in finance, so protocols for quantum-enhanced cryptographic attacks, quantum-safe cryptography, and quantum-enhanced cryptography are all out of the scope of this survey.

\subsection{Financial Use Cases for Quantum Computing}
\label{sec:financial_problems}
As discussed earlier, finance is one of the first industry sectors expected to benefit from quantum computing, largely because financial problems lend themselves to be solved on near-term quantum computers. Since many of the financial problems are based on stochastic variables as inputs, they are, more often than not, relatively more tolerant to imprecisions in the solutions compared with problems in chemistry and physics. Arguably, certain problems in chemistry and physics  also can benefit from quantum computing.
In this survey, however, we aim to review common financial problems that can benefit from quantum computing and to discuss the associated quantum algorithms applicable to solving them.
We group these problems and quantum algorithms into three broad categories: stochastic modeling, optimization, and machine learning.

Stochastic modeling is concerned with the study of the dynamics and statistical characteristics of stochastic processes.
In finance, one of the most commonly seen problems that involves stochastic modeling, using numerical techniques, is estimating the prices of financial assets and their associated risks, whose values may depend on certain stochastic processes. In Section \ref{sec:stochastic_modeling}, we will discuss two of the most used techniques for modeling stochastic processes in finance---Monte Carlo methods and numerical solutions of differential equations, and the corresponding quantum approaches for them, namely, quantum Monte Carlo integration (QMCI) and quantum partial differential equation (PDE) solvers.
We will also demonstrate the applicability of the two quantum approaches to financial problems through examples of derivative pricing (Section \ref{sec:quantumderivspricing}) and risk modeling (Section \ref{sec:quantumriskanaly}).
Stochastic modeling with quantum machine learning (QML) techniques will be discussed in Sections \ref{sec:assetpricing} and \ref{sec:impliedvol}.

Optimization involves finding optimal inputs to a real-valued function so as to minimize or maximize the value of the function.
In Section \ref{sec:optimization} we will review prevailing quantum techniques for combinatorial and convex optimization problems.
We will cover adiabatic and variational quantum algorithms, as well as other quantum-classical hybrid approaches.
A large variety of financial problems that involve optimization could potentially benefit from these quantum techniques, in particular, portfolio optimization (Section \ref{sec:quantportfolioopt}), hedging and swap netting (Section \ref{sec:swapnetting}), optimal arbitrage (Section \ref{sec:optimalarbitrage}), credit scoring (Section \ref{sec:quantcreditscore}), and financial crash prediction (Section \ref{sec:fincrashes}).

Machine learning has become an essential aspect in modern business and research across various industries.
It has revolutionized  data processing and decision-making, empowering organizations large and small with unprecedented ability to leverage the ever-growing amount of easily accessible information.
Using quantum machine learning techniques can potentially speed up the training of the algorithm or certain parts of the whole process.
We will explore these techniques in Section \ref{sec:ML}.
In finance, some of the most common use cases for machine learning are anomaly detection (Section \ref{sec:anomalydetection}), natural language modeling (Section \ref{sec:nlp}), asset pricing (Section \ref{sec:assetpricing}), and implied volatility calculation (Section \ref{sec:impliedvol}).
Applications of quantum machine learning in these use cases will be discussed in the respective sections.

In Table \ref{table:fintable} we present a summary of the quantum techniques and their applicable financial use cases covered in this survey.

\begin{center}

\begin{table}[h]

\caption{
Financial Use Cases with Corresponding Classical and Quantum Solutions}
\label{table:fintable}
\begin{tabular}{p{0.14\textwidth}p{0.23\textwidth}p{0.23\textwidth}p{0.27\textwidth}}

\RaggedRight{\textbf{Problem Category}} & \RaggedRight{\textbf{Example Use Cases}} & \RaggedRight{\textbf{Classical Solutions}} & \RaggedRight{\textbf{Quantum Solutions}}\\ \hline

\RaggedRight{\small{Stochastic Modeling}}
& \RaggedRight{\small{
    Derivative Pricing (Section \ref{sec:quantumderivspricing}),
    \newline Risk Analysis (Section \ref{sec:quantumriskanaly})
}}
& \RaggedRight{\small{
    Monte Carlo Integration,
    \newline Numerical PDE Solver,
    \newline  Machine Learning
}}
& \RaggedRight{\small{
    Quantum Monte Carlo Integration (Section \ref{sec:simulation and pricing}),
    \newline Quantum PDE Solver (Section \ref{sec:quantum_pde}),
    \newline Quantum Machine Learning (Section \ref{sec:ML})
}} \\ \hline

\small{Optimization}
& \RaggedRight{\small{
    Portfolio Optimization (Section \ref{sec:quantportfolioopt}),
    Hedging, Swap Netting (Section \ref{sec:swapnetting}),
    Optimal Arbitrage (Section \ref{sec:optimalarbitrage}),
    Credit Scoring (Section \ref{sec:quantcreditscore}),
    and Financial Crash Prediction (Section \ref{sec:fincrashes})
}} 
& \RaggedRight{\small{
    Branch-and-Bound (with cutting -planes, heuristics, etc.) for non-convex cases \cite{wolsey1999integer}
    \newline and Interior-Point Methods for certain convex cases \cite{boyd2004convex}
}}
& \RaggedRight{\small{
    Quantum Optimization (Section \ref{sec:optimization})
}}  \\ \hline

\small{Machine Learning}
& \RaggedRight{\small{
    Anomaly Detection (Section \ref{sec:anomalydetection}),
    \newline Natural Language Modeling (Virtual Agents, Analyzing Financial Documents, Section \ref{sec:nlp}),
    \newline Risk Clustering (Section \ref{sec:clustering})
}}
& \RaggedRight{\small{
    Deep Learning,
    \newline Cluster Analysis
}}
& \RaggedRight{\small{
    Quantum Machine Learning (Section \ref{sec:ML}),
    \newline Quantum Cluster Analysis (Sections \ref{sec:kmeansclust} and \ref{sec:speccluster})
}}  \\ \hline
\end{tabular}
\end{table}

\end{center}

\section{Quantum Computing Concepts}
\label{sec:quantumconcepts}
Quantum computing is an emerging and promising field of science~\cite{alexeev2021prx}. Global venture capital funds, public and private companies, and governments have invested millions of dollars in this technology. Quantum computing is driven by the need to solve computationally hard problems efficiently and accurately. 
For decades, the advancement of classical computing power has been remarkably following the well-known Moore's law. Initially introduced by Gordon Moore in 1965, Moore's law forecasts that the number of transistors in a classical computer chip will double every two years, resulting in lower prices for manufacturers and consumers. 
Today, transistors are already at the size at which quantum-mechanical effects become apparent and problematic \cite{kumar2017cmos}. These effects will only become more difficult to manage as classical chips decrease in size. This problem is partially circumvented by alternative forms of computing that do not use conventional transistors, such as photonic computing, neuromorphic computing, biocomputing, and special-purpose chips. However, these techniques still use classical algorithms and are subject to the scaling of them.
Thus, it is imperative to investigate quantum computing, which provides scaling advantages no classical architectures can achieve. 

Google has demonstrated an enormous computational speedup using quantum computing for the task of simulating the output of pseudo-random quantum circuits \cite{arute2019quantum}. Specifically, Google claimed that what takes its quantum device, Sycamore, 200 seconds to accomplish
would take Summit, one of the most powerful supercomputers, approximately 10,000 years. Thus,  \emph{quantum supremacy} has been achieved. As defined by John Preskill~\cite{preskill2012quantum}, quantum supremacy is  the promise that certain computational  (not necessarily useful) tasks can be executed exponentially faster on a quantum processor than on a classical one. The decisive demonstration of quantum devices to solve more useful problems faster and/or more accurately than classical computers, such as large-scale financial problems (Section \ref{sec:financial_problems}), is called \emph{quantum advantage}. Quantum advantage is more elusive and, arguably, has not been demonstrated yet. The main challenge is that existing quantum hardware does not yet  seem capable of running algorithms on large enough problem instances. Current quantum hardware (Section \ref{sec:currentquantumhardware}) is in the \emph{noisy intermediate-scale quantum} (NISQ) technology era \cite{2018preskillnisq}, meaning that the current quantum devices are underpowered and suffer from multiple issues. The \emph{fault-tolerant} era refers to a yet unknown period in the future in which large-scale quantum devices that are robust against errors will be present. In the next two subsections we briefly discuss quantum information (Section \ref{sec:quantuminfo}) and models for quantum computation (Section \ref{sec:modelsofquantum}). In the last subsection (Section \ref{sec:currentquantumhardware}) we give an overview of the current state of quantum hardware and the challenges that must be overcome.

\subsection{Fundamentals of Quantum Information}
\label{sec:quantuminfo}
All computing systems rely on the fundamental ability to store and manipulate information. Today's classical computers operate on individual bits, which store information as binary $0$ or $1$ states.  In contrast, quantum computers use the physical laws of quantum mechanics to manipulate information. At this fundamental level, the unit of information is represented by a \textit{quantum bit}, or \textit{qubit}. Physically, a qubit is any two-level quantum system \cite{shankar2012principles, nielsen2002quantum}. Mathematically, the state space of a single qubit can be associated with the complex projective line, denoted $\mathbb{C}\text{P}^1$ \cite{veblen1918projective}.\footnote{Generally, quantum states live in a projective Hilbert space, which can be infinite-dimensional \cite{shankar2012principles}.} However, one commonly  considers qubit states as elements $\psi$, called \textit{state vectors}, of a two-dimensional complex-vector space but restrict consideration to those that satisfy $\lVert\psi\rVert_2 = 1$ and allow for $\psi$ and $e^{i\theta}\psi$ to be used interchangeably (i.e., consider specific elements of the equivalence classes in $\mathbb{C}\text{P}^1$). A state vector is usually denoted by using Dirac's ``bra-ket'' notation: $\psi$ is represented by the ``ket''  $\ket{\psi}$. Examples of two single-qubit kets are the states $\ket{0}$ and $\ket{1}$, which are analogous to the classical bits $0$ and $1$. 

Multiqubit state spaces are expressed through the use of the vector space tensor product \cite{roman2005advanced}, which results in a $2^n$-dimensional complex vector space for $n$ qubits. The tensor product of two state vectors is denoted as $\ket{\psi_1}\otimes\ket{\psi_2}$, $\ket{\psi_1}\ket{\psi_2}$, or $\ket{\psi_1\psi_2}$ and extends naturally to multiple qubits. \emph{Entanglement} is a quantum phenomenon that can be present in multiqubit systems in which the states of the subsystems \emph{cannot} be described independently. This results in correlated observations (measurement results) even when physical communication between qubits, which would correlate them in a classical sense, is impossible \cite{nielsen2002quantum}. Entangled states are mathematically expressed as tensors that are not factorable over the involved subsystems, in other words, non-simple tensors. A quantum state is a  \emph{product state} of certain subsystems if those subsystems \emph{can} be described independently and it is a simple tensor with respect to those subsystems. Qubits are in a \emph{superposition} state with respect to a given basis for the state space if their state vector can be expressed as a nontrivial linear combination of basis states. Thus, a qubit can be in a state that is an arbitrary superposition of $\ket{0}$ and $\ket{1}$, whereas a classical bit can  be in only one of the analogous states $0$ or $1$. The coefficients of the superposition are complex numbers called \emph{amplitudes}. Although multiplying a state by a phase $e^{i\theta}$, called a \emph{global phase}, has no physical significance, the \emph{relative phase} between the amplitudes of two basis states in a superposition does. 

A \emph{measurement} in quantum mechanics consists of probing a system to obtain a numerical result. Measurement of a quantum system is probabilistic. A \emph{projective measurement} of a system with respect to a \emph{Hermitian} operator $A$, called an \emph{observable}, results in the state vector of the system being orthogonally projected onto an eigenspace, with orthogonal projector $\Pi_{\lambda}$, and the observable quantity is the associated eigenvalue, $\lambda$. A potential projective measurement result $\lambda$ is observed with probability equal to $\lVert\Pi_{\lambda}\ket{\psi}\rVert_2^2$. The \emph{expected value} of the measurement is equal to $\expval{A}{\psi}$, where $\bra{\psi}$, called a ``bra,'' is the Hermitian adjoint. In physics, the quantum \emph{Hamiltonian} is the observable for a system's energy. A \emph{ground state} of a quantum Hamiltonian is a state vector in an eigenspace associated with the smallest eigenvalue and thus has the lowest energy. Any physical transformation of a quantum system can be represented by a completely positive non-trace increasing linear operator. Two important special cases of such operations are \emph{unitary} operators and measurements, which are not unitary. The dynamics of a closed quantum system follow the Schr\"{o}dinger equation, $i\hbar\ket{\dot{\psi}(t)}=H\ket{\psi(t)}$, where $\hbar$ is the reduced Planck constant and $H$ is the system's quantum Hamiltonian. The unitary operator  $ e^{-iH\Delta/\hbar}$, which transforms a solution $\ket{\psi(t)}$  to the Schr\"{o}dinger equation at one point in time $t$ into another one at a different point in time $t + \Delta$, is called the \emph{time-evolution} operator or propagator. There can also be a \emph{time-dependent Hamiltonian}, $H(t)$, where the operator $H$ changes over time. However, the propagator is computed by  the product integral \cite{dallard1979product} to take into account that $H(t_i)$ and $H(t_j)$ may not commute for all $t_i,t_j$ in the specified evolution time interval  \cite{shankar2012principles}. The simulation of time evolution on a quantum computer is called \emph{Hamiltonian simulation}. For quantum algorithms, the Hamiltonian can be an arbitrary observable with no connection to a physical system, and $\hbar:=1$ is often assumed. Three important observables on a single qubit are the Pauli operators: $\mathsf{X} := \ketbra{+}{+} - \ketbra{-}{-}$, $\mathsf{Y}:=\ketbra{i}{i}-\ketbra{-i}{-i}$, and $\mathsf{Z}:=\ketbra{0}{0}-\ketbra{1}{1}$, where $\ket{\pm}:=\frac{\ket{0}\pm\ket{1}}{\sqrt{2}}$, and $\ket{\pm i}:=\frac{\ket{0}\pm i\ket{1}}{\sqrt{2}}$, where $\ketbra{\psi}$ denotes the vector outer product.\footnote{The Pauli operators are also sometimes denoted as $\sigma_x$, $\sigma_y$, and $\sigma_z$, respectively.}
For $n$ qubits, the set $\{\ket{k}~|~k \in \{0, 1\}^n\}$ forms the \emph{computational basis}. \emph{Measuring in the computational basis} probabilistically projects the quantum state onto a computational basis state. A measurement in the context of quantum computation, without further clarification, usually refers to a measurement in the computational basis.

A positive semi-definite operator $\rho$ on the quantum state space with $\text{Tr}(\rho)=1$ is called a \emph{density operator}, where $\mathrm{Tr}(\cdot)$ is the trace operator.
The classical probabilistic mixture of \emph{pure states} $\ket{\psi_i}$, in other words, state vectors, with probabilities $p_i$ has density operator $\rho := \sum_ip_i\ketbra{\psi_i}{\psi_i}$ and is called a \emph{mixed state}. In general, a density operator $\rho$ represents a mixed state if $\text{Tr}(\rho^2) < 1$ and a pure state if $\text{Tr}(\rho^2) = 1$. 
\emph{Fidelity} is one of the distance metrics for quantum states $\rho$ and $\sigma$, defined as $F(\rho, \sigma) = \text{Tr}(\sqrt{\rho^{1/2}\sigma\rho^{1/2}})$ \cite{nielsen2002quantum}. The \emph{fidelity of an operation} refers to the computed fidelity between the quantum state that resulted after the operation was performed and the expected state. \emph{Decoherence} is the loss of information about a quantum state to its environment, due to interactions with it. Decoherence constitutes an important source of error for near-term quantum computers.

For further details, the  works of Nielsen and Chuang \cite{nielsen2002quantum} and Kitaev et al.~\cite{kitaev2002classical} are the standard references on the subject of quantum computation. The former  covers a variety of topics from quantum information theory, while the latter focuses on quantum computational complexity theory.

\subsection{Models of Quantum Computation}
\label{sec:modelsofquantum}
A qubit-based\footnote{There are other models of quantum computation that are not based on qubits, such as $n$-level quantum systems, for $n >2$, called \emph{qudits} \cite{2020qudits}, and continuous-variable, infinite-dimensional systems \cite{buck2021continuous}. This article  discusses only quantum models and algorithms based on qubits.} \emph{model of quantum computation} is, simply, a methodology for performing computations on qubits.
A model can realize \emph{universal} quantum computation if it can be used to approximate any $n$-qubit unitary, for all $n$, with arbitrarily small error \cite{nielsen2002quantum, wang2021comparative}. Two universal models are polynomial-time equivalent if they can simulate each other with only a polynomially-bounded overhead in computational resources \cite{aharonov2008adiabatic, wang2021comparative}.
We briefly discuss two polynomial-time-equivalent models that can be used to realize universal quantum computation: gate-based quantum computing and adiabatic quantum computing. There are others based on qubits \cite{wang2021comparative} that we will not review since current quantum devices do not implement them, even in a non-universal fashion. 

When running an algorithm on quantum hardware, one must contend with errors that can occur during computation. In the near term, \emph{quantum error mitigation} techniques are commonly used to reduce (mitigate) errors, rather than eliminate them. However, error correction will be required in the long term. Techniques for \emph{quantum error correction} \cite{lidar2013quantum} or fault tolerance have been developed to construct \emph{logical} qubits and operations that tolerate errors. This strategy has been extensively studied for gate-based quantum computing \cite{2012surfacecode}.

\subsubsection{Gate-Based Quantum Computing}
\label{sec:Universal Gate}

Gate-based quantum computing, also known as the quantum circuit model, composes a countable sequence of unitary operators, called \emph{gates}, into a \emph{circuit} to manipulate qubits that start in a fixed state \cite{nielsen2002quantum,kitaev2002classical}. A circuit can be viewed as a directed acyclic graph tracking the operations applied, which qubits they operate on, and the time step at which they are applied. Typically, the circuit ends with measuring in the computational basis. However, measurements of subsets of qubits before the end of the circuit and operations conditioned on the results of measurements are also possible \cite{rattew2021efficient}. Sequences of gates that act on separate sets of qubits can, if hardware supports it, be applied in parallel. The \emph{circuit depth} is the longest sequence of gates in the circuit, from initial state until measurement, that cannot be made shorter through further parallelization.
The quantum circuit model is analogous to the Boolean circuit model of classical computation \cite{kitaev2002classical}.
An important difference with quantum computation is that because of unitarity all gates are invertible, i.e., computation involving only quantum gates is reversible. A \emph{shot} is a single execution of a quantum circuit followed by measurements. Multiple shots are required if the goal is to obtain statistics.

This model can be utilized to realize universal quantum computation by considering circuits with $n$ qubits built by using any type of single-qubit gate and only one type of two-qubit gate \cite{nielsen2002quantum}. Any $n$-qubit unitary can be approximated in this manner to arbitrarily small error. The two-qubit gates can be used to mediate entanglement. A discrete set of gates that can be used for universal quantum computation is called a \emph{basis gate set}. The basis gate set realized by a quantum device is called its \emph{native gate set}. The native gate set typically consists of a finite number of (parameterized and non-parameterized) single-qubit gates and one two-qubit entangling operation. An arbitrary single-qubit gate can be approximately decomposed into the native single-qubit gates. 

Some common single-qubit gates are the Hadamard gate ($\mathsf{H}$), the Phase ($\mathsf{S}$) and $\frac{\pi}{8}$-Phase ($\mathsf{T}$) gates, the Pauli gates ($\mathsf{X}$, $\mathsf{Y}$, $\mathsf{Z}$), and the rotation gates  generated by Pauli operators ($\mathsf{R}_\mathsf{X}(\theta), \mathsf{R}_\mathsf{Y}(\theta), \mathsf{R}_\mathsf{Z}(\theta)$). Examples of two-qubit gates are the controlled-NOT gate denoted $\mathsf{CX}$ or $\mathsf{CNOT}$ and the $\mathsf{SWAP}$ gate. The $\mathsf{H}$, $\mathsf{S}$, and $\mathsf{CX}$ gates generate, under the operation of composition, the \emph{Clifford group}, which is the normalizer of the group of Pauli gates \cite{nielsen2002quantum}.
The \emph{time complexity} of an algorithm is measured in terms of the depth of the circuit implementing it. This is in terms of native gates when running the circuit on hardware. The majority of commercially available quantum computers implement the quantum circuit model. 

We note that quantum circuits generated for gate-based quantum computers can also be simulated on classical hardware. It is generally a computationally inefficient process, however, with enormous memory requirements scaling exponentially with the number of qubits~\cite{Wu2019,wu2019full,wu2018memory,wu2018amplitude}. But for certain types of computations such as computing the expectation values of observables, these requirements can be dramatically reduced by using tensor network simulators~\cite{lykov2021large,LykovGPU,lykov_diagonal}. Classical simulation is commonly used for the development of algorithms, verification, and benchmarking, to name a few applications.

\subsubsection{Adiabatic Quantum Computing}
\label{sec:AQC}
Adiabatic quantum computing (AQC) is a model for quantum computation that relies on the quantum \emph{adiabatic theorem}. This theorem states that as long as the evolution of a time-dependent Hamiltonian, $H(t)$, is sufficiently slow (defined in references) from an initial $n$th eigenstate of the time-dependent Hamiltonian at $t=0$, it will remain in an instantaneous $n$th eigenstate throughout the evolution for all $t$ \cite{Albash_2018}. A unitary operation can be applied by following an adiabatic trajectory from an initial Hamiltonian $H_i$ whose ground state is easy to prepare (e.g., a product state of qubits) to the final Hamiltonian $H_f$ whose ground state is the result that would be obtained after applying the unitary. A computation consists of evolving a time-dependent Hamiltonian $H(s)$ for time $t_f$ according to a schedule $s(t) : [0, t_f] \mapsto [0, 1]$, which interpolates between the two (non-commuting) Hamiltonians: $H(s) = (1-s)H_i + sH_f$ \cite{Albash_2018}. The time required for an adiabatic algorithm to produce the result to some specified error is related to the minimum \emph{spectral gap} among all instantaneous Hamiltonians. The spectral gap of the Hamiltonian at time $t$ is the difference between the $n$th eigenvalue and the next closest one (e.g., difference between the ground-state energy and the first excited level). This quantity helps define how slow the evolution needs to be in order to remain on the adiabatic trajectory---although it is notoriously difficult to bound in practice \cite{Albash_2018}. One example of where this has been done is for the adiabatic version of Grover's algorithm (Section \ref{sec:grovers}) \cite{Roland_2002}. There exist classes of time-dependent Hamiltonians that through adiabatic evolution can approximate arbitrary unitary operators applied to qubits \cite{Albash_2018, 2008Biamonte}. Thus AQC can realize universal quantum computation. As of this writing, there does not exist a commercially available device for universal AQC.  By polynomial equivalence to the circuit model, however, gate-based devices can efficiently simulate it. There are commercially available devices that implement a specific AQC algorithm for combinatorial optimization, called \emph{quantum annealing}, discussed in Section \ref{sec:aqoalgo}. It is believed to be unlikely that the transverse-field Ising (stoquastic \cite{Albash_2018}) Hamiltonian used for quantum annealing is universal \cite{2008Biamonte}.

\subsection{Quantum Hardware Challenges}
\label{sec:currentquantumhardware}
As mentioned, the quantum computers available today are called NISQ devices. Multiple physical realizations of qubit-based quantum computation  have been developed over the past decade. The majority of efforts to develop quantum technologies at this time are driven by industry and, to a lesser degree, by academia. Superconducting-based quantum computers \cite{superconductingengineers} are manufactured by IBM, Google, D-Wave, and Rigetti, among others.  Quantum computers using trapped atomic ions \cite{trappedion} have been developed, for example, by IonQ, Quantinuum, and AQT. These two technologies are currently the most widely available and thus, have been frequently utilized for research. However, other promising technologies also  are under development, including  photonic systems by PsiQuantum and Xanadu; neutral-atoms by ColdQuanta, QuEra Computing, Atom Computing and Pasqal, spins in silicon by Silicon Quantum Computing; quantum dots; molecular qubits; and topological quantum systems. As mentioned in Section \ref{sec:modelsofquantum}, most of the current quantum devices and those that are in development, with the exception of the D-Wave quantum annealers, follow the quantum circuit model. All of the mentioned quantum devices have system-level attributes, which need to be considered in multiqubit systems. Significant scientific and engineering efforts are  needed to improve and optimize these devices. The most important attributes are discussed in the remainder of this section.

The technical challenges of NISQ devices discussed below have a major impact on the current state of quantum algorithms. The algorithms can be roughly split into two camps: ($1$) near-term algorithms designed to run on NISQ devices and ($2$) algorithms that have a theoretically proven advantage for when hardware advances enough but require a large number of logical qubits with extremely high-fidelity quantum gates. In this work we cover both types of algorithms. One should keep in mind that arguably none of the discussed algorithms implemented on NISQ devices provide a decisive advantage over classical algorithms yet.

    \subsubsection{Noise} Qubits lose their desired quantum state or decohere (Section \ref{sec:quantuminfo}) over time. The decoherence times of each of the qubits are important attributes of a quantum device. Various mechanisms of qubit decoherence exist, such as quantum amplitude and phase relaxation processes and depolarizing quantum noise \cite{nielsen2002quantum}.  One potentially serious challenge for superconducting systems is cosmic rays~\cite{vepsalainen2020impact}. While single-qubit decoherence is relatively well understood, the multiqubit decoherence processes, generally called \textit{crosstalk}, pose more serious challenges \cite{2020crosstalk}. Even two-qubit operations have an order of magnitude higher error rate than do single-qubit operations. This makes it difficult to entangle a large number of qubits without errors. Various error-mitigation techniques \cite{2021bravyierrormitigation, 2010dynamicdecoupling} have been developed for the near term. In the long term, quantum error correction using logical operations, briefly mentioned in Section \ref{sec:modelsofquantum}, will be necessary \cite{kitaev2003fault, 2012surfacecode}. However, this requires multiple orders of magnitude more physical qubits than available on today's NISQ systems, and native operations must have sufficiently low error rates. Thus, algorithms executed on current hardware must contend with the presence of noise and are typically limited to low-depth circuits. In addition, current quantum error correction techniques theoretically combat local errors \cite{kitaev2003fault, 2012surfacecode, 2020crosstalk}. The robustness of these techniques to non-local errors is still being investigated \cite{fowler2014quantifying}.

    \subsubsection{Connectivity} Quantum circuits need to be optimally mapped to the topology of a quantum device in order to minimize errors and total run time. With current quantum hardware, qubits can  interact only with neighboring subsets of other qubits. Existing superconducting and trapped-ion processors have a fixed topology that defines the allowed connectivity. However, trapped-ion devices can change the ordering of the ions in the trap to allow for, originally, non-neighboring qubits to interact \cite{pino2021demonstration}. Since this approach utilizes a different degree of freedom from that used to store the quantum information for computation, technically the device can be viewed as having an all-to-all connected topology. For existing superconducting processors, the positioning of the qubits cannot change. As a result, two-qubit quantum operations between remote qubits have to be mediated by a chain of additional two-qubit $\mathsf{SWAP}$ gates via the qubits connecting them. Moreover, the two-qubit gate error of current NISQ devices is high. Therefore,  quantum circuit optimizers and quantum hardware must be developed with these limitations in mind. Connectivity is also a problem with current quantum annealers, which is discussed in Section \ref{sec:QuantumAnnealing}.

    \subsubsection{Gate Speed} Having fast quantum gates is important for achieving quantum supremacy and quantum advantage with NISQ devices in the quantum circuit model. However, some types of quantum devices are particularly slow, for example trapped-ion quantum processors, compared with superconducting devices, although these devices typically have lower gate error rates. There is a well-known trade-off between space, speed, and fidelity. Execution time is particularly relevant for algorithms that require a large number of circuit execution repetitions, such as variational algorithms (Section \ref{sec:qva}) and sampling circuits.
     Error rates typically  increase sharply if the gate time is reduced below a certain duration. Thus, one must find the right balance, which often requires tedious calibrations and fine-tuning.
    \subsubsection{Native Gates} Another important attribute, specific to gate-based quantum devices, is the set of available quantum gates that can be executed natively, namely, those that map to physical operations on the system (Section \ref{sec:Universal Gate}). The existence of a diverse universal set of native gates is crucial for designing short-depth high-fidelity quantum circuits. Developing advanced quantum compilers is, therefore, critical for efficiently mapping general quantum gates to the native gates of a particular device.

\section{Foundational Quantum Algorithms}
\label{sec:quantum_algorithms}

In this section we discuss foundational quantum algorithms that are the building blocks for more sophisticated algorithms. These algorithms have been extended to address problems in different domains, particularly finance, as described in Sections \ref{sec:stochastic_modeling}, \ref{sec:optimization}, and \ref{sec:ML}. We briefly review the concepts from computational complexity theory that are relevant to the discussions in this survey. We refer the reader to the references for more rigorous discussions on the topic.

\emph{Computational complexity}, the efficiency in time or space with which computational problems can be solved, is presented by using the standard asymptotic ``Big-O'' notation \cite{cormen2009introduction}: the set $\mathcal{O}(f(n))$ for an upper bound on complexity (worst case) and the set $\Omega(g(n))$ for a lower bound on complexity (best case). Also,  $h(n) \in \Theta(f(n))$ if and only if $h \in \mathcal{O}(f(n))$ and $h \in \Omega(f(n))$.\footnote{Let $*$ signify  $\mathcal{O}$, $\Omega$, or $\Theta$. It is common practice to abuse notation and use the symbol $=$ instead of $\in$ to signify $g(n)$ is in $*(f(n))$ \cite{cormen2009introduction}. Similarly, it is common to interchange the phrases ``the complexity is $*(f(n))$'' and ``the complexity is in $*(f(n))$.''} Computational problems, which are represented as functions, are divided into \emph{complexity classes}. The most common classes are defined based on \emph{decision problems}, which can be represented as Boolean functions. $\mathsf{P}$ denotes the complexity class of decision problems that can be solved in \emph{polynomial time} (i.e., the time required is upper bounded, in ``Big-O'' terms, by a polynomial in the input size) by a deterministic Turing machine (TM), in other words, a traditional classical computer \cite{kitaev2002classical, nielsen2002quantum}. An amount of time or space required for computation that is upper bounded by a polynomial is called \emph{asymptotically efficient}, whereas, for example, an amount upper bounded by an exponential function is not. However, asymptotic efficiency does not necessarily imply efficiency in practice; the coefficients or order of the polynomial can be  large. $\mathsf{NP}$ denotes the class of decision problems for which proofs of correctness exist, called \emph{witnesses}, that can be executed in polynomial time on a deterministic TM \cite{kitaev2002classical}.

A problem is \emph{hard} for a complexity class if any problem in that class can be reduced in polynomial time to it, and it is \emph{complete} if it is both hard and contained in the class \cite{arora2009computational}. Problems in $\mathsf{P}$ can be solved efficiently on a classical or quantum device. While $\mathsf{NP}$ contains $\mathsf{P}$, it also contains many problems not known to be efficiently solvable on either a quantum or classical device. \#$\mathsf{P}$ is a set of counting problems. More specifically, it is the class of functions that count the number of witnesses that can be computed in polynomial time on a deterministic TM for $\mathsf{NP}$ problems \cite{arora2009computational}. $\mathsf{BQP}$, which contains $\mathsf{P}$, denotes the class of decision problems solvable in polynomial time on a quantum computer with bounded probability of error \cite{nielsen2002quantum}. A common way to represent the complexity of quantum algorithms is by using \emph{query complexity} \cite{ambainis2017understanding}. Roughly speaking, the problem setup provides the algorithm access to functions that the algorithm considers as ``black boxes.'' These are represented as unitary operators called \emph{quantum oracles}. The asymptotic complexity is computed based on the number of calls, or \emph{queries}, to the oracles.

The goal of quantum algorithms research is to develop quantum algorithms that provide \emph{ computational speedups}, a reduction in computational complexity. In practice, however, one commonly  finds algorithms with improved efficiency  without any proven reduction in complexity. These algorithms typically utilize \emph{heuristics}. As mentioned in Section \ref{sec:currentquantumhardware}, a majority of algorithms for NISQ fall into this category. When one  can theoretically show a reduction in asymptotic complexity, the algorithm has a \emph{provable speedup} for the problem. In the discussions that follow, we emphasize the category into which each algorithm currently falls.

\subsection{Quantum Unstructured Search}
\label{sec:grovers}
 Grover's algorithm \cite{grover1996fast} is a quantum procedure for solving unstructured search with an asymptotic quadratic speedup, in terms of the number of oracle queries made, over the best known classical algorithm. The goal of unstructured search is as follows: Given an oracle $f : \mathcal{X} \mapsto \{0, 1\}$, 
find $w \in \mathcal{X}$ such that $f(w) = 1$. Sometimes $w$ is also called a marked element. More specifically, the algorithm amplifies the probability of measuring the state $\ket{w}$ encoding $w$ such that $f(w) = 1$. In Grover's algorithm, a marked state is identified by utilizing a phase oracle, $O_f|x\rangle = (-1)^{f(x)}|x\rangle$, which can be a composition of an oracle computing $f$ into a register and a conditional phase gate.  One can represent $f$  by a unitary  utilizing techniques for reversible computation \cite{nielsen2002quantum}.  Classically, in the worst case $f$ will be evaluated on all items in the set, $\mathcal{O}(N)$ queries. Grover's algorithm makes $\mathcal{O}(\sqrt{N})$. The complexity stated for this algorithm is said to be in terms of quantum query complexity. In fact, this algorithm is optimal for unstructured search \cite{1997strengths}.

Along with an efficient unitary simulating $f$, the algorithm requires the ability to query uniform superpositions of states representing elements of $\mathcal{X}$. This criterion is typically referred to as \emph{quantum access to classical data} \cite{kerenidis2016quantum}. The general case can be achieved by using quantum random access memory (qRAM) \cite{Giovannetti_2008}. As of this writing, the problems associated with constructing a physical qRAM have not been solved. However, a uniform superposition over computational basis states (i.e., with Hadamard gates, Section \ref{sec:Universal Gate}) can be used for the simple case where $\mathcal{X} = \{0, \dots, N-1\}$. Moreover, hybrid techniques have been developed to deal with larger classical data sets without qRAM \cite{harrow2020small}.  Ambainis \cite{ambainis2010quantum} developed a variable-time version of the algorithm that more efficiently handles cases when queries can take different times to complete. We note that loading data onto a quantum computer is generally exponentially expensive. It is believed that a quantum computer may help with computing small but complex datasets, which are computationally easy to load onto a quantum computer.

\subsection{Quantum Amplitude Amplification}
\label{sec:QuantumAmplitudeAmplification}
The quantum amplitude amplification (QAA) algorithm \cite{QAE} is a generalization of Grover's algorithm (Section \ref{sec:grovers}).
Suppose  an oracle $O_{\phi}$ exists that marks a quantum state $\ket{\phi}$ (e.g., $O_{\phi}$ only acts nontrivially on $\ket{\phi}$, say, by adding a phase shift by $\pi$). Also, assume there is a unitary $U$ and $\ket{\Phi} := U\ket{\psi}$ for some input state $\ket{\psi}$.
In addition, if $\ket{\Phi}$ has nonzero overlap with the marked state $\ket{\phi}$, namely, $ \bra{\Phi}\phi\rangle \neq 0$,
then without loss of generality $\ket{\Phi}= \sin(\theta_a)\ket{\phi} + \cos(\theta_a){\ket{\phi^{\perp}}}$, where $\bra{\phi}\phi^{\perp}\rangle = 0$ and $\sin(\theta_a) = \bra{\Phi}\phi\rangle$. For example, in the case of Grover's algorithm,  $\ket{\Phi}$ is a uniform superposition over states corresponding to the elements of $\mathcal{X}$, and $\ket{\phi}$ is the marked state.
QAA will amplify the amplitude $\sin(\theta_a)$ and as a consequence the probability $\sin^2(\theta_a)$ of measuring $\ket{\phi}$ (using an observable with $\ket{\phi}$ as an eigenstate) to $\Omega(1)$\footnote{$\Omega(1)$ probability means the probability that the event occurs is at least some constant. This is used to signify that the desired output of the algorithm occurs with a probability that is independent of the variables in the algorithm's time complexity or query complexity. In addition, this implies an asymptotically efficient amount of repetitions (classical probabilistic boosting) can be used to increase the probability of success close to $1$ (e.g., see Section \ref{sec:qpesection}).} using $\mathcal{O}(\frac{1}{\sin(\theta_a)})$ queries to $U$ and $O_{\phi}$. This is done utilizing a series of interleaved reflection operators involving $U$, $\ket{\psi}$, and $O_\phi$ \cite{QAE}. 

This algorithm can be understood as a quantum analogue to classical probabilistic boosting techniques and achieves a quadratic speedup \cite{QAE}. Classically, $\mathcal{O}(\frac{1}{\text{sin}^2(\theta_a)})$ samples would be required, in expectation, to measure $\ket{\phi}$ with $\Omega(1)$ probability. QAA is an indispensable technique for quantum computation because of the inherit randomness of quantum mechanics.
Improvements to the algorithm have been  made to handle issues that occur when one cannot  perform the required reflection about $\ket{\psi}$ \cite{berry2014exponential},  the number of queries to make (which requires knowing $\sin(\theta_a)$) is not known \cite{1998tightbounds, grover2005fixed}, and $O_\phi$ is imperfect with bounded error \cite{2003searchboundederror}. The last two improvements are also useful for Grover's algorithm (Section \ref{sec:grovers}). A version also exists for quantum algorithms (i.e., a unitary $U$) that can be decomposed into multiple steps with different stopping times \cite{ambainis2010variable}.

\subsection{Quantum Phase Estimation}
\label{sec:qpesection}
The quantum phase estimation (QPE) algorithm \cite{ kitaev1995quantum} serves as a critical subroutine in many quantum algorithms such as HHL (Section \ref{sec:hhlsection}). Given a  desired additive error of $\delta$, 
QPE makes $\mathcal{O}(\frac{1}{\delta})$ queries to a unitary $U$ to produce an estimate $\Tilde{\theta}_j$ of the phase $\theta_j \in [0, 1)$ of one of the eigenvalues of $U$, $e^{i2\pi\theta_j}$,
such that $|\theta_j - \Tilde{\theta}_j| \leq \delta$. The probability of successfully measuring an estimate $\Tilde{\theta}_j$ to the actual eigenvalue phase $\theta_j$ within error $\delta$ is $\Omega(1)$ and can be boosted to $1- \epsilon$ by adding and then discarding $\mathcal{O}(\log(1/\epsilon))$ ancillary qubits \cite{nielsen2002quantum, Cleve_1998}. Alternatively, it can be boosted to $1 - \frac{1}{\text{poly}(n)}$ by using $\mathcal{O}(\log(n))$ repetitions of QPE and taking the most frequent estimate, according to standard Chernoff bounds \cite{prakash2014quantum}. This brings the overall query complexity to $\mathcal{O}(\frac{1}{\delta \epsilon})$ and $\mathcal{O}(\frac{\log(n)}{\delta})$, respectively. In the cases where the actual eigenvalue phase $\theta_j,$ can be represented with finite precision, QPE can perform the transformation $\ket{v_j}\ket{0} \mapsto \ket{v_j}\ket{\theta_j}$, when $\ket{v_j}$ is an eigenvector of $U$. A more detailed explanation of the algorithm and what happens when the eigenvalue phases cannot be perfectly represented with the specified precision, $\delta$, is provided by Nielsen and Chuang \cite{nielsen2002quantum}.

\subsection{Quantum Amplitude Estimation}
\label{sec:QuantumAmplitudeEstimation}
The quantum amplitude estimation (QAE) algorithm \cite{QAE} estimates the total probability of measuring states marked by a quantum oracle. Consider the state $\ket{\Phi}$, oracle $O_{\phi}$, unitary $U$, and amplitude $\sin(\theta_a)$ as defined in Section \ref{sec:QuantumAmplitudeAmplification}.
QAE utilizes $\mathcal{O}(\frac{1}{\delta \sin(\theta_a)}$) queries to $U$ and $O_{\phi}$ to estimate $\sin^2(\theta_a)$ to relative additive error $\delta \sin^2(\theta_a)$ \cite{prakash2014quantum}.
The algorithm utilizes QPE (Section \ref{sec:qpesection}) as a subroutine, which scales as $\mathcal{O}(\frac{1}{\delta})$, for desired additive error $\delta$. Some variants of QAE do not make use of QPE \cite{suzuki2020amplitude, Grinko_2021, giurgica2020low}, including a variational (Section \ref{sec:qva}) one \cite{varqae}, and are potentially more feasible on near-term devices. An important application of QAE to finance is to perform Monte Carlo integration, discussed in Section \ref{sec:simulation and pricing}. Other applications of QAE include estimating the probability of success of a quantum algorithm and counting. 

\subsection{Quantum Linear System Algorithms}
\label{sec:hhlsection}
The quantum linear systems problem (QLSP) is defined as follows Given an $\mathcal{O}(1)$--sparse invertible Hermitian matrix $A \in \mathbb{R}^{N\times N}$ and a vector $\vec{b} \in \mathbb{R}^N$, output a quantum state $\ket{x}$ that is the solution to $A\vec{x}=\vec{b}$ up to a normalization constant with bounded error probability \cite{dervovic2018quantum}. While this definition requires the sparsity to be independent of the dimensions of the matrix, the quantum linear system algorithms (QLSAs) have a polynomial dependence on sparsity. As can be understood by viewing the time complexities of these algorithms, we can allow for $\mathcal{O}(\text{polylog}(N))$--sparse matrices, which is what is meant when we simply say ``a sparse matrix'' \cite{Childs_2017}. The $\mathcal{O}(1)$-sparsity specification comes from the decision problem version of QLSP that is BQP-complete~\cite{dervovic2018quantum, harrow2009quantum}.

The Harrow--Hassidim--Lloyd (HHL) algorithm \cite{harrow2009quantum} is the first algorithm invented for solving the QLSP. It provides an exponential speedup in the system size $N$ for well-conditioned matrices (for QLSP this implies a condition number in $\mathcal{O}(\text{polylog}(N))$) \cite{aaronson2015read}, over all known classical algorithms for a simulation of this problem. For matrix sparsity $s$ and condition number $\kappa$, HHL runs with worst-case time complexity $\mathcal{O}(s^2\kappa^2\text{log}(N)/\epsilon)$, for desired error $\epsilon$. This complexity was computed under the assumption that a procedure based on higher-order Suzuki--Trotter methods \cite{Berry_2006, 2021trotter} was used for sparse Hamiltonian simulation. However,  a variety of more efficient techniques  have been developed since the paper's inception \cite{Berry_2015, Low_2017}. These potentially reduce the original HHL's quadratic dependence on sparsity. HHL's query complexity, which is independent of the complexity of  Hamiltonian simulation, is $\mathcal{O}(\kappa^2/\epsilon)$ \cite{costa2021optimal}. 

Alternative quantum linear systems solvers also  exist that have  better dependence on some of the parameters, such as almost linear in the condition number from Ambainis \cite{ambainis2010variable} and polylogarithmic dependence on $1/\epsilon$ for precision from Childs, Kothari, and Somma (CKS) \cite{Childs_2017}. In addition, Wossnig et al.~\cite{Wossnig_2018} utilized the quantum singular value estimation (QSVE) algorithm of Kerenedis and Prakash \cite{kerenidis2016quantum} to implement a QLS solver for dense matrices. This algorithm has $\mathcal{O}(\sqrt{N}\text{polylog}(N))$ dependence on $N$ and hence offers no exponential speedup. However, it still obtains a quadratic speedup over HHL for dense matrices. Following this, Kerenedis and Prakash generalized the QSVE-based linear systems solver to handle both sparse and dense matrices and introduced a technique for spectral norm estimation \cite{Kerenidis_2020LS}. In addition, the quantum singular value transform (QSVT) framework provides methods for QLS that have the improved dependencies on $\kappa$ and $1/\epsilon$ mentioned above \cite{chakraborty2018power, gribling2021improving}. The QSVT framework also provides algorithms for a variety of other linear algebraic routines such as implementing singular-value-threshold projectors \cite{gilyen2019, kerenidis2016quantum} and matrix-vector multiplication \cite{chakraborty2018power}. As an alternative to QLS based  on the QSVT, Costa et al.~\cite{costa2021optimal} devised a discrete-time adiabatic approach \cite{dranov1998discrete} to the QLSP  that has  optimal \cite{harrow2009quantum} query complexity: linear in $\kappa$ and logarithmic in $1/\epsilon$. Since QLS algorithms invert Hermitian matrices, they can also be used to compute the Moore--Penrose pseudoinverse of an arbitrary matrix. This requires computing the Hermitian dilation of the matrix and  filtering out singular values that are near zero \cite{harrow2009quantum, gilyen2019}.

While solving the QLSP does not  provide classical access to the solution vector, this can still be done by utilizing methods for vector-state tomography \cite{kerenidis2018quantum}. Applying tomography would no longer allow for an exponential speedup in the system size $N$. However, polynomial speedups using tomography for some linear-algebra-based algorithms are still possible (Section \ref{sec:convex}).
In addition, without classical access to the solution, certain statistics, such as the expectation of an observable with respect to the solution state, can still be obtained without losing the exponential speedup in $N$. Providing quantum access to $A$ and $\vec{b}$ is a difficult problem that can, in theory, be dealt with by using quantum models for data access. Two main models of data access for quantum linear algebra routines exist \cite{chakraborty2018power}: the sparse-data access model and the quantum-accessible data structure. The sparse-data access model provides efficient quantum access to the nonzero elements of a sparse matrix. 
The quantum-accessible data structure,  suitable for fixed-sized, non-sparse inputs, is a classical data structure stored in a quantum memory (e.g., qRAM) and provides efficient quantum queries to its elements \cite{chakraborty2018power}.

For problems that involve only low-rank matrices, classical Monte Carlo methods for numerical linear algebra (e.g., the FKV algorithm \cite{frieze2004fast}) can be used. These  techniques have been used to produce classical algorithms that have an asymptotic exponential speedup in dimension for various problems involving low-rank linear algebra (e.g., some machine learning problems) \cite{Tang_2019, Chia_2020}. As of this writing, these classical \emph{dequantized} algorithms have impractical polynomial dependence on other parameters, making the quantum algorithms still useful. In addition, since these results do not apply to sparse linear systems, there is still the potential for provable speedups for problems involving sparse, high-rank matrices.

\subsection{Quantum Walks}
\label{sec:quantumwalks}

Formally, random walks are discrete-time stochastic processes formed through the summation of independent and identically distributed random variables \cite{lawler2010random} and play a fundamental role in finance \cite{ black2019pricing}. In the limit, such discrete-time processes approach the continuous-time Wiener process \cite{klebaner2012introduction}. One type of stochastic process whose logarithm follows a Wiener process is geometric Brownian motion (GBM), commonly used to model market uncertainty.
Random walks have also been generalized to the vertices of graphs \cite{lovasz1993random}. In this case they can be viewed as finite-state Markov chains over the set of vertices \cite{norris1998markov}. 
Quantum walks are a quantum mechanical analogue to the processes mentioned above \cite{Kempe_2003}.
Generally, quantum walks in discrete time can be viewed as a Markov chain over its \emph{edges} (i.e., a quantum analogue of a classical bipartite walk) \cite{szegedy2004quantum, 2011Magniez} and consist of a series of interleaved reflection operators. This approach generalizes QAA (Section \ref{sec:QuantumAmplitudeAmplification}) and Grover's algorithm  (Section \ref{sec:grovers}) \cite{2003walksearchgrover, ambainis2004coins}. Discrete-time quantum walks have be shown to provide polynomially faster mixing times \cite{ambainis2001one, aharonov2001quantum, moore2002quantum} as well as polynomially faster hitting times for Markov-chain-based search algorithms \cite{szegedy2004quantum, ambainis2007quantum, 2011Magniez, apers2019unified, somma2007quantum, 2005ChildsSubset, ambainis2019quadratic}. Alternatively, there are quantum walks in continuous time \cite{Farhi_1998, Childs_2002}, which have certain provable advantages \cite{Childs_2003}, using a different notion of hitting time. A similar result, using this notion, has also been shown for discrete-time quantum walks \cite{kempe2002quantum}. In addition, the various mixing operators used in the Quantum Approximate Optimization Algorithm (QAOA, Section \ref{sec:QAOA}) can be viewed  as continuous-time quantum walks connecting feasible solutions \cite{ruan2020quantum}. There have been multiple proposed unifications of quantum walks in continuous and discrete time \cite{Childs_2009, 2012walkscomp}, something that is true for the classical versions \cite{klebaner2012introduction}. For a comprehensive review of quantum walks, see the survey by Venegas--Andraca \cite{2012walkscomp}; and for an overview and a discussion of connections between the various quantum-walk-based search algorithms, see the work of Apers et al. \cite{apers2019unified}.

\subsection{Variational Quantum Algorithms}
\label{sec:qva}
Variational quantum algorithms (VQAs) \cite{Cerezo_2021}, also known as classical-quantum hybrid algorithms, are a class of algorithms, typically for gate-based devices, that modify the parameters of a quantum (unitary) procedure based on classical information obtained from running the procedure on a quantum device. This information is typically in the form of a cost function dependent on the expectations of observables with respect to the state that the quantum procedure produces. Generally, the training of quantum variational algorithms is $\mathsf{NP}$-hard \cite{bittel2021training} which withholds us from reaching arbitrarily good approximate local minima. In addition, these methods are heuristic. The primordial quantum-circuit-based variational algorithm is called the variational quantum eigensolver (VQE) \cite{Peruzzo_2014} and is used to compute the minimum eigenvalue of an observable (e.g., the ground-state energy of a physically realizable Hamiltonian). However, it is also applicable to optimization problems (Section \ref{sec:vqe}) and both linear (e.g., quantum linear systems \cite{bravoprieto2020variational, huang2019nearterm}) and nonlinear problems (Section \ref{sec:var_pde}). VQE utilizes one of the quantum variational principles based on the Rayleigh quotient of a Hermitian operator (Section \ref{sec:vqe}). The quantum procedure that provides the state, called the ansatz, is typically a parameterized quantum circuit (PQC) built from Pauli rotations and two-qubit gates (Section \ref{sec:Universal Gate}). For VQAs in general, finding the optimal set of parameters is a non-convex optimization problem \cite{2021vqaloss}. A variation of this algorithm, specific to combinatorial optimization problems, known as the quantum approximate optimization algorithm (QAOA) \cite{farhi2014quantum}, utilizes the alternating operator ansatz \cite{Hadfield_2019} and is typically used for observables that are diagonal in the computational basis. There are also variational algorithms for approximating certain dynamics, such as real-time  and imaginary-time quantum evolution that utilize variational principles for dynamics \cite{Yuan_2019} (e.g., McLachlan's principle \cite{mclachlan1964variational}). Furthermore, variational algorithms utilizing PQCs and various cost functions have been applied to machine learning tasks (Section \ref{sec:qnn}) \cite{Mitarai_2018}. Problems using VQAs are seen as one of the leading potential applications of near-term quantum computing.

\subsection{Adiabatic Quantum Optimization}
\label{sec:aqoalgo}
Adiabatic quantum optimization \cite{farhi2000quantum}, commonly referred to as quantum annealing (QA), is an AQC (Section \ref{sec:AQC}) algorithm for quadratic unconstrained binary optimization (QUBO). QUBO is an $\mathsf{NP}$-hard combinatorial optimization problem. These problems are discussed in Section \ref{sec:combopt}. The main reason behind the interest in QA is that it provides a nonclassical heuristic, called quantum tunneling, that is potentially helpful for escaping from local minima. Its closest classical analogue is simulated annealing \cite{van1987simulated}, a Markov chain Monte Carlo method that was inspired by classical thermodynamics and uses temperature as a heuristic to escape from local minima.\footnote{Somma et al.~\cite{somma2007quantum} produced a quantum algorithm with provable speedup, in terms of the spectral gap, for simulated annealing using quantum walks (Section \ref{sec:quantumwalks}).} So far QA has proven useful for solving QUBOs with tall yet narrow peaks in the cost function \cite{2016narrowqa, Denchev_2016}. The overall benefit of QA is still a topic of ongoing research \cite{Denchev_2016, farhi2002quantum}. However, the current scale of quantum annealers allows for potential applicability in the near term \cite{Ajagekar_2020}.

\section{Stochastic Modeling}
\label{sec:stochastic_modeling}
Stochastic processes are commonly used to model phenomena in physical sciences, biology, epidemiology, and finance.
In finance, stochastic modeling is often used to help make investment decisions, usually with a goal of maximizing return and minimizing risks.
Quantities that are descriptive of the market condition, including stock prices, interest rates, and their volatilities, are often modeled by stochastic processes and represented by random variables.
The evolution of such stochastic processes is governed by stochastic differential equations (SDEs), and stochastic modeling aims to solve the SDEs for the expectation value of a certain random variable of interest, such as the expected payoff of a financial derivative at a future time, which determines the price of the derivative.

Although analytical solutions for SDEs are available for a few simple cases, such as the well-known Black--Scholes equation for European options \cite{black2019pricing}, a vast majority of financial models involve SDEs of more complex forms that have to resort to numerical approaches.

In the following subsections, we review two commonly used numerical methods for solving SDEs, namely, Monte Carlo integration (MCI, Section \ref{sec:simulation and pricing}) and numerical solutions of differential equations (ODEs/PDEs, Section \ref{sec:quantum_pde}), and we discuss respective quantum solutions and potential advantages.
Then, in Section \ref{sec:finapplications}, we show example financial applications for these quantum solutions.

\subsection{Monte Carlo Integration}
\label{sec:simulation and pricing}
Monte Carlo methods utilize sampling to approximate the solutions to problems that are intractable to solve analytically or with numerical methods that scale poorly for high-dimensional problems. Classical Monte Carlo methods have been used for inference \cite{neal1993probabilistic}, integration \cite{robertmontecarlo}, and optimization \cite{homem2014monte}. Monte Carlo integration (MCI), the focus of this subsection, is critical to finance for pricing and risk predictions \cite{glasserman2004monte}. 
These tasks often require large amounts of computational power and time to achieve the desired precision in the solution. Therefore, MCI is another key technique, heavily used in finance, for which a quantum approach is appealing. Interestingly, there exists a quantum algorithm with a proven advantage over classical Monte Carlo methods for numerical integration.

In stochastic modeling, MCI typically is used to estimate the expectation of a quantity that is a function of other random variables. 
The methodology usually starts with a stochastic model (e.g. SDEs) for the underlying random variables from which samples can be taken and the corresponding values of the target quantity subsequently evaluated given the samples drawn.
For example, consider the following sequence of random variables, $X_0, \dots, X_T$.
These random variables are often assumed to follow a diffusion process \cite{klebaner2012introduction}, where each $X_t$ with $t \in \{0, 1, \dots, T\}$ represents the value of the quantity of interest at time $t$, and the collective values of the random variables from the outcome of a single drawing are known as a sample path. Suppose the quantity whose expectation we want to compute is a function of this process at various time points: $g(X_0, \dots, X_T)$.
In order to estimate the expectation, many sample paths are drawn, and sample averages of $g$ are computed.
The estimation error decays as $\mathcal{O}(\frac{1}{\sqrt{N_s}})$ independent of the problem dimension, where $N_s$ is the number of paths taken. This is in accordance with the law of large numbers and Chebyshev's inequality \cite{robertmontecarlo}.

Quantum MCI (QMCI) \cite{brassard2011optimal, heinrich2002quantum, Montanaro_2015} provides a quadratic speedup for Monte Carlo integration by making use of the QAE algorithm (Section \ref{sec:QuantumAmplitudeEstimation}).\footnote{To disambiguate, there are classical computational techniques called \emph{quantum Monte Carlo} methods used in physics to classically simulate quantum systems  \cite{ceperley1986quantum}. These are not the topic of this discussion.}  With QAE, using the example from the previous paragraph, the error in the expectation computation decays as $\mathcal{O}(\frac{1}{N_q})$, where $N_q$ is the number of queries made to a quantum oracle that computes $g(X_0, \dots, X_T)$.  This is in contrast to the complexity in terms of the number of \emph{samples} mentioned earlier for MCI. Thus, if samples are considered as classical queries, QAE requires quadratically fewer queries than classical MCI requires in  order to achieve the same desired error.

The general procedure for Monte Carlo integration utilizing QAE is outlined below \cite{Chakrabarti_2021}. Let $\Omega$ be the set of potential sample paths $\omega$ of a stochastic process that is distributed according to $p(\omega)$, and $f : \Omega \mapsto A$ is a real-valued function on $\Omega$, where $A \subset \mathbb{R}$ is bounded. The task is to compute $\mathbb{E}[f(\omega)]$  \cite{Montanaro_2015}, which can be achieved with QAE in three steps as follows.

 \begin{enumerate}
    \item Construct a unitary operator $\mathcal{P}_l$ to load a discretized and truncated version of $p(\omega)$. The probability value $p(\omega)$ translates to the amplitude of the quantum state $\ket{\omega}$ representing the discrete sample path $\omega$. In mathematical form, it is
    \begin{equation}
        \mathcal{P}_l\ket{0} = \sum_{\omega \in \Omega}\sqrt{p(\omega)}\ket{\omega}.
    \end{equation}
    
    \item Convert $f$ into a normalized function $\tilde{f}: \Omega \mapsto [0, 1]$, and construct a unitary $\mathcal{P}_f$ that computes $\tilde{f}(\omega)$ and loads the value onto the amplitude of $\ket{\omega}$.~\footnote{In general, quantum arithmetic methods \cite{haner2018optimizing} can be used to load $\arcsin(\sqrt{\tilde{f}(\omega)})$ into a register and perform controlled rotations to load $\sqrt{\tilde{f}(\omega)}$ onto the amplitudes \cite{Chakrabarti_2021}.} The resultant state after applying $\mathcal{P}_l$ and $\mathcal{P}_f$ is
    \begin{equation}
        \label{eqn:probability_oracle}
        \mathcal{P}_f\mathcal{P}_l\ket{0} =\sum_{\omega \in \Omega}\sqrt{(1-\tilde{f}(\omega))p(\omega)}\ket{\omega}\ket{0} + \sqrt{\tilde{f}(\omega)p(\omega)}\ket{\omega}\ket{1}.
    \end{equation}
    
    \item Using the notation from Section \ref{sec:QuantumAmplitudeEstimation}, perform quantum amplitude estimation with $U = \mathcal{P}_f\mathcal{P}_l$ and an oracle, $O_{\phi}$, that marks states with the last qubit being $\ket{1}$. The result of QAE will be an approximation to $\sum\limits_{\omega \in \Omega} \tilde{f}(\omega)p(\omega) = \mathbb{E}[\tilde{f}(\omega)]$. This value can be estimated to a desired error $\epsilon$  utilizing $\mathcal{O}(\frac{1}{\epsilon})$ evaluations of $U$ and its inverse \cite{Montanaro_2015}. Then scale $\mathbb{E}[\tilde{f}(\omega)]$ to the original bounded range, $A$, of $f$ to obtain $\mathbb{E}[f(\omega)]$.
\end{enumerate}
Variants \cite{Montanaro_2015} of QMCI also exist that can be applied when $f(\omega)$ has bounded $L^2$ norm and bounded variance (both absolutely and relatively). The problem of estimating a multidimensional random variable is called multivariate Monte Carlo estimation. Quantum algorithms that obtain similar quadratic speedups in terms of error $\epsilon$ have been proposed by Cornelissen and Jerbi \cite{cornelissen2021}. They solved this problem in the framework of Markov reward processes, and thus it has applicability to reinforcement learning (Section \ref{sec:rl}).

As mentioned earlier (Section \ref{sec:grovers}), preparing an arbitrary quantum state is a difficult problem. In the case of QMCI, we need to prepare the state in \eqref{eqn:probability_oracle}. The Grover--Rudolph algorithm \cite{grover2002creating} is a promising technique for loading efficiently integrable distributions (e.g., log-concave distributions). It requires the ability to compute the probability mass over subregions of the sample space. However, it has been shown that when numerical methods such as classical MCI are used to integrate the distribution, QMCI does not provide a quadratic speedup when using this state preparation technique \cite{Herbert_2021}. 
Additionally, variational methods (Section \ref{sec:qva}) have been developed to load distributions \cite{Zoufal_2019}, such as normal distributions \cite{Chakrabarti_2021}. Geometric Brownian motion (GBM) is an example of a diffusion process whose solution is a series of log-normal random variables. These variables can be implemented by utilizing procedures for loading standard normal distributions, followed by quantum arithmetic \cite{haner2018optimizing, Chakrabarti_2021}. 

While GBM is common for modeling stock prices, it is often considered unrealistic \cite{Dupire94pricingwith} because it assumes a constant volatility. In addition, unlike realistic models, the SDE of GBM can be solved in closed form; that is, the distribution at every point in time is known, which avoids the need to use approximations involving the SDE to simulate the process. The local volatility (LV) model is one approach for accounting for a changing volatility by making it a function of the current price and time step. However, the corresponding SDE does not have a closed-form solution, and  hence this means numerical methods such as Euler--Maruyama schemes \cite{maruyama1954transition} need to be used to simulate paths of the SDE with discretized time steps. Kaneko et al.  \cite{Kaneko_2020} proposed quantum circuits implementing the required arithmetic for simulating the LV model, namely, the operation $\mathcal{P}_{l}$, based on improved techniques for simulating stochastic dynamics on a quantum device \cite{Miyamoto_2020}.  However, the discretized simulation of the SDE introduces additional error and complexities when computing expectations of functions of the stochastic process the SDE defines. Popular  classical approaches to this problem are known as  Multilevel Monte Carlo (MLMC) methods \cite{giles2008multilevel}, which can be used to approximately recover the error scaling of classical single-level MCI. An et al.~\cite{An_2021} proposed a quantum-enhanced version of MLMC that utilizes  QMCI as subroutine to compute all expectations involved. This allowed them to approximately recover the error scaling of  QMCI, and the approach  benefits from being applicable to a wider variety of stochastic dynamics beyond GBM and LV. 
With regard to implementing payoff functions, Herbert \cite{herbert2021quantum} developed a near-term technique for functions that can be approximated well by truncated Fourier series. Furthermore, if fault tolerance (Section \ref{sec:quantumconcepts}) is to be taken into account, advancements in the field of quantum error correction are required in order to realize quantum advantage \cite{Babbush_2021}. 

Also worth noting are non-unitary methods for efficient state preparation \cite{gomes2021adaptive,rattew2021efficient, rattewkoczor}. These methods are usually incompatible with QAE, however, because of the non-invertibility of the operations involved. The methods would be more useful in a broader context where the inverse of the state preparation operation is not required; this topic is out of the scope of this survey.

\subsection{Numerical Solutions of Differential Equations}
\label{sec:quantum_pde}

While Monte Carlo methods are widely used for modeling stochastic processes, other numerical methods are also commonly used to calculate properties of stochastic systems.
In fact, according to the Feynman--Kac formula \cite{kac1949distributions,feynman2005principle}, the expectation of certain random variables, whose evolution is described by SDEs, can be formulated as the solution to parabolic partial differential equations (PDEs).
This connection between SDEs and PDEs allows one to study stochastic processes using deterministic methods.
Specifically, this enables an alternative to Monte Carlo methods for studying SDEs. A prime example of an alternative to MCI, applicable to certain problems, is  the Black--Scholes PDE, which forms the keystone for pricing financial derivatives.

Classical numerical methods \footnote{For details on numerical methods for PDEs, see, for example,  \cite{grossmann2007numerical}.} for solving PDEs and ordinary differential equations (ODEs) usually require discretization of the domain on which solutions are sought.
Specifically, the finite difference method (FDM) solves a PDE by approximating the solution and its derivatives on a predefined grid in both space and time and advancing it in time on the grid.
Similarly, the finite volume method (FVM) employs a grid to divide the space into volumes, on which the function to be solved is integrated and the volume integrals are converted to surface integrals using the divergence theorem.
These surface integrals are then connected by the original PDE.
Another grid-based method, the finite element method (FEM), uses functions to form a basis of solutions on the subdomains divided by the grid.
These basis functions are then assembled into an approximate for the solution in the entire domain.

In addition to grid based methods mentioned above, spectral methods may also be used to solve PDEs.
Spectral methods \cite{shen2011spectral} expand the solution into a linear combination of basis functions in the entire domain, whose spatial derivatives can be computed exactly.
Although spectral methods do not require a grid when computing the spatial derivatives, numerical solution on a computer still requires evaluation of the functions on a grid.
Moreover, spectral methods are often expensive to implement for PDEs with nonlinear terms.
The reason is that spectral methods often require Fourier transforms, which have a computational cost of $O(N \log N)$, with $N$ being the number of points on which the solution is evaluated.
In contrast, grid-based spatial differentiation methods usually have a complexity of $O(N)$.

All  these techniques can easily become computationally intractable for complex and high-dimensional problems, since they often require extremely large grid sizes to achieve desired accuracy and numerical stability in the solution.
Specifically, these algorithms have complexities that scale at least linearly with the number of points $N$ on which the solution is to be evaluated and exponentially in the number of dimensions $d$ of the spatial variable.

On the other hand, quantum algorithms often address the same problem by representing a vector of size $N$ using $O(\log N)$ space, and operating on it in potentially $O(\mathrm{poly}(\log N))$ time.

\subsubsection{Quantum-linear-system-based algorithms}
\label{sec:qlsa_pde}
Since classical algorithms for numerical solutions of PDEs such as the ones mentioned above often resort to the solution of linear systems, a natural candidate for quantum algorithms for PDEs is to employ the quantum linear system algorithms  as mentioned in Section \ref{sec:hhlsection}.

Because  of the linear nature of QLS, QLSAs have been used to solve linear PDEs and ODEs \cite{cao2013quantum,berry2014high,berry2017quantum,childs2021high,montanaro2016quantum}.
These algorithms often consider differential equations with the following form:
\[
\mathcal{L}(u(\bm{x})) = f(\bm{x}),
\]
where $\bm{x} \in \mathbb{C}^d$ is a $d$-dimensional vector, $f(\bm{x}) \in \mathbb{C}$ is a scalar function that accounts for the inhomogeneity in the PDE, $u(\bm{x}) \in \mathbb{C}$ is the scalar function that we are solving, and $\mathcal{L}$ is a linear differential operator.
The differential equations are transformed into a set of linear equations by using the same approaches as the classical algorithms mentioned above  for computing numerical derivatives given a discretization. 
The resultant linear system can then be solved by QLSAs.
This approach will potentially give the PDE solver algorithm a complexity of $O(\mathrm{poly}(d, \log(1/\epsilon)))$, where $\epsilon$ is the error tolerance, which is an exponential improvement in $d$ compared with that of the best-known classical algorithms.

In particular, FDM-based quantum algorithms \cite{cao2013quantum,berry2014high,berry2017quantum,costa2019quantum,childs2021high} use quantum states $\ket{a} = \sum_{\bm{x}}u(\bm{x})\ket{\bm{x}}$ and $\ket{b} = \sum_{\bm{x}}f(\bm{x})\ket{\bm{x}}$ to encode the functions $u(\bm{x})$ and $f(\bm{x})$ (up to a normalization factor), with the $N$ computational basis states representing the $N$ discretized points in $\bm{x}$ and amplitudes proportional to the scalar values of the functions at the respective points.
A finite-difference scheme that defines how the derivatives are approximated using values from neighboring grid points is then used to convert $\mathcal{L}$ into an $N \times N$ matrix operator $A$ acting on $\ket{a}$, and hence the solution of the differential equations becomes a QLSP:
\[
A\ket{a} = \ket{b},
\]
where $\ket{a}$ is the solution we would like to obtain.

Similarly, quantum versions of FVM \cite{fillion2019simple} and FEM \cite{montanaro2016quantum} also use amplitude-encoded quantum states to represent the solution or the basis functions used to approximate the solution, and generate the matrix $A$ based on the relation of the grid points as defined by differential equations and the respective methods.
The same applies to quantum algorithms based on spectral methods, in which the discretization is done in the spectral space \cite{childs2021high}.

To extend the QLSA-based algorithms to the solution of nonlinear differential equations, linearization of the equations is usually required.
One approach is Carleman linearization  \cite{carleman1932application,kowalski1991nonlinear}, which approximates the nonlinear functions in an ODE with truncated Taylor expansions of them, so that the nonlinear ODE can be approximated by a system of linear ODEs of the different powers of the solution.
Specifically, Liu et al.~\cite{liu2021efficient} utilize Carleman linearization to convert dissipative quardratic ODEs into linear ODEs, which are then solved by QLSA.
The overall complexity achieved by this algorithm is $O(qT^2 \mathrm{poly}(\log T, \log d, \log 1/\epsilon)/\epsilon)$, where $T$ is the evolution time requested for the solution, $q$ measures the decay of the solution, and $d$ and $\epsilon$ are the same as defined earlier.

Another approach for implementing nonlinearity is to build multiple copies of the quantum state representing the solution, apply linear transformations in the expanded space to simulate the nonlinearities in the differential equations, and then trace out the additional copies to achieve the effective nonlinear transformation on a single copy of the quantum state \cite{leyton2008quantum,lloyd2020quantum}.
In particular, Lloyd et al. \cite{lloyd2020quantum} combined the methods for simulating the dynamics of the nonlinear Schr\"{o}dinger equation with QLSA-based quantum linear differential equation solvers to solve nonlinear ODEs with an $m$-th order polynomial nonlinear term.
It was shown that by using a sufficiently large number of copies, $n$, of the quantum state that scales quadratically with the evolution time $T$, the approximation error $\epsilon$ is suppressed by a factor of $1/n$.

While the aforementioned linearization approaches all involve approximations, the Koopman--von Neumann approach \cite{joseph2020koopman} is another linearization method that converts nonlinear ODEs to linear transport equations, and the conversion is exact.
Specifically, Jin et al.~\cite{jin2022quantum} showed that a system of $d$ nonlinear ODEs can be exactly converted into a $d+1$ dimensional linear transport PDE, which can be solved with QLSA-based linear PDE solvers.
Although this approach applies generally to all nonlinear ODEs, it was shown that the only possible quantum advantage one may get in the general case is when we have multiple ($M \gg 1$) initial data points on which an ensemble average of some physical variables are to be estimated.

In the same work, it is also demonstrated that for certain types of nonlinear PDEs such as the Hamilton--Jacobi and hyperbolic PDEs, an exact linearization exists using the level set formalism \cite{jin2003multi}, which converts the $d+1$-dimensional ($d$ spatial dimensions and time) nonlinear PDE to a $2d+1$- and a $d+2$-dimensional PDE, respectively.

For a general nonlinear PDE with $d+1$ dimensions, one may discretize the nonlinear PDE and convert it to a system of nonlinear ODEs \cite{jin2022quantum}, and then use the linearization techniques mentioned above to further convert the nonlinear ODEs to linear PDEs or ODEs.
Therefore, same as the approach for general nonlinear ODEs, this algorithm will have a quantum advantage only when $M$ is large for general nonlinear PDEs.

We note that in all of the QLSA-based algorithms discussed above, the output will be in the form of a quantum state, which is generally difficult to be accessed classically without losing information or voiding the overall quantum speedup.
Nevertheless, as discussed in Section \ref{sec:hhlsection}, one can still efficiently obtain certain statistics of the solution without classical access to the quantum output state.
This is usually the assumption of these QLSA-based algorithms.

\subsubsection{Variational algorithms}
\label{sec:var_pde}

In addition to QLSA-based PDE solvers, variational algorithms have  been developed for solving linear and nonlinear PDEs.
One approach is to transform the PDE to a Wick-rotated Schr\"{o}dinger equation with an imaginary time variable and then use variational quantum imaginary time evolution (VarQITE; see Section \ref{sec:varite}) to solve the equation along the imaginary time axis.
This approach has been demonstrated in solving linear PDEs such as the Black--Scholes equation \cite{fontanela2021quantum}, the Feynman--Kac equation, and the Ornstein--Uhlenbeck equation \cite{alghassi2021variational}, and nonlinear PDEs such as the Hamilton-Jacobi and the Bellman equation
 \cite{alghassi2021variational}.

In another approach proposed by Lubasch et al. \cite{lubasch2020variational}, a variational circuit is built to construct the quantum state representing the solution of the PDE in an amplitude encoding.
Nonlinearities are implemented by creating multiple copies of the variational circuit and hence the quantum state.
The algorithm then employs tensor networks to efficiently calculate the linear operations acting on the solution functions, and the variational parameters are optimized based on cost functions built on the expectations of these linear operators on the quantum state, obtained by measurements at the end of the circuit.

While these two variational algorithms  both encode the solution into a quantum state such that the computational basis states denote the discretized values of the spatial variable and the amplitudes are proportional to the corresponding values of the function, Kyriienko et al. \cite{kyriienko2021solving, kyriienko2022protocols} proposed a different approach to encode the function $u(\bm{x})$.
Quantum circuits in this approach start with a quantum feature map in which values of the function variables $\bm{x}$ are used as rotation angles in the circuit.
The feature map is followed by a variational circuit parameterized by $\bm{\theta}$, which produces the quantum state $\ket{u_{\bm{\theta}}(\bm{x})}$.
The encoded function value $u(\bm{x})$ is then given by the expectation value of a cost operator $\mathcal{C}$, namely,
\[
u(\bm{x}) = \bra{u_{\bm{\theta}} (\bm{x})}\mathcal{C} \ket{u_{\bm{\theta}}(\bm{x})}.
\]
Derivatives of $u$ may be approximated by shifting the $\bm{x}$ parameters following a finite-difference scheme or computed exactly by using quantum circuit differentiation with respect to $\bm{x}$.
Using the outputs of the quantum circuits, a classical loss function is built to quantify how well the solution satisfies the PDE, and hence the best solution can be obtained by minimizing the loss function through adjusting $\bm{\theta}$ in the variational circuit.

Paine et al. \cite{Paine2021} utilized the differentiable quantum circuit approach for simulating the quantile function of an SDE. The quantile function, whose evolution satisfies a PDE \cite{steinbrecher2008quantile}, is the inverse of the cumulative distribution function of the solution to the SDE and can be used for generating samples \cite{gentle2003random}. Kubo et al. \cite{Kubo_2021} considered the trinomial-tree model \cite{Boyle1986OptionVU} for SDEs. The evolution of the probability distribution can be represented as the solution to the Schr\"{o}dinger equation with a non-Hermitian Hamiltonian and thus simulated with VarQITE. The authors also proposed methods for computing expectation values of functions of the SDE variables.

\subsection{Financial Applications}
\label{sec:finapplications}
Two main financial applications  can be solved via the quantum algorithms presented for stochastic modeling: derivative pricing and risk modeling.
\subsubsection{Derivative Pricing}
The mathematical framework used for pricing financial assets uses techniques from measure-theoretic probability theory \cite{follmer2016stochastic}. Specifically, the market is modeled as a filtered probability space $(\Omega, \mathfrak{F}, \mathbb{P})$, where the sample space $\Omega$ corresponds to potential states of the market, the filtration $\mathfrak{F} := \{\mathcal{F}_t\}_{t=0}^{\infty}$ corresponds to information and events that become observable over time (i.e., sub-$\sigma$-algebras), and $\mathbb{P}$ is the real-world market probability measure. The time $t$ can be continuous or discrete. Nondeterministic financial quantities such as stock prices and interest rates are represented as $\mathfrak{F}$-adapted stochastic processes. Powerful mathematical properties can be derived when the market is assumed to be arbitrage free and complete. An arbitrage-free market is one in which there does not exist a riskless profit with no initial investment. The arbitrage-free assumption is equivalent to the existence of at least one probability measure, equivalent to the real-world one $\mathbb{P}$, under which the discounted (stochastic) value process, $\{V_t\}_{t=0}^{T}$, of a portfolio of financial instruments is a martingale. Such a  measure is called an equivalent martingale measure or \emph{risk-neutral measure} $\mathbb{Q}$. The value process being a $\mathbb{Q}$ martingale, also called the fair value, implies that the current value of the portfolio is equal to the expected value at maturity:
\begin{equation}
\label{eqn:risk_neutral}
    V_t = \mathbb{E}_{\mathbb{Q}}[V_{T} | \mathcal{F}_t] = \mathbb{E}_{\mathbb{Q}}[C(t, T) | \mathcal{F}_t].
\end{equation}  The second equality follows since the definition of the fair value at maturity $V_T$, as seen from time $t$, is the discounted accumulated cash flows realized from time $t$ until $T$, which from now on we denote $C(t, T)$. If the stronger assumption of a complete market is included, namely, all participants have access to all information that could be obtained about the market, then the risk-neutral measure is unique.

An important type of pricing problem in finance is the pricing of derivatives. A derivative is a contract that derives its value from another source (e.g., collection of assets, financial benchmarks) called the \emph{underlying}, which is modeled by  $\{X_{t}(\omega)\}_{t=0}^{T}$, where $\omega \in \Omega$. The value of a derivative is typically calculated by simulating the dynamics of the underlying and computing the payoff accordingly. The payoff is a sequence of Borel-measurable functions $z_t$, which map the evolution of the underlying to the (discounted) payoff process $Z_{t} := z_t(X_0, \dots, X_t)$. The functions $z_t$ are based on the contract. Thus, for derivatives, $C(t, T)$ is the accumulation of all $Z_{t}$ occurring from $t$ until maturity $T$. We can apply Equation \eqref{eqn:risk_neutral} under a risk-neutral measure. Thus, the value of the derivative at $t$, $V_t$, is equal $\mathbb{E}[C(t, T)| \mathcal{F}_t]$. This justifies the usage of MCI for derivative pricing \cite{glasserman2004monte,weinzierl2000introduction}, assuming sample paths follow the risk-neutral measure.

We next discuss the application of  QMCI and quantum differential equation solvers to the pricing of financial derivatives. Specifically, we consider two types of derivatives: options and collateralized debt obligations (CDOs).

\label{sec:quantumderivspricing}
    
\subsubsubsection{Options} 
An option gives the holder the right, but not the obligation, to purchase (call option) or sell (put option) an asset 
at a specified price (strike price) and within a given time frame (exercise window) in the future.
EOne of the most well-known models for options is the Black--Scholes model \cite{black2019pricing}. This model assumes that the price of the underlying (risky) asset, $\{X_t\}_{t=0}^{T}$, evolves like a GBM with a constant volatility. The Black--Scholes PDE, which governs the evolution of the price of an option on the underlying asset, has a closed-form solution, called the Black--Scholes formula, for the simple European option. A European call option has a payoff that satisfies $z_{t} = 0, \forall t < T$, and $z_{T}(X_T) = e^{-rT}(X_{T} - K)^{+}$.\footnote{$(A)^+ := \max\{A,0\}$} This payoff depends only on the state of the underlying at maturity; in other words, it is path independent.  The interest rate $r$ is the return of a risk-free asset, which is included in the Black--Scholes model. According to Equation \eqref{eqn:risk_neutral}, the fair value of the European option at time $t$ is $V_t = e^{rt}\mathbb{E}_{\mathbb{Q}}[z_{T}(X_T)|X_t]$, and the Black--Scholes formula provides $V_t$ in closed form for all $t$. However, many pricing tasks involve more complicated options that are often path dependent, and hence information about the asset at multiple time steps is required; that is,  $z_{T}$ depends on previous $X_t$. This is one of the reasons  MCI-based approaches \cite{boyle1977options} are more widely used than those based on directly solving PDEs, since the path-dependent PDE can be complicated.

Stamatopoulos et al.~\cite{Stamatopoulos_2020} constructed quantum procedures for the uncertainty distribution loading, $\mathcal{P}_l$, and the payoff function, $\mathcal{P}_f$, for European options. The operator $\mathcal{P}_l$ loads a distribution over the computational basis states representing the price at maturity. The operator $\mathcal{P}_f$ approximately computes the rectified linear function, characteristic of the payoff for European options. QMCI has also been applied to path-dependent options such as barrier and Asian options \cite{Stamatopoulos_2020, Rebentrost_2018}, and to multi-asset options \cite{Stamatopoulos_2020}.
For these options, prices at different points on a sample path or for different assets are represented by separate quantum registers, and the payoff operator $\mathcal{P}_f$ is applied to all of these registers.
Pricing of more complicated derivatives with QAE can be found in the literature. For example, Chakrabarti et al.~performed an in-depth resource estimation and error analysis for QAE applied to the pricing of autocallable options and target accrual redemption forwards (TARFs)~\cite{Chakrabarti_2021}.

Miyamoto and Kubo \cite{Miyamoto_derivatives} utilized the FDM-based approach to solve the multi-asset Black--Scholes PDE accelerated by using QLSA, as mentioned in Section \ref{sec:qlsa_pde}. However, since applying the quantum FDM method to the Black--Scholes PDE produces a quantum state encoding the value of the option at exponentially many different initial prices, the quantum speedup is lost if we need to read out a single amplitude. The authors avoided this issue by reading out the expected value of the option at a future time point under the probability distribution of the price of the underlying at the same time point, which is the current price of the option as a consequence of the value process being a martingale under the risk-neutral measure. This approach results in an exponential speedup in terms of the dimension of the PDE over the classical FDM-based approach. Linden et al.~\cite{linden2020} compared various quantum and classical approaches for solving the heat equation, which the Black--Scholes PDE reduces to. The quantum approaches discussed make use of the QLSA-based PDE solvers and quantum walks (Section \ref{sec:quantumwalks}). Fontanela et al.~\cite{fontanela2021quantum} transformed the Black--Scholes PDE into the Sch\"{o}dinger equation with a non-Hermitian Hamiltonian. They utilized VarQITE to simulate the Wick-rotated Sch\"{o}dinger equation. Similarly, Alghassi et al.~\cite{alghassi2021variational} applied imaginary time evolution to the more general Feynmann--Kac formula for linear PDEs. Alternatively, Gonzalez--Conde et al.~\cite{gonzalez2021} used unitary dilation, that is, unitary evolution in an expanded space, and postselection to simulate the nonunitary evolution associated with the non-Hermitian Hamiltonian that results from mapping the Black--Scholes PDE to the Schr\"{o}dinger. They also mentioned the potential applicability beyond constant-volatility models.

Another prominent option  is the American-style option \cite{follmer2016stochastic}. While European, barrier, and Asian options  allow the holder to exercise their right only at maturity, American options allow the holder to buy/sell the underlying asset at any point in time up to maturity, $T$. Thus, the buyer of the option needs to determine the optimal exercise time that maximizes the expected payoff. More generally, finding the optimal execution time is an optimal stopping problem \cite{shiryaev2007optimal}. In modern probability theory, a \emph{stopping time} is an $\mathfrak{F}$-adapted stochastic process $\tau : \Omega \mapsto [0, \infty)$. The adaptedness of $\tau$ implies that making the decision to stop at time $t$ only requires information obtained at time steps up to and including $t$. The corresponding stopped process at time $t$, using stopping time $\tau$, is defined as $(Z_{\tau})_t := Z_{\min(t, \tau(\omega))}$. This definition is used because the option payoff cannot change once the buyer has decided to exercise the option. 

The American option pricing problem involves a sequence of stopping times. An optimal stopping time $\tau_{t}$ starting at time $t$ consists of choosing to exercise at the current time when the payoff is greater than or equal to the expected payoff that could be obtained from continuing, that is, $\mathbb{E}[Z_{\tau_{t+1}} | X_{t}]$, and is called a continuation value. This recursive definition implies that finding the optimal stopping time can be formulated as solving a dynamic programming problem. The value of the American option is the maximum expected payoff; and since an optimal stopping time maximizes the expected payoff, this price is equal to $\mathbb{E}[Z_{\tau_{0}} | X_0]$, where the current time is $0$. This is can be computed by using Monte Carlo integration with sampled payoffs, with discretized time steps, that have optimally stopped.\footnote{Technically, the discretization of time converts the American option into a Bermudan option.} However, this also requires computing the continuation values. A popular approach to estimate continuation values is through least-squares regression using a fixed set of basis functions. The combined technique of regressing the conditional expectation values and sampling optimally-stopped payoffs to solve the dynamic programming problem is called the least-squares Monte Carlo method (LSM) \cite{longstaff2001valuing}.

Doriguello et al.~\cite{Doriguello_2021} proposed a quantum version of LSM that makes use of  QMCI. The authors presented unitary operators for computing the stopping time at time step $t$, $\tau_{t}$, which was done by backtracking from time step $T$. The stopping time $\tau_{t-1}$ can be computed from the stopping time $\tau_{t}$ by using the definition of the dynamic program for the optimal stopping time. Note that although in the classical case, we can store all previously computed $\tau_{t}$ as we compute them backward in time, in the quantum case, we need to compute $\tau_{t}$ for each linear regression step by starting at the maturity date $T$. As a result, while the quantum LSM has an improved error dependence, it now depends quadratically on the maturity time $T$. Miyamoto \cite{Miyamoto_options} performed a similar quantization of LSM for Bermudan options, similar to American options but the exercise times are a discrete subset of time steps until maturity. Alghassi et al.~\cite{alghassi2021variational} also mentioned the potential applicability of their VarQITE approach, mentioned in Section \ref{sec:var_pde}, to the pricing of American options and options with a stochastic volatility.

\subsubsubsection{Collateralized Debt Obligations} A collaterialized debt obligation (CDO) is a derivative that is backed by a pool of loans and other assets that serve as collateral if the loan defaults. The tool is used to free up capital and shift risk. A typical CDO pool is divided into three tranches: equity, mezzanine, and senior. The equity tranche is the first to bear loss; the mezzanine tranche investors bear loss if the loss is greater than the first attachment point; and the senior tranche investors lose money if the loss is greater than the second attachment point.  Although the senior tranche is protected from loss, other default events can cause the CDO to collapse, such as events that caused the 2008 financial crisis~\cite{tang2021quantum}. 

    Conditional independence models \cite{li2000default} are often used for estimating the chances of parties defaulting on the credit in the CDO pool. In such models, given the systemic risk $W$, the default risks $B_1, \dots, B_n$ of the assets involved are independent. The random variable $W$ is then used as a latent variable to introduce correlations between the default risks. The distributions of the default risks and the systemic risk are usually either Gaussian \cite{li2000default} or normal-inverse Gaussian \cite{barndorff1997normal}.

    Tang et al.~\cite{tang2021quantum} presented quantum procedures for both the $\mathcal{P}_l$ and $\mathcal{P}_f$ operators used for QMCI. The goal was to utilize QAE to estimate the expected loss of a given tranche. $\mathcal{P}_l$ is a composition of rotations to load the uncorrelated probabilities of default (i.e., Bernoulli random variables). Then a procedure is used for loading the distribution of $W$, and controlled rotations are used to apply the correlations. $\mathcal{P}_f$ computes the loss under a given realization of the variables $B_1, \dots, B_n$ and uses a quantum comparator to determine whether the loss falls within the specified tranche range. QAE is then used, instead of MCI, to compute the expectation of the total loss given that it falls within the tranche range.

\subsubsection{Risk Modeling}
\label{sec:quantumriskanaly}
Risk modeling is another crucial problem for financial institutions, since risk metrics can determine the probability and amount of loss under various financial scenarios. Risk modeling is also important in terms of compliance with regulations. Because of the Basel III regulations mentioned earlier, financial institutions need to calculate and meet requirements for their VaR, CVaR, and other metrics. However, the current classical approach of Monte Carlo methods \cite{glasserman2004monte} is computationally intensive. Similar to the pricing problem mentioned above, a quantum approach could also benefit this financial problem. 
In this subsection we  explore the application of the presented quantum algorithms to computing risk metrics such as VaR and CVaR. We  also review quantum approaches for sensitivity analysis and credit valuation adjustments.

\subsubsubsection{Value at Risk} As first mentioned in Section \ref{sec:regulations}, VaR and CVaR are common metrics used for risk analysis.
Mathematically, VaR at confidence $\alpha \in [0, 1]$ is defined as $\text{VaR}_\alpha[X] = \text{inf}_{x \geq0}\{x~|~\mathbb{P}[X \leq x] \geq \alpha \}$, usually computed by using Monte Carlo integration \cite{hong}. CVaR is defined as $\text{CVaR}_\alpha[X] = \mathbb{E}[X~|~0 \leq X \leq \text{VaR}_\alpha[X]]$ \cite{Woerner_2019}; $\alpha$ is generally around 99.9\% for the finance industry.
QAE can be used to evaluate $\text{VaR}_{\alpha}$ and $\text{CVaR}_{\alpha}$ faster than with classical MCI. The procedure to do so, by Woerner and Egger \cite{Woerner_2019, egger2019credit}, is as follows. 
    
Similar to the CDO case above (Section \ref{sec:quantumderivspricing}), the risk can be modeled by using a Gaussian conditional independence model. Quantum algorithms for computing $\text{VaR}_\alpha$ and $\text{CVaR}_\alpha$ via QAE can utilize the same realization of $\mathcal{P}_l$ as for the CDO case \cite{Woerner_2019}. Note that $\mathbb{P}[X \leq x] = \mathbb{E}[h(X)]$, where $h(X) =1$ for $X \leq x$ and $0$ otherwise. Thus $\mathcal{P}_f$ implements a quantum comparator similar to both of the previous applications discussed for derivatives. A bisection search can be used to identify the value of $x$ such that $\mathbb{P}[X \leq x] \geq \alpha$ using multiple iterations of QAE. $\text{CVaR}_\alpha$ uses a comparator to check against the obtained value for $\text{VaR}_\alpha$ when computing the conditional expectation of $X$.
    
\subsubsubsection{Economic Capital Requirement} 
Economic capital requirement (ECR) summarizes the amount of capital required to remain solvent at a given confidence level and time horizon \cite{egger2019credit}. Egger et al.~\cite{egger2019credit} presented a quantum method using  QMCI to compute ECR. Consider a portfolio of $K$ assets with $L_1, \dots ,L_K$ denoting the losses associated with each asset. The expected value of the total loss, $\mathcal{L}$, is $\mathbb{E}[\mathcal{L}] = \sum_{m = 1}^K \mathbb{E} [L_m]$. The losses can be modeled by using the same Gaussian conditional independence model discussed above. ECR at confidence level $\alpha$ is defined as

\begin{equation*}
    \text{ECR}_{\alpha}[\mathcal{L}] = \text{VaR}_{\alpha}[\mathcal{L}] - \mathbb{E}[\mathcal{L}], 
\end{equation*}
where $\alpha$ is generally around 99.9\% for the finance industry. The expected total loss $\mathbb{E}[\mathcal{L}]$ can be efficiently computed classically, so only the estimate of $\text{VaR}_{\alpha}[\mathcal{L}]$ is computed by using QAE \cite{egger2019credit}.

\subsubsubsection{The Greeks} 
Sensitivity analysis is an important component of financial risk modeling. The stochastic model of the underlying assets of a financial derivative, which is the process $X_t$, typically contains multiple variables representing the market state. Some common examples are the spot prices of the underlying assets $S^{(i)}_{t}$, current volatilities $\sigma^{(i)}_{t}$, and the maturity times $T^{(i)}$. In mathematical finance, the partial derivatives, or sensitivities, of $V_{t}$ with respect to these quantities are called \emph{Greeks}. For example, $\frac{\partial V_{t}}{\partial S^{(i)}_{t}}$ are called \emph{Deltas},  $\frac{\partial V_{t}}{\partial \sigma^{(i)}_{t}}$ are called \emph{Vegas},
$\frac{\partial^2 V_{t}}{(\partial \sigma^{(i)}_{t})^2}$ are called \emph{Gammas}, and $\frac{\partial V_{t}}{\partial T^{(i)}}$ are called \emph{Thetas}. Classically, this can be computed utilizing finite-difference methods and MCI. The optimal classical method, which uses variance reduction techniques, requires $\mathcal{O}(k/\epsilon^2)$ samples to compute $k$ Greeks. Stamatopoulos et al.~\cite{stamatopoulos2021} proposed using quantum finite-difference methods, which were first proposed by Jordan \cite{jordan2005}, to accelerate the computation of Greeks. 

The quantum setting uses  what is called a probability oracle. In the context of financial sensitivity, this makes use of the operation in \eqref{eqn:probability_oracle} from the QMCI algorithm, namely,  operators $\mathcal{P}_{l}$ and $\mathcal{P}_{f}$. For computing the $k$-dimensional numerical gradient, Jordan's original algorithm utilized a single call to a quantum binary oracle, which computes a fixed-point approximation to $V_t$ into a register. The derivative of $V_t$ is then encoded in the relative phases of quantum state utilizing phase kickback. In the case described above, however, we are provided an analog quantity encoded in the amplitude of a quantum state, and thus we would require an analog to digital, e.g. QAE, to analog conversion. The query complexity of QAE scales exponentially in the number of bits in the digital representation. Gily\'{e}n, Arunachalam, and Wiebe (GAW) \cite{gaw} proposed a conversion technique that avoids the digitization step. Their approach takes as input a probability oracle and uses block-encoding-based Hamiltonian simulation \cite{low2019hamiltonian, chakraborty2018power} to perform the phase encoding. In other words, it converts the probability oracle to a phase oracle. The GAW algorithm makes use of higher-order central difference methods \cite{li2005}; and under certain smoothness conditions of the function whose derivative we are taking \cite{cornelissen2019quantum},  the algorithm uses $\mathcal{O}(\sqrt{k}/\epsilon)$ queries to a probability oracle to compute the $k$-dimensional gradient. Stamatopoulos et al. showed that this can be used to produce a quadratic speedup for computing $k$ Greeks over the classical MCI and finite-difference-based methods. They also proposed an alternative technique that does not attain the full quadratic speedup but avoids the Hamiltonian simulation required for the probability to phase oracle conversion. As mentioned by Stamatopoulos et al., the number of Greeks, $k$, can be large in practice, which makes the quadratic speedup significant.

\subsubsubsection{Credit Value Adjustments}
Value adjustments \cite{green2015xva}, or XVAs, are a set of corrections to the value of a financial instrument due to  factors that are not included in the pricing model. Specifically, credit value adjustments (CVAs) modify the price of a financial contract to take into account the risk of loss associated with a counterparty defaulting or failing to meet their obligation, namely, the counterparty-credit risk (CCR). The usual risk-neutral valuation does not take CCR into account.  For example, from the perspective of a holder of an option, if the seller of the contract defaults at time $t$, the loss to the holder is the value of the option at $t$. For certain contracts, a proportion of the lost value can be recovered. This fraction is called the recovery rate, $R$. The value that is lost after recovery is called the loss given default (LGD). The LGD is computed as $1-R$.

We consider a finite set of time points: $t_0, t_1, \dots, t_N$, where $t_N$ is the largest maturity date of all contracts in the portfolio. If we are performing a valuation of the CCR at time $t_0$, the expected loss if a default occurs at $t_i > t_0$ requires determining the fair value of the instrument at time $t_i$. Since $t_i$ is in the future, however, the expected value lost if default occurs at time $t_i$, called the expected exposure profile, is $\mathbb{E}_{\mathbb{Q}}[(V_{t_i})^{+} | \mathcal{F}_{t_0}]$, where $V_{t_i}$ is the fair price of a portfolio of contracts at $t_i$. Since default can occur before or at maturity probabilistically, the CVA at $t_0$ is defined to be
\begin{equation}
    \text{CVA}_{t_0} := \mathbb{E}[(1- R)(V_{\tau})^{+} | \mathcal{F}_{t_0}] = (1- R)\sum_{i=0}^{N} \mathbb{P}(t_i < \tau < t_{i+1})\mathbb{E}_{\mathbb{Q}}[(V_{t_i})^{+} | \mathcal{F}_{t_0}],
\end{equation}
where measure $\mathbb{P}$ denotes the probability of defaulting in a given time period. The maximum is to ensure we  consider only obligations of the counterparty, and $\tau$ is a random stopping time representing the time of default. The adjusted price is
\begin{equation}
    V'_{t_0} := V_{t_0} - \text{CVA}_{t_0}.
\end{equation}

As mentioned by Han and Rebentrost \cite{han2022}, if the joint distribution can be represented as a simple product, namely,
\begin{equation}
    q^{(t_i)}_{j} := \mathbb{P}(t_i < \tau < t_{i+1})\mathbb{Q}(j),
\end{equation}
where  $j$ is a series of economic events, for example a path of the underlying assets, leading to a price of $V_{t_i}^{(j)}$ at time $t_i$ occurring with risk-neutral probability $\mathbb{Q}$, then the CVA computation can be viewed as the following inner product:
\begin{equation}
    \text{CVA}_{t_0} = \sum_{i, j} q^{(t_i)}_{j}V_{t_i}^{(j)} : = \vec{q} \cdot \vec{V}.
\end{equation}
Han and Rebentrost presented a variety of quantum inner product estimation algorithms that make use of QMCI to obtain a quadratic speedup over the classical approach of sampling the entries of $\vec{V}$ according to $\vec{q}$ and taking sample averages. Their approach also applies to the problem of pricing a portfolio of derivatives. Along similar lines, Alcazar et al.~\cite{Alcazar_2022} proposed a more near-term approach to solving this problem that makes use of a Bayesian approach to QAE \cite{Wang_2021, Koh_2022}. This Bayesian QAE approach has a complexity that varies according to the expected hardware noise and allows for interpolating between the complexities of MCI and QMCI. They also performed an end-to-end resource analysis of their approach.

\section{Optimization}
\label{sec:optimization}

Solving optimization problems (e.g., portfolio optimization, arbitrage) presents the most promising commercially relevant applications on NISQ hardware (Section \ref{sec:currentquantumhardware}).
In this section we discuss the potential of using quantum algorithms to help solve various optimization problems. 
We start by discussing quantum algorithms for the $\mathsf{NP}$-hard\footnote{By definition (Section \ref{sec:quantum_algorithms}), optimization problems are not in $\mathsf{NP}$, but they can be $\mathsf{NP}$-hard. However, there is an analogous class for optimization problems called $\mathsf{NPO}$ \cite{hromkovivc2013algorithmics}.} \emph{combinatorial} optimization problems \cite{wolsey1999integer, hromkovivc2013algorithmics} in Section \ref{sec:combopt}, with  particular focus on those with quadratic cost functions. Neither classical nor quantum algorithms are believed to be able to solve these $\mathsf{NP}$-hard problems asymptotically efficiently (Section \ref{sec:quantum_algorithms}). However, the hope is that quantum computers can solve these problems faster than classical computers on realistic problems such that it has an impact in practice.
 Section \ref{sec:convex} focuses on \emph{convex} optimization. In its most general form convex optimization is $\mathsf{NP}$-hard. However, in a variety of situations  such  problems can be solved efficiently \cite{nesterov1994interior} and have a lot of nice properties and a rich theory \cite{boyd2004convex}.
 There exist quantum algorithms that utilize QLS (Section \ref{sec:hhlsection}), and other techniques to  provide polynomial speedups for certain convex, specifically conic, optimization problems. These can have a significant impact in practice.  Section \ref{sec:largescale} discusses \emph{large-scale} optimization, which utilizes hybrid approaches that decompose large problems into smaller subtasks solvable on a near-term quantum device. The more general mixed-integer programming problems, in which there are both discrete and continuous variables, can potentially be handled by using quantum algorithms for optimization (both those  mentioned in Section \ref{sec:combopt} and, in certain scenarios, Section \ref{sec:convex}) as subroutines in a classical-quantum hybrid approach (i.e., decomposing the problem in a similar manner to the ways mentioned in Section \ref{sec:largescale}) \cite{braineMBOtransacitons, Ajagekar_2020}. Many quantum algorithms for combinatorial optimization are NISQ friendly and heuristic (i.e., with no asymptotically proven benefits). The convex optimization algorithms are usually for the fault-tolerant era of quantum computation.

\subsection{Combinatorial Optimization}
\label{sec:QUBO, Ising}
\label{sec:combopt}
In the general case, integer programming (IP) problems, which involve variables restricted to integers (including those with integer variable constraints), are $\mathsf{NP}$-hard \cite{wolsey1999integer}. This section focuses on \emph{combinatorial optimization}, sometimes used synonymously with discrete or integer optimization. In this paper, combinatorial optimization refers specifically to integer optimization problems that only consist of binary variables, namely, binary integer programs \cite{wolsey1999integer}. Problems that consist of selecting objects from a finite set to optimize some cost function, potentially including constraints, fit this formalism. However, integer programs can be reduced to binary programs by representing integers in binary or using one-hot encoding. Most of the financial optimization problems presented in this article can be reduced to combinatorial problems with quadratic cost functions called binary quadratic programming problems \cite{wolsey1999integer}.

A binary quadratic program without constraints, which is also $\mathsf{NP}$-hard, is called a quadratic unconstrained binary optimization  problem \cite{kochenberger2014unconstrained} and lends itself naturally to many quantum algorithms. A QUBO problem on $N$ binary variables can be expressed as
\begin{equation}\label{eq:qubo}
    \min_{\vec{x} \in \mathbb{B}^N}
    \vec{x}^\mathsf{T}Q\vec{x} + \vec{x}^\mathsf{T}\vec{b},
\end{equation}
where $\vec{b} \in \mathbb{R}^N$, $Q \in \mathbb{R}^{N \times N}$, and $\mathbb{B} = \{0, 1\}$.
Unconstrained integer quadratic programming (IQP) problems can be converted to QUBO in the same way that IP can be reduced to general binary integer programming \cite{okada2019efficient}.
Constraints can be accounted for by optimizing the Lagrangian function \cite{boyd2004convex} of the constrained problem, where each dual variable is a hyperparameter that signifies a penalty for violating the constraint \cite{IsingNP}.

QUBO has a connection with finding the ground state of the generalized Ising Hamiltonian from statistical mechanics \cite{IsingNP}. A simple example of an Ising model is a two-dimensional lattice under an external longitudinal magnetic field that is potentially site-dependent; in other words,  the strength of the field at a site on the lattice is a function of location.
At each site $j$, the direction of the magnetic moment of spin $z_j$ may take on values in the set $\{-1, +1\}$. The value of $z_j$ is influenced by both the moments of its neighboring sites and the strength of the external field.
This model can be generalized to an arbitrary graph where the vertices on the graph represent the spin sites and weights on the edges signify the interaction strength between the sites, hence allowing for long-range interactions.
The classical Hamiltonian for this generalized model is
\begin{equation*}
\mathcal{H} = -\sum_{ij} J_{ij} z_iz _j  -\sum_{j} h_jz _j,
\end{equation*}
where the magnitude of $J_{ij}$ represents the amount of interaction between sites $i$ and $j$ and its sign is the desired relative orientation; $h_j$ represents the external field strength and direction at site $j$ \cite{Kadowaki_1998}.
Using a change of variables $z_j = 2x_j - 1$, a QUBO cost function as defined in Equation \eqref{eq:qubo} can be transformed into a classical Ising Hamiltonian, with $x_j$ being the $j$th element of $\vec{x}$.
The classical Hamiltonian can then be ``quantized'' by replacing the classical variable $z_j$ for the $j$th site with the Pauli-$\mathsf{Z}$ operator $\sigma^z_j$ (Section \ref{sec:quantuminfo}). 
Therefore, finding the optimal solution for the QUBO is equivalent to finding the ground state of the corresponding Ising Hamiltonian.

Higher-order combinatorial problems can be modeled by including terms for multispin interactions. Some of the quantum algorithms presented in this section can handle such problems (e.g., QAOA in Section \ref{sec:QAOA}). From the quantum point of view, these higher-order versions are observables that are diagonal in the computational basis. The following subsections will more fully show  why the quantum Ising or more generally the diagonal observable formulation results in quantum algorithms for solving combinatorial problems. In Section \ref{sec:optfinapp} we discuss a variety of financial problems that can be formulated as combinatorial optimization and solved by using the quantum methods described in this section.
\subsubsection{Quantum Annealing}
\label{sec:QuantumAnnealing}
Quantum annealing, initially mentioned in Section \ref{sec:aqoalgo},  consists of adiabatically evolving according to the time-dependent transverse-field Ising Hamiltonian \cite{Kadowaki_1998}, which is mathematically defined as
\begin{equation}
    H(t) = A(t)\left( -\sum_i \sigma^x_i \right) + B(t) \left(- \sum_{ij}J_{ij}\sigma^z_i\sigma^z_j - \sum_i h_i\sigma^z_i\right).
    \label{eqn:QAhamiltonian}
\end{equation}
This formulation natively allows for solving QUBO. As discussed earlier, however, with proper encoding and additional variables, unconstrained IQP problems can be converted to QUBO. Techniques also  exist for converting higher-order combinatorial problems into QUBO problems, at the cost of more variables \cite{chancellor2017circuit, domino2021quadratic}. QUBO, as mentioned earlier, can, through a transformation of variables, be encoded in the site-dependent longitudinal magnetic field strengths $\{h_i\}$ and coupling terms $\{J_{ij}\}$ of an Ising Hamiltonian.
The QA process initializes the system in a uniform superposition of all states, which is the ground state of $-\sum_i \sigma^x_i$. $\sigma^x_i$ is the Pauli-$\mathsf{X}$ (Section \ref{sec:modelsofquantum}) operator applied to the $i$th qubit. This initialization implies the presence of a large transverse-field component, $A(t)$ in Equation \eqref{eqn:QAhamiltonian}.
The annealing schedule, controlled by tuning the strength of the transverse magnetic field, is defined by the functions $A(t)$ and $B(t)$ in Equation \eqref{eqn:QAhamiltonian}. If the evolution is slow enough, according to the adiabatic theorem (Section \eqref{sec:AQC}), the system will remain in the instantaneous ground state for the entire duration and end in the ground state of the Ising Hamiltonian encoding the problem. This process  is called \emph{forward} annealing. There  also exists \emph{reverse} annealing \cite{king2019quantum}, which starts in a user-provided classical state,\footnote{The term ``classical state'' refers to any computational basis state. The quantum Ising Hamiltonian has an eigenbasis consisting of computational basis states.} increases the transverse-field strength, pauses and then, assuming an adiabatic path, ends in a classical state. The pause is useful  when reverse quantum annealing is used in combination with other classical optimizers.

The D-Wave devices are examples of commercially available quantum annealers \cite{Ohzeki_annealer}. These devices have limited connectivity resulting in extra qubits being needed to encode arbitrary QUBO problems (i.e., Ising models with interactions at arbitrary distances). Finding such an embedding, however, is in general a hard problem \cite{lobe2021minor}. Thus heuristic algorithms \cite{cai2014practical}  or precomputed templates \cite{goodrich2017optimizing} are typically used. Assuming the hardware topology is fixed, however, an embedding can potentially be efficiently found for QUBOs with certain structure \cite{venturelli2015quantum}.
Alternatively, the embedding problem associated with restricted connectivity can be avoided if the hardware supports three- and four-qubit interactions, by using the LHZ encoding \cite{lechner2015quantum} or its extension by Ender et al.~\cite{ender2021parity}. The extension by Ender et al.~even supports higher-order binary optimization problems \cite{driebschon2021parity, fellner2021parity}.
Moreover, the current devices for QA provide no guarantees on the adiabaticity of the trajectory that the system follows. Despite the mentioned limitations of today's annealers,  a number of financial problems still remain  that can be solved by using these devices (Section \ref{sec:optfinapp}).

\subsubsection{Quantum Approximate Optimization Algorithm}
\label{sec:QAOA}
QAOA is a variational algorithm (Section \ref{sec:qva}) initially inspired by the adiabatic evolution of quantum annealing. The initial formulation of QAOA can be seen as a truncated or Trotterized \cite{nielsen2002quantum} version of the QA evolution to a finite number of time steps.\footnote{In the case of standard QAOA, however, the adiabatic trajectory would move between instantaneous eigenstates with the \emph{highest} energy \cite{farhi2014quantum}. This still follows from the adiabatic theorem (Section \ref{sec:AQC}).} QAOA follows the adiabatic trajectory in the limit of infinite time steps, which provides evidence of good convergence properties \cite{farhi2014quantum}. This algorithm caters to gate-based computers. However, it has  been tested on an analog quantum simulator \cite{silverio2021pulser, Henriet2020quantumcomputing, ebadi20}.

QAOA can be used to solve general unconstrained combinatorial problems. This capability allows QAOA to solve, without additional variables,  a broader class of problems than QA can. As for QA, constraints can be encoded by using the Lagrangian of the optimization problem (i.e., penalty terms). Consider an unconstrained combinatorial maximization problem with $N$ binary variables $\{x_i\}_{i=1}^{N}$ concatenated into the bit string $z:=x_1\dots x_N$ and cost function $f(z)$. The algorithm seeks a string $z^*$ that maximizes $f$. This is done by first preparing a parameter-dependent $N$-qubit quantum state (realizable as a PQC, Section \ref{sec:qva}):
\begin{equation}
\label{eqn:altop}
    \ket{\bm{\gamma}, \bm{\beta}} = U(B, \beta_p)U(C, \gamma_p) \cdots U(B, \beta_1)U(C, \gamma_1)\ket{s},
\end{equation}
where $\bm{\gamma} = (\gamma_1, \dots, \gamma_p), \bm{\beta} = (\beta_1, \dots, \beta_p)$. The unitary  $U(C, \gamma)$ is called the phase operator or phase separator and defined as $U(C, \gamma) = e^{-i\gamma C}$, where $C$ is a diagonal observable that encodes the objective function $f(z)$. In addition, $U(B, \beta)$ is the mixing operator defined as $U(B, \beta) = e^{-i\beta B}$, where, in the initial QAOA formulation, $B = \sum_{i=1}^N \sigma_i^x$. The initial state, $\ket{s}$, is a uniform superposition, which is an excited state of $B$ with maximum energy. However,  other choices for $B$ exist that allow  QAOA to incorporate constraints without penalty terms \cite{Hadfield_2019}. Equation \eqref{eqn:altop}, where the choice of $B$ can vary, is called the alternating operator ansatz \cite{Hadfield_2019}. Preparation of the state is then followed by a measurement in the computational basis (Section \ref{sec:quantuminfo}), indicating the assignments of the $N$ binary variables. The parameters are updated such that the expectation of $C$, that is, $\bra{\bm{\gamma}, \bm{\beta}}C\ket{\bm{\gamma}, \bm{\beta}}$, is maximized. Note that we can multiply the cost function by $-1$ to convert it to a minimization problem. The structure of the QAOA ansatz allows for finding good parameters purely classically for certain problem instances \cite{farhi2014quantum}. In general, however, finding such parameters is a challenging task. As mentioned earlier (Section \ref{sec:qva}), the training of quantum variational algorithms is $\mathsf{NP}$-hard. This does in fact withhold us from reaching arbitrarily good approximate or local minima and serves as merely an indication of the complexity of finding the exact global minimum. However, a lot of recent research has been proposing practical approaches for effective parameter and required circuit depth estimation \cite{Shaydulin2021,shaydulin2019multistart,galda2021transferability,streif2020training,khairy2019learning,shaydulin2019evaluating} as well as reducing the phase operator's complexity \cite{wang2022quantum,liu2022sparsification} and mixer's \cite{ gulania2022dynamics} complexity. Recent demonstrations of QAOA on gate-based quantum hardware have revealed the capability of start-of-the-art quantum hardware technologies in handling unconstrained \cite{harrigan2021quantum, otterbach2017unsupervised} and constrained optimization problems \cite{niroula2022}.

\subsubsection{Variational Quantum Eigensolver}
\label{sec:vqe}
The VQE algorithm, mentioned in Section \ref{sec:qva}, is a prominent algorithm in the NISQ era because it can be run on current hardware \cite{fedorov2022vqe}. Essentially, the algorithm solves the problem of approximating the smallest eigenvalue of some observable and  gives a prescription for reconstructing an approximation of a corresponding eigenstate. Given an observable, the expected value or equivalently the Rayleigh quotient (Section \ref{sec:qva}) with respect to any state is an upper bound of the smallest eigenvalue. There is equality only when the state is an eigenstate with the smallest eigenvalue. The variational procedure based on minimizing the Rayleigh quotient is known as the Rayleigh--Ritz method \cite{Yuan_2019}.

VQE utilizes a parameterized ansatz to represent a space of candidate wave functions. A classical optimizer modifies the parameters to minimize the expectation of the observable (i.e., the cost function) and thus tries to approach the value of the minimum eigenvalue. The hybrid system works in an iterative loop where the trial wave function is loaded onto the quantum processor and then the classical computer feeds the quantum processor new sets of parameters to improve the initial trial state based on the computed expected value. Like QAOA, VQE can also be used to find a minimum eigenvalue state of an observable, which is diagonal in the computational basis, used to encode a combinatorial optimization problem. VQE is more general than QAOA (Section \ref{sec:QAOA}). The  reason  is that QAOA specifically uses the alternating operator ansatz, while, VQE can utilize any PQC as an ansatz. However, certain ans\"{a}tze are more suitable for particular problems because they take into account prior information about the ground-state wave function(s) \cite{Peruzzo_2014}. Alternatively, there exist evolutionary, noise-resistant techniques for optimizing the structure of the ansatz and significantly increasing the space of candidate wave functions \cite{evqe}.  Amaro et al.~\cite{filtervar} introduced quantum variational filtering, which can be used to increase the probability of sampling low-eigenvalue states of an observable when the filtering operator is applied to an arbitrary quantum state. Their algorithm, Filtering VQE, was applied to combinatorial optimization problems and outperformed both standard VQE and QAOA. A rigorous analysis of the convergence of VQE with an overparameterized ansatz has been performed by Xuchen et al.~\cite{xuchen2022}.

\subsubsection{Variational Quantum Imaginary Time Evolution (VarQITE)}
\label{sec:varite}
In relativistic terms, real-time evolution of a quantum state happens in Minkowski spacetime, which, unlike Euclidean space, has an indefinite metric tensor \cite{naber2012geometry}.
Imaginary-time evolution (ITE) \cite{osterwalder1973axioms} transforms the evolution to Euclidean space and is performed by making the replacement $t \mapsto it = \tau$, known as a \emph{Wick rotation}~\cite{wick1954properties}. The imaginary time dynamics follow the Wick-rotated Schr\"{o}dinger equation
    \begin{equation}
    \label{eqn:wick}
        \ket{\dot{\psi}(\tau)} = -(H - E_\tau)\ket{\psi(\tau)},
    \end{equation}
where $H$ is a quantum Hamiltonian and $E_\tau = \langle\psi(\tau)|H\ket{\psi(\tau)}$. The time-evolved state (solution to Equation \eqref{eqn:wick}) is $\ket{\psi(\tau)} = C(\tau)e^{-H\tau}\ket{\psi(0)}$, where $C(\tau)$ is a time-dependent normalization constant \cite{McArdle_2019}. As $\tau \xrightarrow[]{} \infty$, the smallest eigenvalue of the non-unitary operator $e^{-H\tau}$ dominates, and the state approaches a ground state of $H$. This is assuming the initial state has nonzero overlap with a ground state. Thus, ITE presents another method for finding the minimum eigenvalue of an observable \cite{McArdle_2019, 2019ITE}.

For variational quantum ITE (VarQITE) \cite{Yuan_2019}, the time-dependent state $\ket{\psi(\tau)}$ is represented by a parameterized ansatz (i.e., PQC), $\ket{\phi(\bm{\theta}(\tau))}$, with a set of time-varying parameters, $\bm{\theta}(\tau)$. The evolution is projected onto the circuit parameters, and the convergence is also dependent on the expressibility of the ansatz \cite{Yuan_2019}. However, quantum ITE  using McLachlan's principle is typically expensive to implement because of the required metric tensor computations and matrix inversion \cite{Yuan_2019, Mitarai_2019}. Benedetti et al. developed an alternative method for VarQITE that avoids these computations and is gradient free \cite{effvarte}. Their variational time evolution approach is also applicable to real-time evolution. VarQITE  can be used to solve a combinatorial optimization problem by letting $H$ encode the combinatorial cost function, in a similar manner to QAOA and VQE, and is also not restricted to QUBO. In addition, VarQITE has been used to prepare states that do not result from unitary evolution in real time, such as the quantum Gibbs state \cite{Zoufal_2021}.

We note that the variational principles for real-time evolution, mentioned in Section \ref{sec:qva}, can be used to approximate adiabatic trajectories \cite{Chen_2020}. This can potentially be used for combinatorial optimization as well.

\subsubsection{Optimization by Quantum Unstructured Search}
D\"{u}rr and H{\o}yer
\label{sec:durrhoyer}
\cite{durrhoyer1996GroverMin} presented an algorithm that applies quantum unstructured search (Section \ref{sec:grovers}) to the problem of finding the \emph{global} \cite{baritompa2005grover} minimum of a black-box function. This approach makes use of the work of Boyer et  al.~\cite{1998tightbounds} to apply Grover's search when the number of marked states is unknown. Unlike the metaheuristics of quantum annealing and gate-based variational approaches, the D\"{u}rr-H{\o}yer algorithm has a provable quadratic speedup in query complexity. That is, given a search space $\mathcal{X}$ of size $N$, the algorithm requires $\mathcal{O}(\sqrt{N})$ queries to an oracle that evaluates the function $f: \mathcal{X} \mapsto \mathbb{R}$ to minimize. A classical program would require $\mathcal{O}(N)$ evaluations.

The requirements to achieve quadratic speedup are the same as for Grover's algorithm. The quantum oracle in this case is $O_{g_y}|x\rangle = (-1)^{g_y(x)}|x\rangle$; $g_y: \mathcal{X} \mapsto \{0, 1\}$ is a classical function mapping $g_y(x)$ to $1$ if $f(x) \leq y$, otherwise $0$; and $y$ is the current smallest value of $f$ evaluated so far \cite{baritompa2005grover}. The Grover adaptive search framework of Bulger et al.~\cite{bulger2003implementing} generalizes and extends the D\"{u}rr--H{\o}yer algorithm.
Gilliam et al.~\cite{Gilliam2021groveradaptive} proposed a method for efficiently implementing the oracle $O_{g_y}$  by converting to Fourier space. This applies particularly to binary polynomials, and thus this method can be used to solve arbitrary constrained combinatorial problems with a quadratic speedup over classical exhaustive search.

\subsection{Convex Optimization}
\label{sec:convex}
Quantum algorithms have been invented for common convex optimization problems \cite{boyd2004convex} such as linear programming, second-order cone programming (SOCP), and semidefinite programming (SDP). These conic programs have a variety of financial applications, such as portfolio optimization, discussed in Section \ref{sec:quantportfolioopt}.
The quantum linear systems algorithms from Section \ref{sec:hhlsection} have been used as subroutines to improve algorithms for linear programming such as the simplex method \cite{bertsimas1997introduction} in the work by Nannicini \cite{nannicini2021fast} and interior-point methods (IPMs) \cite{nesterov1994interior} in the work by Kerenedis and Prakash \cite{kerenidis2018quantum}. The quantum IPM utilizes QLS solvers with tomography to accelerate the computation of the Newton linear system.
Kerenedis and Prakash \cite{Kerenidis_2021} also applied quantum IPM to provide, under certain conditions, small polynomial speedups for the primal-dual IPM method for SOCP \cite{monteiro2000polynomial}. Furthermore, various approaches using quantum computing to solve SDPs have been presented. Brand\~{a}o \cite{brandao2017quantum}, Brand\~{a}o et al.~\cite{brandao2019quantum}, and van Apeldoorn et al.~\cite{2020sdpvanApeldoorn} proposed quantum improvements to the multiplicative-weights method of Arora and Kale \cite{arora2007combinatorial} for SDP. Alternatively, Kerenedis and Prakash applied their quantum IPM to SDP \cite{kerenidis2018quantum}. 

\subsection{Large-Scale Optimization}
\label{sec:largescale}
Near-term quantum devices are expected to have a limited number of qubits in the foreseeable future. This is a major obstacle for real applications in finance that are large scale in many domains. The hybridization of quantum and classical computers using decomposition-based approaches is a prominent solution for this problem \cite{shaydulin2019hybrid}. The idea behind this class of approaches is similar to what is done with numerical methods and high-performance computing. The main driving routine that is executed on a classical machine decomposes the problem into multiple parts. These subproblems are then each solved separately on a quantum device. The classical computer then combines solutions received from the quantum computer. For example, in quantum local search for modularity optimization \cite{shaydulin2019network} and an extension of it to a multilevel approach \cite{ushijima2021multilevel}, the main driving routine identifies small subsets of variables and solves each subproblem on a quantum device. A similar decomposition approach for finding maximum graph cliques was developed by Chapuis et  al. \cite{chapuis2019finding}. 

\subsection{Financial Applications}
\label{sec:optfinapp}
In this section we  explore multiple applications of the quantum optimization algorithms presented above to financial problems.

\subsubsection{Portfolio Optimization}
\label{sec:quantportfolioopt}
In finance, one of the most commonly seen optimization problems is \emph{portfolio optimization}.
Portfolio optimization is the process of selecting the best set of assets and their quantities, from a pool of assets being considered, according to some predefined objective. The objective can vary depending on the investor's preference regarding financial risk and expected return. Modern portfolio theory \cite{markowitzmpt1952} focuses on the trade-offs between risk and return to produce what is known as an efficient portfolio, which maximizes the expected return given a certain amount of risk. This trade-off relationship is represented by a curve known as the efficient frontier. The expected return and risk of a financial portfolio can often be modeled respectively by looking at the mean and variance of static portfolio returns. The problem setup for portfolio optimization can be formulated as constrained utility maximization \cite{markowitzmpt1952}. 

In portfolio optimization, each asset class, such as stocks, bonds, futures, and options, is assigned a weight. Additionally, all assets within the class are allocated in the portfolio depending on their respective risks, returns, time to maturity, and liquidity. The variables controlling how much the portfolio should invest in an asset can be continuous or discrete. However, the overall problem can contain variables of both types. Continuous variables are suitable for representing the proportion of the portfolio invested in the asset positions when non-discrete allocations are allowed. Discrete variables are used in situations where assets can  be purchased or sold only in fixed amounts. The signs of these variables can also be used to indicate long or short positions. When risk is accounted for, the problem is typically a quadratic program, usually with constraints. Depending on the formulation, this problem can be solved with techniques for convex \cite{markowitzmpt1952} or mixed-integer programming \cite{benati2007mixed, mansini2015linear}. Speeding up the execution of and improving the quality of portfolio optimization is particularly important when the number of variables is large. 

This section first focuses on combinatorial formulations using the algorithms mentioned in Section \ref{sec:combopt}. As mentioned there, the algorithms presented can be generalized to integer optimization problems. The second part of this section focuses on convex formulations of the problem solved by using algorithms from Section \ref{sec:convex}. The handling of mixed-integer programming
was briefly mentioned at the beginning of Section \ref{sec:optimization}.
\subsubsubsection{Combinatorial Formulations} 
The first combinatorial formulation considered is  risk minimization or hedging \cite{cornuejols1977exceptional}. The risk term is quadratic and is derived from the covariance or correlation matrix computed by  using historical pricing information. A simplified version of this problem can be formulated as the following QUBO:
  \begin{equation}
  \label{eqn:risk}
    \underset{\Vec{x}\in\mathbb{B}^N}{\min} \Vec{x}^\mathsf{T} \Sigma \Vec{x},
    \end{equation}
where $\Sigma \in \R^{N \times N}$ is the covariance or correlation matrix. This problem can be modified to include budget constraints.
With regard to this problem, Kalra et al. used a time-indexed correlation graph such that the vertices represent assets and an edge between vertices represents the existence of a significant correlation between them \cite{Portfolio_asset}. The price changes are modeled by data collected daily. Instead of continuous correlation values, Kalra et  al. utilized a threshold on the correlations to decide whether to create edges between assets. This was to create sparsity in the graph. One approach they took to solve risk minimization was formulating the problem as finding a maximum-independent set in the graph constructed. This is similar to Equation \eqref{eqn:risk}. They also presented a modified formulation suitable for the minimum graph coloring problem. Instead of only choosing whether to pick an asset or not, Rosenberg and Rounds \cite{Rounds_risk} used binary variables to represent whether to hold long or short positions. This approach is known as long-short minimum risk parity optimization. The authors represented the problem by an Ising model, which is equivalent to a QUBO problem, as follows:

\begin{equation*}
    \min_{\vec{s} \in \{-1, 1\}^N} \vec{s}^\mathsf{T} W^\mathsf{T} \Sigma W \vec{s},
\end{equation*}
where $W$ is an $N \times N$ diagonal matrix with the entries of $\vec{w} \in [0, 1]^{N}$, such that $\lVert\vec{w}\rVert_1 = 1$, on the diagonal; $\vec{w}$, contains a fixed weight for each asset $j$ indicating the proportion of the portfolio that should hold a long or short position in $j$. In both cases, the authors utilized quantum annealing  (Section \ref{sec:QuantumAnnealing}).

Instead of just risk minimization, the problem can be reformulated to specify a desired fixed expected return,  $\mu \in \mathbb{R}$,
  \begin{equation}
  \label{eqn:portoptnolagrange}
    \underset{\Vec{x}\in\mathbb{B}^N}{\min} \Vec{x}^\mathsf{T} \Sigma \Vec{x}:
    ~\Vec{r}^\mathsf{T}\vec{x} = \mu,~
    ~\Vec{x}^\mathsf{T}\Vec{1} = \beta 
    \end{equation}
    or to maximize the expected return,
   
    \begin{equation}
    \label{eqn:portoptconstr}
    \underset{\Vec{x}\in\mathbb{B}^N}{\min} q\Vec{x}^\mathsf{T} \Sigma \Vec{x} - \Vec{r}^\mathsf{T}\Vec{x}~:
    ~\Vec{x}^\mathsf{T}\Vec{1} = \beta ,
    \end{equation}
    where $\vec{x}$ is a vector of Boolean decision variables, $\vec{r} \in \R^N$ contains the expected returns of the assets,  $\Sigma$ is the same as above, $q>0$ is the risk level, and $\beta \in \mathbb{N}$ is the budget. In both cases, the Lagrangian of the problem, where the dual variables are used to encode user-defined penalties, can be mapped to a QUBO (Section \ref{sec:QUBO, Ising}). These formulations are known as mean-variance portfolio optimization problems \cite{markowitzmpt1952}. More constraints can also be added to these problems.
 
 Hodson et al.~\cite{hodson2019portfolio} and Slate et al.~\cite{slate2020quantum} both utilized QAOA to solve a problem similar to Equation \eqref{eqn:portoptconstr}. They allowed for three possible decisions: long position, short position, or neither. In addition, they utilized mixing operators (Section \ref{sec:QAOA}) that accounted for constraints. Alternatively, Gilliam et al.~\cite{Gilliam2021groveradaptive} utilized an extension of Grover adaptive search (Section \ref{sec:durrhoyer}) to solve Equation \eqref{eqn:portoptconstr} with a quadratic speedup over classical unstructured search. Rosenberg  et al.~\cite{Rosenberg_2016} and Mugel et al.~\cite{mugel2020dynamic} solved dynamic versions of Equation \eqref{eqn:portoptconstr}, where portfolio decisions are made for multiple time steps. Additionally, an analysis of benchmarking quantum annealers for portfolio optimization can be found in a paper by Grant et al.~\cite{Grant_2021}. Furthermore, an approach using reverse quantum annealing (Section \ref{sec:QuantumAnnealing}) was proposed by Venturelli and Kondratyev (Section \ref{sec:reverseannealingportopt}). Note that any of the methods from Section \ref{sec:combopt} can be applied whenever quantum annealing can. Currently, however, because  quantum annealers possess more qubits than existing gate-based devices possess, more experimental results with larger problems have been obtained with annealers.

\subsubsubsection{Convex Formulations} 
This part of the section focuses on formulations of the portfolio optimization problem that are convex. The original formulation of mean-variance portfolio optimization by Markowitz \cite{markowitzmpt1952} was convex, and thus it did not contain integer constraints, as were included in the discussion above. 
Thus, if the binary variable constraints are relaxed, the optimization problem, represented by Equation \eqref{eqn:portoptnolagrange}, can be reformulated as a convex quadratic program with linear equality constraints with both a fixed desired return and budget (Equation \ref{opt_equation}):

\begin{equation}
\label{opt_equation}
\underset{\vec{w} \in \mathbb{R}^N}{\text{min }}  \vec{w}^\mathsf{T} \Sigma \vec{w}~:~ \vec{p}^\mathsf{T} \vec{w} = \xi,~ \vec{r}^\mathsf{T} \vec{w} = \mu,
\end{equation}
    where $\vec{w} \in \mathbb{R}^N$ is the allocation vector. The $i$th entry of the vector $\vec{w}$ represents the proportion of the portfolio that should invest in asset $i$. This is in contrast to the combinatorial formulations where the result was a Boolean vector. The vectors $\Vec{p} \in \mathbb{R}^N$ and $\Vec{r} \in \mathbb{R}^N$ contain the assets' prices and returns, respectively; $\xi \in \mathbb{R}$ is the budget; and $\mu \in \mathbb{R}$ is desired expected return. In contrast to Equation \eqref{eqn:portoptnolagrange}, Equation \eqref{opt_equation} admits a closed-form solution \cite{kerenidis2019quantumPortOpt}. However, more problem-specific conic constraints and cost terms (assuming these terms are convex) can be added, increasing the complexity of the problem, in which case it can potentially be solved with more sophisticated algorithms (e.g., interior-point methods). As mentioned, there exist polynomial speedups provided by quantum computing for common cone programs (Section \ref{sec:convex}). Equation \eqref{opt_equation} with additional positivity constraints on the allocation vector can be represented as an SOCP and solved with quantum IPMs \cite{kerenidis2019quantumPortOpt}.
    
    Considering the exact formulation in Equation \eqref{opt_equation}, however, we  might be able to obtain superpolynomial speedup if we relax the problem even further. Rebentrost and Lloyd \cite{rebentrost2018quantum} presented a relaxed problem where the goals were to sample from the solution and/or compute statistics. Their approach was to apply the method of Lagrange multipliers to Equation \eqref{opt_equation} and obtain a linear system. This system can be formulated as a quantum linear systems problem and solved with QLS methods (Section \ref{sec:hhlsection}). The result of solving the QLSP (Section \ref{sec:hhlsection}) is a normalized quantum state corresponding to the solution found with time complexity that is polylogarithmic in the system size, $N$. Thus, if only partial information about the solution is required, QLS methods potentially provide an exponential speedup, in $N$, over all known classical algorithms for this scenario, although the sparsity and well-conditioned matrix assumptions mentioned in Section \ref{sec:hhlsection} have to be met to obtain this speedup. As mentioned in Section \ref{sec:hhlsection}, however, in scenarios involving low-rank matrices randomized numerical linear algebra algorithms can be applied to obtain the same exponential speedup in dimension \cite{gilyen_stoch_regr2018}. However, these methods typically have a high polynomial dependence on the rank and precision, which still allows for a potential polynomial quantum advantage. This dequantized approach was benchmarked in  \cite{Arrazola_2020}.
\subsubsection{Swap Netting}
\label{sec:swapnetting}
A \emph{swap} is a contract where two parties agree to exchange cash flows periodically for a specific term \cite{Swap_netting}. Common types of swaps are credit default swaps \cite{stulz2010credit}, foreign exchange currency swaps, and interest rate swaps \cite{hu1989swaps}. The simplest example is the fixed-to-floating interest rate swap, where two parties exchange the responsibility of paying fixed-interest rate and floating-interest rate payments based on a principal amount called the notional value \cite{bicksler1986economic}. Examples of potential reasons to enter into such a contract could be to hedge or take advantage of the opposite party's comparative advantage \cite{bicksler1986economic}. Once an agreement between the parties has been made, a clearing house converts an agreement between two parties into two separate agreements with the clearing house. The clearing house potentially has multiple swaps with multiple parties. In the case that multiple cash flows cancel, the clearing house would like to \emph{net} all of the contracts that cancel into a new one for only the net flow \cite{duffie1996swap}. This is to reduce the risk exposure associated with having multiple contracts. 
 
 Rosenberg et al.~\cite{Swap_netting} formulated the task of finding a set of ``nettable'' swaps associated with a single counterparty that maximizes the total notional value as a QUBO. This optimization problem can be solved in parallel for each subset of potentially nettable swaps. Since swaps can be netted for potentially many reasons,  the set of combinations can grow significantly. The procedure proposed by the authors is as follows. First, partition the set of swaps associated with a counterparty into subsets that can be netted. Second, for a subset $\mathcal{M}$ in the partition, with $N := |\mathcal{M}|$, construct the following QUBO:
\begin{equation*}
    \max_{\vec{x} \in \mathbb{B}^{N}}~\vec{x}^\mathsf{T}\vec{p} - \alpha\left(\sum\limits_{i \in \mathcal{M}}\vec{x}_i\vec{d}_i\vec{p}_i\right)^2 - \beta \vec{x}^{\mathsf{T}}V\vec{x},
\end{equation*}
where $\vec{x} \in \mathbb{B}^{N}$ is a vector of decision variables indicating whether to include the swap in the netted subset, $\vec{p} \in \mathbb{R}^{N}$ is the vector of notional values associated with the swaps, and $\vec{d} \in \{+1, -1\}^N$ indicates the direction of the fixed-rate interest payments: $+1$ if the clearing house pays the counterparty and $-1$ if the opposite is true. $V \in \mathbb{R}_{\geq 0}^{N \times N}$ signifies how incompatible two potentially ``nettable'' swaps are, and  $\alpha$ and $\beta$ weigh the importance of the second and third terms.
The combined goal of the three terms is to maximize the total notional value netted (first term) such that the notional values of the selected swaps cancel (second term) and the swaps are compatible (third term). The QUBO problems were solved by utilizing quantum annealing (Section \ref{sec:QuantumAnnealing}). 
\subsubsection{Optimal Arbitrage}
\label{sec:optimalarbitrage}

An \emph{arbitrage} opportunity occurs when one can profit from buying and selling similar financial assets in different markets because of price differences \cite{shleifer1997limits}. Arbitrage is due to the lack of market equilibrium;  acting on arbitrage opportunities should shift the market back into equilibrium  \cite{delbaen2006mathematics}. Thus, it is to one's advantage to be able to identify these opportunities as quickly and efficiently as possible. An example is currency arbitrage, where one can buy and sell currencies in foreign exchange (FX) markets such that when one ends up holding the initial currency again, one  has profited because of the differences in FX rates (conversion rates) \cite{shleifer1997limits}.

One way to formulate the problem of arbitrage detection is to  view the different assets as the vertices of a graph with weighted edges between them indicating the negative logarithm of the conversion rate. The problem then reduces to identifying negative cycles \cite{Rosenberg_arbitrage}, for which multiple classical algorithms exist \cite{cherkassky1999negative}. However, the arbitrage opportunity detected is not necessarily the \emph{optimal} one.

Soon and Ye \cite{soon2011currency} formulated the task of identifying the optimal currency arbitrage opportunity in the graph structure mentioned above as a binary linear programming problem. Thus this problem is $\mathsf{NP}$-hard (Section \ref{sec:combopt}). Rosenberg \cite{Rosenberg_arbitrage} reformulated this as a QUBO and solved it using quantum annealing. Alternatively, this can  be solved by  utilizing the other techniques for combinatorial optimization from Section \ref{sec:combopt}. The QUBO resulted from converting the constrained binary linear program of Soon and Ye into an unconstrained problem. In addition, Rosenberg 
provided an alternative model that allows for arbitrage decisions that revisit the same asset (i.e., the same vertex). By definition, this is not allowed in a cycle. Solving this formulation of the problem effectively consists of selecting the assets to buy or sell and the point in the arbitrage loop at which the buying or selling occurs. However, this results in a problem with significantly more binary variables \cite{Rosenberg_arbitrage}.

\subsubsection{Identifying Creditworthiness}
\label{sec:quantcreditscore}

Creditworthiness identification is an important problem in finance. With large sums of money being loaned globally, it is important to be able to identify the key \emph{independent} features that are \emph{influential} in determining the creditworthiness of the requester. The task of identifying features that balance independence and influence can be expressed as a QUBO problem. 

The problem setup for this \emph{combinatorial} feature selection task is as follows. Assume that from an initial set of $N$ features, we want to select $K$ to use. The data is cleaned and set up in a matrix $U \in \mathbb{R}^{M \times N}$ where each column represents a feature and each row contains the values of the features for each of the $M$ past credit recipients. The past decisions that were made regarding creditworthiness are represented in a column vector $\vec{v} \in \mathbb{B}^{M}$. The goal is to find a set of columns in $U$ that are most correlated with $\vec{v}$ but not with each other. The next step is to construct a quadratic objective function that is a convex combination of terms evaluating the influence and independence of the features. Milne et al.~\cite{Optimal_Feature} utilized quantum annealing to solve the QUBO. Thus the constraint of selecting $K$ features can be accounted for by adding a penalty term to the QUBO. However, the problem can be solved by using any of the techniques mentioned in Section \ref{sec:combopt}. Classical or quantum machine learning algorithms (Section \ref{sec:ML}) can then be used to classify new applicants based  on the features selected \cite{Optimal_Feature}.

\subsubsection{Financial Crashes}
\label{sec:fincrashes}
A financial network can viewed as a collection of entities whose valuations are interconnected to the valuations of other members of the network. The analysis of financial networks is important for being able to predict crises. Elliot et al.~\cite{financial_model} developed a problem formulation for counting the number of intuitions that fail in a financial network. Their proposed problem was found to be $\mathsf{NP}$-hard \cite{hemenway2016sensitivity}. If solved, however, it allows for analyzing how small changes in the network propagate through a web of complex interdependencies, something that is difficult \cite{grabel2003predicting} with alternative techniques for crash prediction \cite{Classical_crash}.
By modeling a financial system, mapping the system to a QUBO problem, and perturbing the system to find the ground state it reaches, we can determine the effect of small perturbations on the market. 

We briefly discuss the formulation of Elliott et al.~\cite{financial_model}.
First, suppose the financial network consists of $N$ institutions and $M$ assets, with prices $\vec{p} \in \mathbb{R}_{\geq 0}^M$, that the institutions can own. $D \in \mathbb{R}_{\geq 0}^{N \times M}$ contains the percentage of each asset owned by each institution. In addition, the institutions can  have a stake, cross holdings, in each other represented by the matrix $C \in \mathbb{R}_{\geq 0}^{N \times N}$. The diagonal entries of $C$ are set to $0$, and the self-ownership is represented by the diagonal matrix $\Tilde{C}$ such that $\Tilde{C}_{jj} := I - \sum_{i}C_{ij}$. The equity valuations, $\vec{v}$, at equilibrium satisfy 

\begin{equation}
\label{eqn:equity}
\vec{v} = \Tilde{C}(I - C)^{-1}(D\vec{p} - \vec{b}(\vec{v}, \vec{p})),
\end{equation}
where  $\vec{b}(\vec{v}, \vec{p})$ is a term that decreases the equity value further if it falls past a certain threshold. This signifies a loss of investor confidence or the inability to pay operating costs, signifying failure \cite{financial_model}.

Or\'us et al.~\cite{Financial_crash} considered the problem of minimizing a  quadratic function of the difference between the left and right sides of Equation \eqref{eqn:equity}. This allowed for a variational method to be used. Thus the minimum is attained when Equation \eqref{eqn:equity} is satisfied. They then formulated this as a QUBO. Thus, after perturbing the system, the new equilibrium can be found by solving the QUBO. The next step is to count the number of institutions that go into failure as a function of the perturbation to assess the stability. The valuations $\vec{v}$ are continuous variables that can be approximated by using finite binary expansions. This method converts the initial problem into a binary optimization problem. Or\'us et al.~solved the QUBO using quantum annealing, and Fellner et al.~\cite{fellner2021parity} solved it with QAOA.
However, the right side of Equation \eqref{eqn:equity} contains terms that are not quadratic. Thus, the resulting diagonal observable encoding it is not two-local (more specifically, not Ising). Or\'us et al.~utilized the approach of Chancellor et al.~\cite{chancellor2017circuit} to convert the higher-order terms into two-body interaction terms, at the cost of extra variables. For the approach of Fellner et al.~the problem Hamiltonian did not need to be mapped to an Ising Hamiltonian (Section \ref{sec:QAOA}).

\section{Machine Learning}
\label{sec:ML}
The field of machine learning has become a crucial part of various applications in the finance industry. Rich historical financial data and advances in machine learning make it possible, for example, to train sophisticated models to detect patterns in stock markets, find outliers and anomalies in financial transactions, automatically classify and categorize financial news, and optimize portfolios.

Quantum algorithms for machine learning can be further classified by whether they follow a fault-tolerant approach, a near-term approach, or a mixture of the two approaches. The fault-tolerant approach typically requires a fully error-corrected quantum computer, and the main goal is to find speedups compared with their classical counterparts, achieved mostly by applying the quantum algorithms introduced in Section \ref{sec:hhlsection} as subroutines. Thus the applicability of these algorithms may be limited since most of them begin with first loading the data into a quantum system, which can require exponential time \cite{aaronson2015read, plesch2011quantum}. This issue can be addressed in theory by having access to qRAM (Section \ref{sec:grovers}); as of today, however, no concrete implementation exists. On the other hand, with the development of more NISQ devices, near-term approaches aim to explore the capability of these small-scale quantum devices. Quantum algorithms such as VQAs (Section \ref{sec:qva}) consider another type of enhancement for machine learning: the quantum models might be able to generate correlations that are hard to represent classically \cite{mcclean2016theory,huang2021power}. Such models face many challenges, however, such as trainability, accuracy, and efficiency, that need to be addressed in order to maintain the hope of achieving quantum advantage when scaling up these near-term quantum devices \cite{Cerezo_2021}. Substantial efforts are needed before we can answer the question of whether quantum machine learning algorithms can bring us practical and useful applications, and it might take decades. Nevertheless, quantum machine learning shows great promise to find improvements and speedups to elevate the current capabilities vastly.

\subsection{Regression}
\label{sec:regression}
Regression is the process of fitting a numeric function from the training data set. This process is often used to understand how the value changes when the attributes vary, and it is a key tool for economic forecasting. Typically, the problem is solved by applying least-squares fitting, where the goal is to find a continuous function that approximates a set of $N$ data points $\{x_i, y_i\}$. The fit function is of the form \begin{eqnarray*}
f(x, \vec{\lambda}) := \sum_{j=1}^Mf_j(x)\lambda_j, \end{eqnarray*}
where $\vec{\lambda}$ is a vector of parameters and $f_j(x)$ is function of $x$ and can be nonlinear. The optimal parameters can be found by minimizing the least-squares error: \begin{eqnarray*}
\min_{\vec{\lambda}} \sum_{i=1}^N \left|f(x_i, \vec{\lambda}) - y_i\right|^2 = |F\vec{\lambda} - y|^2, \end{eqnarray*}
where the $N\times M$ matrix $F$ is defined through $F_{ij} = f_j(x_i)$. Wiebe et al.~\cite{wiebe2012quantum} proposed a quantum algorithm to solve this problem, where they encoded the optimal parameters of the model into the amplitudes of a quantum state and developed a quantum algorithm for estimating the quality of the least-squares fit by building on the HHL algorithm (Section \ref{sec:hhlsection}). The algorithm consists of three subroutines: a quantum algorithm for performing the pseudoinverse of $F$, an algorithm that estimates the fitting quality, and an algorithm for learning the parameter $\vec{\lambda}$. Later, Wang \cite{wang2017quantum} proposed using the CKS matrix-inversion algorithm (Section \ref{sec:hhlsection}), which has better dependence on the precision in the output than HHL has, and used amplitude estimation (Section \ref{sec:QuantumAmplitudeEstimation}) to estimate the optimal parameters.  Kerenedis and Prakash developed algorithms utilizing QSVE (Section \ref{sec:hhlsection}) to perform a coherent gradient descent for normal and weighted least-squares regression \cite{Kerenidis_2020LS}. Moreover, quantum annealing (Section \ref{sec:QuantumAnnealing}) has been used to solve least-squares regression when formulated as a QUBO (Section \ref{sec:combopt}) \cite{date2020adiabatic}.

A Gaussian process (GP) is another widely used model for regression in supervised machine learning. It has been used, for example, in predicting the price behavior of commodities in
financial markets. Given a training set of $N$ data points $\{x_i, y_i\}$, the goal is to model a latent function $f(x)$ such that \begin{eqnarray*}
y = f(x) + \epsilon_{\text{noise}}, \end{eqnarray*}
where $\epsilon_{\text{noise}} \sim \mathcal{N}(0, \sigma_n^2)$ is independent and identically distributed Gaussian noise. A practical implementation of the Gaussian process regression (GPR) model needs to compute the Cholesky decomposition and therefore requires $O(N^3)$ running time. Zhao et  al.~\cite{zhao2019quantum} proposed using the HHL algorithm to speed up computations in GPR. By repeated sampling of the results of specific quantum measurements on the output states of the HHL algorithm, the mean predictor and the associated variance can be estimated with bounded error with potentially an exponential speedup over classical algorithms.

Another recent development is to use a feedforward quantum neural network with a quantum feature map, i.e., a unitary with configurable parameters used to encode  the classical input data, and a Hamiltonian cost function evaluation for the purposes of continuous-variable regression in a technique called quantum circuit learning (QCL)~\cite{PhysRevA.98.032309}. This technique allows for a low-depth circuit and has important application opportunities in quantum algorithms for finance.

\subsection{Classification}\label{sec:qmlclassification}
Classification is the process of placing objects into predefined groups. This type of process is also called pattern recognition. This area of machine learning can be used effectively in risk management and large data processing when the group information is of particular interest, for example, in creditworthiness identification and fraud detection.

\subsubsection{Quantum Support Vector Machine}
In machine learning, support vector machines (SVMs) are supervised learning models used to analyze data for classification and regression analysis. Given $M$ training data elements of the form $\{(\vec{x}_j, y_j): \vec{x}_j\in \mathbb{R}^N, y_j = \pm 1\}$, the task for the SVM is to classify a vector into one of two classes; it finds a maximum margin hyperplane with normal vector $\vec{w}$ that separates the two classes. In addition, the margin is given by the two hyperplanes that are parallel and separated by the maximum possible distance $2/\lVert\vec{w}\rVert_2$. Typically, we use the Lagrangian dual formulation to maximize the function \begin{eqnarray*}
L(\alpha) = \sum_{j=1}^M y_j\alpha_j -\frac{1}{2}\sum_{j,k=1}^M\alpha_j K_{jk}\alpha_k,
\end{eqnarray*}
where $\vec{\alpha} = (\alpha_1, \cdots, \alpha_M)^T$, subject to the constraints $\sum_{j=1}^M \alpha_j = 0$ and $y_j\alpha_j \geq 0$. Here, we have a key quantity for the supervised machine learning problem \cite{muller2001introduction}, the kernel matrix $K_{jk} = k(\vec{x}_j, \vec{x}_k) = \vec{x}_j \cdot \vec{x}_k$. Solving the dual form involves evaluating the dot products using the kernel matrix and then finding the optimal parameters $\alpha_j$ by quadratic programming, which takes $\mathcal{O}(M^3)$ in the non-sparse case. A quantum SVM was first proposed by Rebentrost et al.~\cite{rebentrost2014quantum}, who showed that a quantum SVM can be implemented with $\mathcal{O}(\log MN)$ runtime in both training and classification stages, under the assumption that there are oracles for the training data that return quantum states $\ket{\vec{x}_j} = \frac{1}{\lVert\vec{x}_j\rVert_2}\sum_{k=1}^N(\vec{x}_j)_k\ket{k}$, the norms $\lVert\vec{x}_j\rVert_2$ and the labels $y_j$ are given, and the states are constructed by using qRAM \cite{lloyd2013quantum,Giovannetti_2008}. The core of this quantum classification algorithm is a non-sparse matrix exponentiation technique for efficiently performing a matrix inversion of the training data inner-product (kernel) matrix. Lloyd et al.~\cite{lloyd2013quantum} showed that, effectively, a low-rank approximation to kernel matrix is used due to the eigenvalue filtering procedure used by HHL \cite{harrow2009quantum}. As mentioned in Section \ref{sec:convex}, Kerenedis et al.~proposed a quantum enhancement to second-order cone programming that was subsequently used to provide a small polynomial speedup to the training of the classical $\ell_1$ SVM \cite{Kerenidis_2021}. In addition, classical SVMs using quantum-enhanced feature spaces to construct quantum kernels have been proposed by Havl{\'\i}{\v{c}}ek et al.~\cite{Havl_ek_2019}. One benefit of quantum kernels is that there exist metrics for testing for potential quantum advantage \cite{huang2021power}. In addition, there have been techniques proposed to enable generalization  \cite{shaydulin2021bandwidth, canatar2022}, which has been observed to be an issue with standard quantum-kernel methods \cite{Kublerkernels2021}.

\subsubsection{Quantum Nearest-Neighbors Algorithm}
The $k$-nearest neighbors technique is an algorithm that can be used for classification or regression. Given a data set, the algorithm assumes that data points closer together are more similar and uses distance calculations to group close points together and define a class based on the commonality of the nearby points. Classically, the algorithm works as follows: ($1$) determine the distance between the query example and each other data point in the set, ($2$) sort the data points in a list indexed from closest to the query example to farthest, and ($3$) choose the first $k$ entries, and, if regression, return the mean of the $k$ labels and, if classification, return the mode of $k$ labels. The computationally expensive step is to compute the distance between elements in the data set. The quantum nearest-neighbor classification algorithm was proposed by Wiebe et al.~\cite{wiebe2014quantum}, who used the Euclidean distance and the inner product as distance metrics.  The distance between two points is encoded in the amplitude of a quantum state. They used amplitude amplification  (Section \ref{sec:QuantumAmplitudeAmplification}) to amplify the probability of creating a state to store the distance estimate in a register, without measuring it.  They then applied the D\"{u}rr--H{\o}yer algorithm (Section \ref{sec:durrhoyer}). Later, Ruan et al.~proposed using the
Hamming distance as the metric \cite{ruan2017quantum}. More recently, Basheer et al. have proposed using a quantum $k$ maxima-finding algorithm to find the $k$-nearest neighbors and use the fidelity and inner product as measures of similarity \cite{basheer2021quantum}.

\subsection{Clustering}
\label{sec:clustering}
Clustering, or cluster analysis, is an unsupervised machine learning task. It explores and discovers the grouping structure of the data. In finance, cluster analysis can be used to develop a trading approach that helps investors build a diversified portfolio. It can also be used to analyze different stocks such that the stocks with high correlations in returns fall into one basket.

\subsubsection{Quantum $k$-Means Clustering}
\label{sec:kmeansclust}
The $k$–means clustering algorithm (also known as Lloyd’s algorithm) \cite{macqueen1967some} clusters training examples into $k$ clusters based on their distances to each of the $k$ cluster centroids. The algorithm for $k$–means clustering  is as follows: ($1$) choose initial values for the $k$ centroids randomly or by some chosen method, ($2$) assign each vector to the closest centroid, and ($3$) recompute the centroids for each cluster; then repeat steps ($2$) and ($3$) until the clusters converge. The problem of optimally clustering data is $\mathsf{NP}$–hard, and thus finding the optimal clustering using this algorithm can be computationally expensive. Quantum computing can be leveraged to accelerate a single step of $k$-means. For each iteration, the quantum Lloyd's algorithm \cite{lloyd2013quantum} estimates the distance to the centroids in $\mathcal{O}(M\log(MN))$, while classical algorithms take $\mathcal{O}(M^2N)$, where $M$ is the number of data points and $N$ is the dimension of the vector. Wiebe et al.~\cite{wiebe2014quantum} showed that, with the same technique they used for the quantum nearest-neighbor algorithm described in Section \ref{sec:qmlclassification}, a step for $k$–means can be performed by using a number of queries that scales as $\mathcal{O}(M\sqrt{k}\log(k)/\epsilon)$. While a direct classical method requires $\mathcal{O}(kMN)$, the quantum solution is substantially better if $kN \gg M$. Kerenidis et al.~\cite{kerenidis2018qmeans} proposed $q$-means, which is the quantum equivalent of $\delta$-$k$-means. The $q$-means algorithm has a running time that depends polylogarithmically on the number of data points. A NISQ version of $k$-means clustering using quantum computing has been proposed by Khan et al.~\cite{khan2019kmeans}. Another version of $k$-means clustering has been proposed by Miyahara et al.~\cite{PhysRevA.101.012326} based on a quantum expectation-maximization algorithm for Gaussian mixture models.

\subsubsection{Quantum Spectral Clustering}
\label{sec:speccluster}
Spectral clustering methods \cite{ng2002spectral} have had great successes in clustering tasks but suffer from a fast-growing running time of $\mathcal{O}(N^3)$, where $N$ is the number of points in the data set. These methods obtain a solution for clustering by using the principal eigenvectors  of the data matrix or the Laplacian matrix. Daskin \cite{daskin2017quantum} employed QAE (Section \ref{sec:QuantumAmplitudeEstimation}) and QPE (Section \ref{sec:qpesection}) for spectral clustering. Apers et al.~\cite{apers2020quantum} proposed quantum algorithms using the graph Laplacian for machine learning applications, including spectral $k$-means clustering. More recently, Kerenidis and Landman~\cite{kerenidis2021quantum} proposed a quantum algorithm to perform spectral clustering. They first created a quantum state corresponding to the projected Laplacian matrix, which is efficient assuming qRAM access, and then applied the $q$-means algorithm \cite{kerenidis2018qmeans}. Both steps depend polynomially on the number of clusters $k$ and polylogarithmically on the dimension of the input vectors.

More discussions on unsupervised quantum machine learning techniques can be found in the works of Otterbach et al.~\cite{Otterbach_ML} and Aïmeur et al.~\cite{Aimeur_Clustering, Aimeur_ML}. Kerenidis et al.~\cite{Kerenidis} discussed a quantum version of expectation-maximization, a common tool used in unsupervised machine learning.  Kumar et al. described a technique for using quantum annealing for combinatorial clustering~\cite{Kumar_2018}.

\subsection{Dimensionality Reduction}
\label{sec:dimensionalityreduc}
The primary linear technique for dimensionality reduction, principal component analysis, performs a linear mapping of the data to a lower-dimensional space in such a way that the variance of the data in the low-dimensional representation is maximized. Generally, principal component analysis mathematically amounts to finding dominant eigenvalues and eigenvectors of a very large matrix. The standard context for PCA involves a data set with observations on $M$ variables for each of $N$ objects. The data defines an  $N\times M$ data matrix $X$, in which the $j$th column is the vector $\vec{x}_j$ of observations on the $j$th variable. The aim is to find a linear combination of the columns of the matrix $X$ with maximum variance. This process boils down to finding the largest eigenvalues and corresponding eigenvectors of the covariance matrix. Lloyd et al.~\cite{lloyd2014quantum} proposed quantum PCA. In addition, they showed that multiple copies of
a quantum system with density matrix $\rho$ can be used to construct the unitary transformation $e^{-i\rho t}$, which leads to revealing the eigenvectors corresponding to the large eigenvalues in quantum form for an unknown low-rank density matrix. He et al. introduced a low-complexity quantum principal component analysis algorithm \cite{He_2020}. Other quantum algorithms for PCA were proposed in \cite{Yu_2019,lin2019improved,Bellante}. Lastly, Li et al.~\cite{Li_2020} presented a quantum version of kernel PCA, Cong~\cite{Cong_2016} presented quantum discriminant analysis, and Kerenidis and Luongo~\cite{Kerendis_2020} developed quantum slow feature analysis.

For data sets that are high dimensional, incomplete, and noisy, the extraction of information is generally challenging. Topological data analysis (TDA) is an approach to study qualitative features and analyze and explore the complex topological and geometric structures of data sets. Persistent homology (PH) is a method used in TDA to study qualitative features of data that persist across multiple scales. Lloyd et al.~\cite{lloyd2016quantum} proposed a  quantum algorithm to compute the Betti numbers, that is, the numbers of connected components, holes, and voids in the data set at varying scales. Essentially, the algorithm operates by finding the eigenvectors and eigenvalues of the combinatorial Laplacian and estimating Betti numbers to all orders and to accuracy $\delta$ in time $\mathcal{O}(n^5/\delta)$. The most significant speedup is for dense clique complexes \cite{gyurik2020quantum}. An improved version as well as a NISQ version of the quantum algorithm for PH has been proposed by Ubaru et al.~\cite{ubaru2021quantum}. Further improvements for implementing the Dirac operator, required for PH, have been made by Kerenedis and Prakash \cite{kerendis2022}.

\subsection{Generative Models}
Unsupervised generative modeling is at the forefront of
deep learning research. The goal of generative modeling is to model the probability distribution of observed data and generate new samples accordingly. One of the most promising aspects of achieving potential quantum advantage lies in the sampling advantage, and many applications in finance require generating samples from complex distributions. Therefore, further investigations of generative quantum machine learning for finance are needed.

\subsubsection{Quantum Circuit Born Machine}
The quantum circuit Born machine (QCBM) \cite{Marcello2019} directly exploits the inherent probabilistic interpretation of quantum wave functions and represents a probability distribution using a quantum pure state instead of the thermal distribution. Liu 
et al.~\cite{liu2018differentiable} developed an efficient gradient-based learning algorithm to train the QCBM.
Numerical simulations suggest that in the task of learning the distribution, QCBM at least matches the performance of the restricted Boltzmann machine (Section \ref{sec:quantumboltzmann}) and demonstrates superior performance as the model scales \cite{coyle2021quantum}. QCBM was also used to model copulas \cite{zhu2021generative}, a family of multivariate distributions with uniform marginals. Recently, Kyriienko et al.~\cite{kyriienko2022protocols} proposed separating the training and sampling stages. In the training stage, they build a model in the latent space, followed by a variational circuit. They showed how probability distributions can be trained and sampled efficiently, and how SDEs can act as differential constraints on such trainable quantum models.

\subsubsection{Quantum Bayesian Networks}
Bayesian networks are probabilistic graphical models \cite{koller2009probabilistic} representing random variables and conditional dependencies via a directed acyclic graph, where each edge corresponds to a conditional dependency and each node corresponds to a unique random variable. Bayesian inference on this graph has many applications, such as prediction, anomaly detection, diagnostics, reasoning, and decision-making with uncertainty. Despite the large complications, exact inference is \#$\mathsf{P}$-hard, but a quadratic speedup in certain parameters can be obtained by using quantum technologies \cite{Low_2014}. Mapping this problem to a quantum Bayesian network seems plausible since quantum mechanics naturally describes a probabilistic distribution. Tucci \cite{tucci1995quantum} introduced the quantum Bayesian network as an analogue to classical Bayesian networks, using quantum complex amplitudes to represent the conditional probabilities in a classical Bayesian network. Borujeni et al.~\cite{borujeni2021quantum} proposed a procedure to design a quantum circuit to represent a generic discrete Bayesian network: ($1$) map each node in the Bayesian network to one or more qubits, ($2$) map conditional probabilities of each node to the probability amplitudes associated with qubit states, and ($3$) obtain the desired probability amplitudes through controlled rotation gates. In a later work they tested it on IBM quantum hardware~\cite{borujeni2020experimental}. The applications in finance include portfolio simulation~\cite{KLEPAC2017391} and decision-making modeling~\cite{Moreira_2016}.

\subsubsection{Quantum Boltzmann Machines}
\label{sec:quantumboltzmann}
A Boltzmann machine is an undirected probabilistic graphical model (i.e., a Markov random field) \cite{koller2009probabilistic} inspired by thermodynamics. Inference, classically, is usually performed by utilizing Markov chain Monte Carlo methods to sample from the model's equilibrium distribution (i.e., the Boltzmann distribution) \cite{goodfellow2016deep}. Because of the intractability of the partition function in the general Boltzmann machine, the graph structure is typically bipartite, resulting in a restricted Boltzmann machine \cite{hinton2002training, salakhutdinov2009deep}. These methods were more popular prior to the current neural-network-based deep learning revolution \cite{goodfellow2016deep}. Quantum Boltzmann machines have been implemented by utilizing quantum annealing (Section \ref{sec:QuantumAnnealing}) \cite{benedetti2016estimation, dixit2021training, amin2018quantum}. This is possible because of the annealer's ability to approximately sample ground states of Ising models. In addition, a gate-based variational approach using VarQITE (Section \ref{sec:varite}) has been designed by Zoufal et al.~\cite{Zoufal_2021}. Since quantum states are always normalized, this approach allows for implementing Boltzmann machines with more complex graph structures.

\subsubsection{Quantum Generative Adversarial Networks}
Generative adversarial networks (GANs) represent a powerful tool for classical machine learning:
a generator tries to create statistics for data that mimic those of the true data set, while a discriminator tries to discriminate between the true and fake data. Lloyd and Weedbrook ~\cite{lloyd2018quantum} introduced the notion of quantum generative adversarial networks (QuGANs), where the data consists either of quantum states or of classical data and the generator and discriminator are equipped with quantum information processors. They showed that when the data consists of samples of measurements made on high-dimensional spaces, quantum adversarial networks may exhibit an exponential advantage over classical adversarial networks. QuGAN has been used to learn and load random distribution and can facilitate financial derivative pricing~\cite{Zoufal_2019}.

\subsection{Quantum Neural Networks}
\label{sec:qnn}
Neural networks (NNs) have achieved many practical successes over the past decade, due to the novel computational power of GPUs. On the other side, the growing interest in quantum computing has led researchers to develop different variants of quantum neural networks (QNNs). However,  many challenges arise in developing quantum analogues to the required components of NNs, such as the perceptron, an optimization algorithm, and a loss function. A few different approaches to QNNs have been developed. Most, however,  follow the same steps \cite{Beer_2020}: initialize a network architecture; specify a learning task,  implement a training algorithm, and simulate the learning task.

Current proposals for QNNs, discussed in the following subsections, involve a  diverse collection of ideas with varying degrees of proximity to classical neural networks.

\subsubsection{Quantum Feedforward Neural Network}
\label{sec:quantumffnn}
Cao et al.~\cite{cao2017quantum} proposed  a quantum neuron and demonstrated its applicability as a building block of quantum neural networks. Their approach uses repeat-until-success techniques for quantum gate synthesis. Kerenedis et al.~developed a quantum analogue to the orthogonal feedforward neural network \cite{jia2019orthogonal}; the orthogonality of the weight matrix is naturally preserved by using the reconfigurable beam splitter gate \cite{kerenidis2021classical}. In addition, quantum algorithms can potentially speed up the operations used by classical feedforward neural networks \cite{allcock2020quantum}. Another quantum feedfoward NN has been developed by Beer et al.~\cite{Beer_2020}.

A different proposed path to designing quantum analogues to feedforward NNs on near-term devices is by using parameterized quantum circuits (Section \ref{sec:qva}) \cite{farhi2018classification,henderson2020quanvolutional, Havl_ek_2019, killoran2019continuous, liu2022embeddig}. Various attempts have been made to analyze the expressive power of the circuit-based QNN \cite{schuld2021effect, huang2021power, abbas2021power, schuld2021supervised, herman2022}. Interestingly, the circuit-based QNN can be viewed as a kernel method with a quantum-enhanced feature map \cite{Havl_ek_2019, huang2021power,schuld2021supervised, jerbibeyond2022}.
Furthermore,  analysis has been performed in the neural tangent kernel framework \cite{jacot2020neural} to show the benefits of using a quantum-enhanced feature map \cite{Havl_ek_2019} in conjunction with a classical NN \cite{nakaji2021quantumenhanced}.

\subsubsection{Quantum Convolutional Neural Network}
\label{sec:quantumcnns}
Convolutional neural networks (CNNs) were originally developed by LeCun  et al.~\cite{lecun1998gradient} in the 1980s. Recently, Kerenidis et al.~\cite{kerenidis2019quantum} proposed a quantum convolutional neural network as a shallow circuit, reproducing completely the classical CNN, by allowing nonlinearities and pooling operations. They used a new quantum tomography algorithm with $\ell_{\infty}$ norm guarantees and new applications of probabilistic sampling in the context of information processing. Quantum Fourier convolutional neural networks have been developed by Shen and Liu \cite{shen2021qfcnn}. Cong et al.~\cite{cong2019quantum} developed circuit-based quantum analogues to CNNs.

\subsubsection{Quantum Graph Neural Networks} Graph neural networks have recently attracted a lot of attention from the machine learning community. Effective neural networks are designed by modeling interactions (edges) between objects of interest (nodes) and assigning features to nodes and edges. The approach  exploits real-world interconnectedness in finance. Graph neural networks are particularly helpful for training low-dimensional vector space representations of nodes that encode both complex graph structure and node/edge features. Verdon et al.~\cite{verdon2019quantum} introduced quantum graph neural networks (QGNNs). This is a new type of quantum neural network ansatz that takes advantage of the graph structure of quantum processes and makes them effective on distributed quantum systems. In addition, quantum graph recurrent neural networks \cite{quatgraphrecugnn}, quantum graph convolutional neural networks, and quantum spectral graph convolutional neural networks are introduced as special cases. A networked quantum system is described with an underlying graph $G=(V,E)$  in which each node $v$ corresponds to a quantum subsystem with its Hilbert space $\mathcal{H}_v$ such that the entire system forms a Hilbert space $\mathcal{H} = \bigotimes_{v\in V} \mathcal{H}_v$. This can be extended to edges $\mathcal{H}_E$, and each node can represent more complicated quantum systems than a single qubit (e.g., several qubits or the entire part of a distributed quantum system). With this representation, the edges of a graph describe communication between vertex subspaces. The QGNN ansatz is a parameterized quantum circuit on a
network, which consists of a sequence of different Hamiltonian evolutions, with the whole sequence repeated several times with variational parameterization. In this setting, the parameterized Hamiltonians have a graph topology of interactions.

\subsection{Quantum Reinforcement Learning}
\label{sec:rl}
Reinforcement learning is an area of machine learning that considers how agents ought to take actions in an environment in order to maximize their reward. It has been applied to  many financial applications, including pricing and hedging of contingent claims, investment and portfolio allocation, buying and selling of a portfolio of securities subject to transaction costs, market making, asset liability management, and optimization of tax consequences.

The standard framework of reinforcement learning is based on
discrete-time, finite-state Markov decision processes (MDPs). It comprises five components: $\{S, A_i, p_{ij}(a), r_{i, a}, V, i, j\in S, a\in A_i\}$, where $S$ is the state space; $A_i$ is the action space for state $i$; $p_{ij}(a)$ is the transition probability; $r$ is the reward function defined as $r: \Gamma \to \{-\infty, +\infty\}$, where $\Gamma = \{(i, a) | i\in S, a \in A_i\}$; and $V$ is a criterion function or objective function. The MDP has a history of successive states and decisions, defined as $h_n = (s_0, a_0, s_1, a_1, \cdots, s_{n-1}, a_{n-1}, s_n)$, and the policy $\pi$ is a sequence $\pi = (\pi_0, \pi_1, \cdots)$. When the history at $n$ is $h_n$, the agent will make a decision according to the probability distribution $\pi_n(\cdot|h_n)$ on $A_{s_n}$. At each step $t$, the agent observes the state of the environment $s_t$ and then chooses an action $a_t$. The agent will then receive a reward $r_{t+1}$. The environment then will change to the next state $s_{t+1}$ under action $a_t$, and the agent will again observe and act. The goal of reinforcement learning is to learn a mapping from states to action that maximize the expected sum of discounted reward or, formally, to learn a policy $\pi: S\times \bigcup_{i\in S}A_i \to [0, 1]$, such that \begin{eqnarray*}
\max V_s^{\pi} &=& \mathbb{E}[r_{t+1} + \gamma r_{t+2} + \cdots|s_t=s, \pi] \\
&=& \sum_{a\in A_s} \pi(s, a)[r_s^a + \gamma \sum_{s'}p_{ss'}V_{s'}^{\pi}],
\end{eqnarray*}
where $\gamma \in [0,1]$ is a discount factor and $\pi(s, a)$ is the probability of the agent to selecting action $a$ at state $s$ according to policy $\pi$.

Dong et al.~\cite{dong2008quantum} proposed representing the state set with a quantum superposition state;  the eigenstate obtained by randomly observing the quantum state is the action. The probability of the eigen action is determined by the probability amplitude, which is updated in parallel according to the rewards. The probability of the ``good'' action is amplified by using iterations of Grover's algorithm (Section \ref{sec:grovers}). In addition, they showed that the approach makes a good trade-off between exploration and exploitation using the probability amplitude and can speed up learning through quantum parallelism. Paparo et  al.~\cite{paparo2014quantum} showed that the computational complexity of a particular model, projective simulation, can be quadratically reduced. In addition, quantum neural networks (Section \ref{sec:qnn}) and quantum Boltzmann machines (Section \ref{sec:quantumboltzmann}) have  been used for approximate RL \cite{chen2020variational, chen2021variational, lockwood2020reinforcement, jerbi2021quantum, crawford2016reinforcement}. Cherret et al. described quantum reinforcement learning via policy iteration \cite{Cherrat_2022}. Additionally, Cornelissen explored how to apply quantum gradient estimation to quantum reinforcement learning \cite{Cornelissen_2018}. 

\subsection{Natural Language Modeling}
\label{sec:nlp}
Natural language processing (NLP) has flourished over the past couple of years as a result of advancements in deep learning \cite{otternlp}. One important component of NLP is building language models \cite{naseem2021comprehensive}. A language model is a statistical model used to predict what word will likely appear next in a sentence helpful. Today, deep neural networks using transformers \cite{vaswani2017attention} are utilized to construct language models \cite{devlin2019bert, sybrandt2021cbag, floridi} for a variety of tasks. Concerns remain, however, such as bias \cite{blodgettlanguage}, controversial conclusions \cite{sybrandt2020agatha}, energy consumption, and environmental impact \cite{ dhar2020carbon}. Thus it makes sense to look to quantum computing for potentially better language models for NLP.

The DisCoCat framework, developed by Coecke et al.~\cite{coecke2010mathematical}, combines both meaning and grammar, something not previously done, into a single language model utilizing category theory. For grammar, they utilized pregroups \cite{Lambek1968}, sets with relaxed group axioms, to validate sentence grammar. Semantics is represented by a vector space model, similar to those commonly used in NLP \cite{mikolov2013efficient}. However, DisCoCat also allows for higher-order tensors to distinguish different parts of speech.

Coecke et al.~noted that pregroups and vector spaces are asymmetric and symmetric strict monoidal categories, respectively. Thus, they can be represented by utilizing the powerful diagrammatic framework associated with monoidal categories \cite{coecke2009categories}.
The revelation made by this group \cite{coecke2020foundations} was that this same diagrammatic framework is utilized by categorical quantum mechanics \cite{abramsky2008categorical}. Because of the similarity between the two, the diagrams can be represented by quantum circuits 
\cite{coecke2020foundations} utilizing the algebra of ZX-calculus \cite{Coecke_2011}. Thus, because of this connection and the use of tensor operations, the group argued that this natural language model is ``quantum-native,'' meaning it is more natural to run the DisCoCat model on a quantum computer instead of a classical one \cite{coecke2020foundations}.
Coecke et al. ran multiple experiments on NISQ devices for a variety of simple NLP tasks \cite{de_Felice_2021, lorenz2021qnlp, meichanetzidis2020grammaraware}.

In addition, there are many ways in which QML models, such as QNNs (Section \ref{sec:qnn}), could potentially enhance the capacity of existing natural language neural architectures \cite{bausch2020recurrent, chen2020quantum, Takaki_2021}.

\subsection{Financial Applications}
The field of quantum machine learning is still in its infancy, and many exciting developments are under way. We list here a few representative or potential financial applications. We also point readers to a recent survey on quantum machine learning for finance \cite{pistoia2021quantum}.
\subsubsection{Anomaly Detection}
\label{sec:anomalydetection}
Anomaly detection is a critical task for financial intuitions,  from both a financial point of view and a technological one. In finance, fraud detection is an extremely important form of anomaly detection. Some examples are identifying fraudulent credit card transactions and financial documents. Also  important is anomaly detection on communication networks used by the financial industry \cite{AHMED201619}.
These tasks are typically performed, classically, by utilizing techniques from statistical analysis and  machine learning \cite{fraudDetectionWest}. The machine learning technique commonly used is unsupervised deep learning \cite{pang}.

Researchers have  also  proposed  quantum versions of these approaches, such as quantum GANs (Section \ref{sec:qnn}). Herr et al.~\cite{Herr_2021} used parameterized quantum circuits to model the generator used in AnoGan \cite{schlegl2017unsupervised} for anomaly detection. Quantum amplitude estimation (Section \ref{sec:QuantumAmplitudeEstimation}) was utilized to enhance density-estimation-based anomaly detection \cite{guo2021quantum}. Quantum variants of common clustering methods also exist,  such as $k$-means (Section \ref{sec:clustering}), that could be useful for anomaly detection. Furthermore, quantum-enhanced kernels (Section \ref{sec:qmlclassification}) could be utilized for kernel-based clustering methods \cite{xu2015comprehensive}. Quantum Boltzmann machines (Section \ref{sec:quantumboltzmann}) have also been used for fraud detection \cite{Zoufal_2021}.

\subsubsection{Asset Pricing}
\label{sec:assetpricing}
The fundamental goal of asset pricing is to understand the behavior of risk premiums. Classically, machine learning has shown great potential to improve our empirical understanding of asset returns. For example, Gu et al.~\cite{gu2020empirical} studied the problem of predicting returns in cross-section and time series. They studied a collection of machine learning methods, including linear regression, generalized linear models, principal component regression, and neural networks. Chen et al.~\cite{chen2020deep} proposed an asset pricing model for individual stock returns using deep neural networks. 

The quantum version of these machine learning methods could potentially be applied to asset pricing, and investigations of the applicability are needed. Nagel \cite{Nagel_2021} goes into detail about classical machine learning methods applied to asset pricing. Chen et al.~\cite{chen2020quantum} proposed a quantum version of the long short-term memory architecture, which is a variant of the recurrent neural network that has been demonstrated to be effective for temporal and sequence data modeling. Both make use of parameterized quantum circuits, and these QML models may offer an advantage over classical models in terms of training complexity and prediction performance.

\subsubsection{Implied Volatility}
\label{sec:impliedvol}
Implied volatility is a metric that captures the market's forecast of a likely movement in the price of a security. It is one of the deciding factors in the pricing of options. Sakuma \cite{sakuma2020application} investigated the use of the deep quantum neural network proposed by Beer et al.~\cite{Beer_2020} in the context of learning implied volatility. Numerical results suggest that such a QML model is a promising candidate for developing powerful methods in finance~\cite{pistoia2021quantum}.
\section{Hardware Implementations of Use Cases}
\label{sec:qubit tech}

In this section we discuss experiments that were performed on quantum hardware to  solve financial problems using some of the methods that have been presented. The hardware platforms include superconducting and trapped-ion universal quantum processors, as well as superconducting quantum annealers.

\subsection{Quantum Risk Analysis on a Transmon Device}
In this section we present the experimental results collected by Woerner and Egger \cite{Woerner_2019}, using their method for risk analysis mentioned in Section \ref{sec:quantumriskanaly}. They utilized the IBM transmon \cite{superconductingengineers} quantum processors~\cite{Woerner_2019}.

\begin{figure}[ht]
    \centering
    \includegraphics[width=\columnwidth]{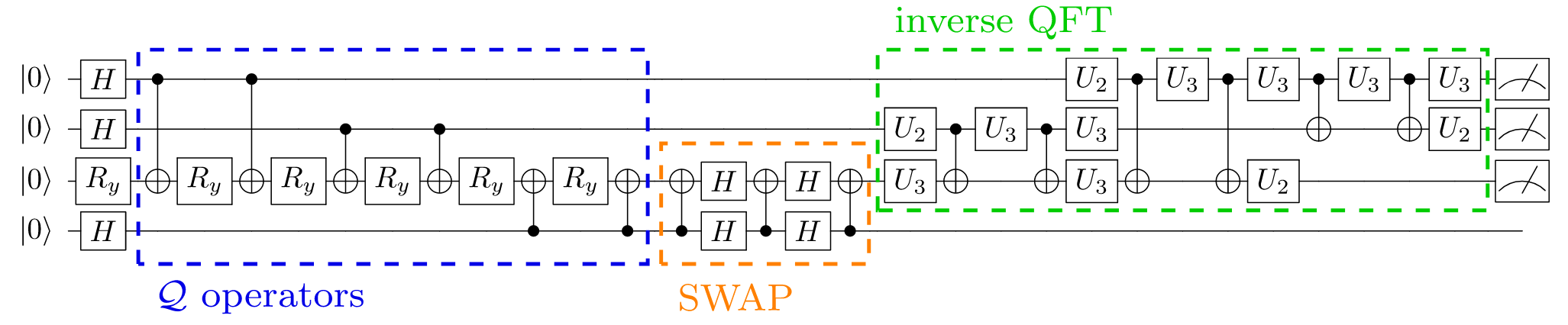}
    \caption{Quantum amplitude estimation circuit for approximating the expected value of a portfolio made up of one T-Bill. The algorithm uses 3 evaluation qubits, corresponding to 8 samples. (Image source:~\cite{Woerner_2019}; use permitted under the Creative Commons Attribution 4.0 International License.)}
    \label{fig:QAE}
\end{figure}

\begin{figure}[ht]
    \centering
    \includegraphics[width=.5\textwidth]{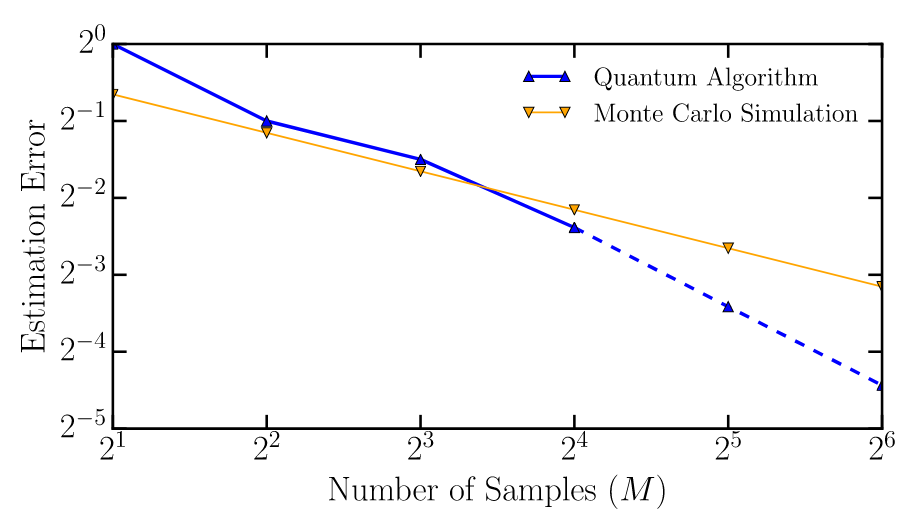}
    \caption{Results of executing the QAE algorithm on the IBMQ 5 Yorktown (ibmqx2) chip for a range of evaluation qubits, with 8,192 shots each. Because of the better convergence rate of the quantum algorithm (blue line), it achieves lower error rates than does Monte Carlo (orange line) for the version with 4 qubits, with the performance gap estimated to grow even more for larger instances on 5 and 6 qubits (blue dashed line). (Image source:~\cite{Woerner_2019}; use permitted under the Creative Commons Attribution 4.0 International License, http://creativecommons.org/licenses/by/4.0/)}
    \label{fig:AE_results}
\end{figure}

The authors tested the algorithm on two toy models: pricing a Treasury-bill, executed on quantum hardware, and determining the financial risk of a two-asset portfolio, performed in simulations that included modeling quantum noise on classical hardware. The QAE algorithm is used to estimate quantities related to a random variable represented as a quantum state, usually using $n$ qubits to map it to the interval $\{0, ..., N - 1\}$, with $N = 2^n$ being the number of possible realizations. The authors presented a method for efficiently applying an approximation to $f(X)$, where QAE computes $\mathbb{E}[f(X)]$, given that $f(x) = \sin^2(p(x))$ for some polynomial $p$. They used this approach to compute statistical moments, such as the mean. The $f(X)$ used for VaR, and similarly CVaR, requires using a quantum comparator, as mentioned in Section \ref{sec:quantumriskanaly}. In addition, for computing VaR, a bisection search is then performed utilizing multiple executions of QAE to find a threshold $x$ such that $\mathbb{E}[f(X)] = \mathbb{P}[X \leq x] \geq \alpha$, for the chosen confidence $\alpha$. For the comparator, the authors referenced the quantum ripple-carry adder of Cuccaro et al.~\cite{cuccaro2004new}. For each of these methods, they discussed the trade-off between circuit depth, qubit count, and convergence rate. An example circuit for $3$ evaluation qubits, corresponding to $2^3 = 8$ samples (i.e., quantum queries, Section \ref{sec:simulation and pricing}), is shown in Figure~\ref{fig:QAE} for computing the value of a T-Bill. Increasing the number of evaluation qubits above $3$ yields smaller estimation errors than with classical Monte Carlo integration (Section \ref{sec:simulation and pricing}), demonstrating the predicted quantum speedup; see Figure~\ref{fig:AE_results}.

Despite the observed speedup of the QMCI-based algorithm, compared with classical Monte Carlo integration, practical quantum advantage cannot be achieved by using this algorithm on NISQ devices because of the relatively small qubit count, qubit connectivity and quality, and decoherence limitations present in modern quantum processors (Section \ref{sec:currentquantumhardware}). Additionally, Monte Carlo for numerical integration can be efficiently parallelized, imposing even stronger requirements on quantum algorithms in order to outperform standard classical methods.

\subsection{Portfolio Optimization with Reverse Quantum Annealing}
\label{sec:reverseannealingportopt}

As explained in Section \ref{sec:quantportfolioopt}, quantum annealing (Section \ref{sec:QuantumAnnealing}) is highly amenable to solving portfolio optimization problems. The commercially available D-Wave flux-qubit-based \cite{superconductingengineers} quantum annealers have been successfully used to solve portfolio optimization instances. Determining the optimal portfolio composition is one of the most studied optimization problems in finance. A typical portfolio optimization problem can be cast into a QUBO or, equivalently, an Ising Hamiltonian suitable for a quantum annealer; see Section~\ref{sec:QUBO, Ising}. The following presents the work of Venturelli and Kondratyev \cite{Venturelli_2019}, who used a reverse-quantum-annealing-based approach.

Portfolio optimization was defined in Section \ref{sec:quantportfolioopt}. The pairwise relationships between the asset choices give rise to the quadratic form of the objective function to be optimized. The following technical challenges need to be overcome in order to perform simulations on the D-Wave quantum annealer. First, the standard way of enforcing the number of selected assets by introducing a high-energy penalty term can be problematic with D-Wave devices because of precision issues and  the physical limit on energy scales that can be programmed on the chip. The authors observed that this constraint can be enforced by artificially shifting the Sharpe ratios, used when computing the risk-adjusted returns of the assets, by a fixed amount.  Second, embedding the Ising optimization problem in the Chimera topology of D-Wave quantum annealers presents a strong limitation on the problems that can be solved on the device (Section \ref{sec:QuantumAnnealing}). To overcome this limitation, one can employ the minor-embedding compilation technique for fully connected graphs ~\cite{boothby2016fast, venturelli2015quantum}, allowing the necessary embedding to be performed using an extended set of variables.

\begin{figure}[ht]
    \centering
    \includegraphics[width=0.9\columnwidth]{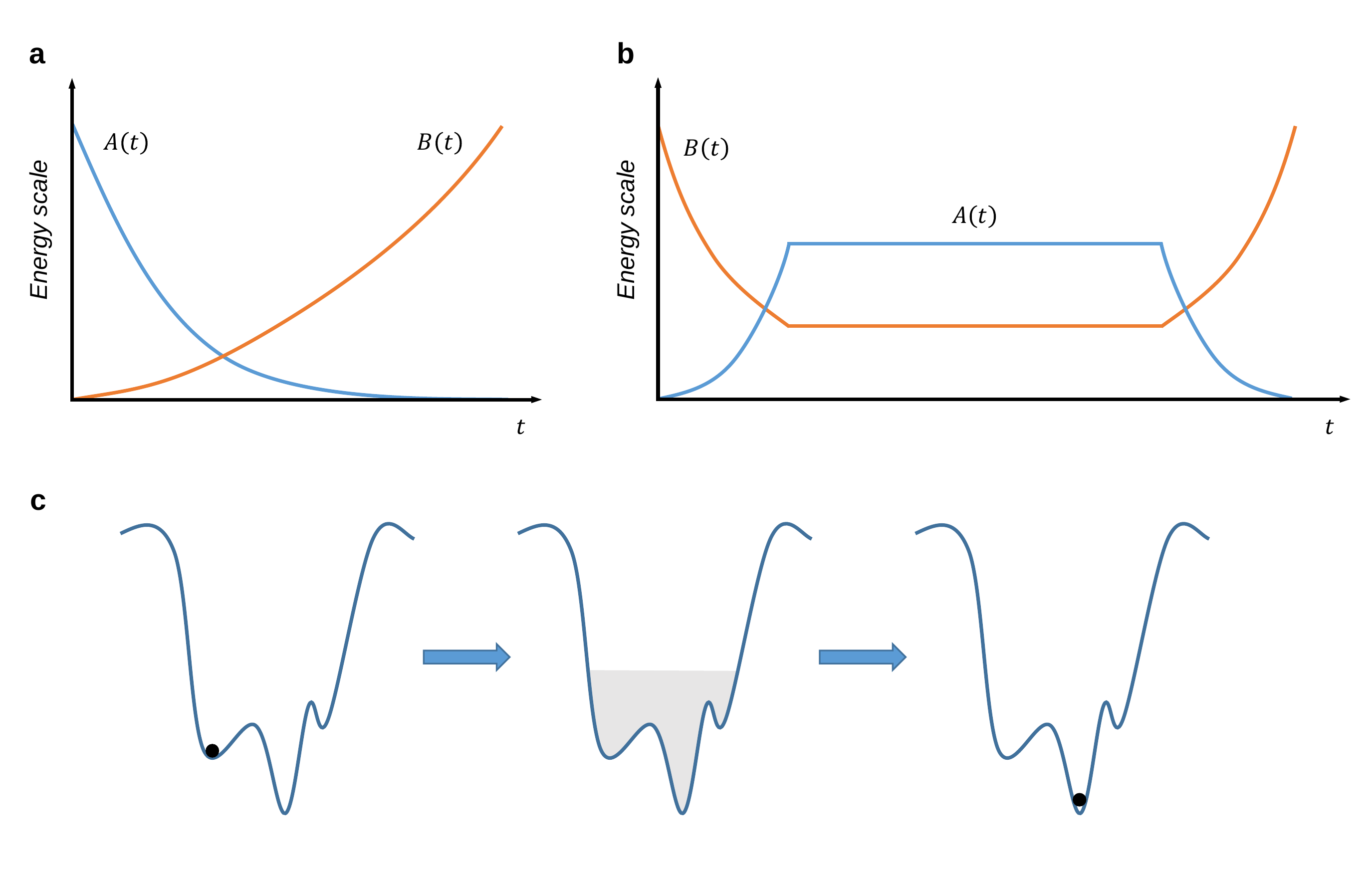}
    \caption{Forward (a) and reverse (b) quantum annealing protocols. During reverse quantum annealing, the system is initialized in a classical state (c, left), evolved following a ``backward'' annealing schedule to some quantum superposition state. At that point the evolution is interrupted for a fixed amount of time, allowing the system to perform a global search (c, middle). The system's evolution then is completed by the forward annealing schedule (c, right). Adapted from~\cite{king2019quantum}.}
    \label{fig:reverse}
\end{figure}
The novelty of this work by Venturelli and Kondratyev is the use of the reverse annealing protocol~\cite{king2019quantum}, which was found to be on average $100+$ times faster than forward annealing, as measured by the time to solution (TTS) metric. Reverse quantum annealing, briefly mentioned in Section \ref{sec:QuantumAnnealing}, is illustrated in Figure~\ref{fig:reverse}. It is a new approach enabled by the latest D-Wave architecture designed to help escape local minima through a combination of quantum and thermal fluctuations. The annealing schedule starts in some classical state provided by the user, as opposed to the quantum superposition state, as in the case of a forward annealing. The transverse field is then increased, driving backward annealing from the classical initial state to some intermediate quantum superposition state. The annealing schedule functions $A(t)$ and $B(t)$ (Section \ref{sec:QuantumAnnealing}) are then kept constant for a fixed amount of time, allowing the system to perform the search for a global minimum. The reverse annealing schedule is completed by removing the transverse field, effectively performing the forward annealing protocol with the system evolving to the ground state of the problem Hamiltonian.

The DW2000Q device used in~\cite{Venturelli_2019} allows  embedding up to 64 logical binary variables on a fully-connected graph, and the authors considered portfolio optimization problems with 24--60 assets, with 30 randomly generated instances for each size, and calculating the TTS. The results for both forward and reverse quantum annealing were then compared with the industry-established genetic algorithm (GA) approach used as a classical benchmark heuristic. The TTS for reverse annealing with the shortest annealing and pause times was found to be several orders of magnitude smaller than for the forward annealing and GA approaches, demonstrating a promising step toward practical applications
of NISQ devices. 

The development of new quantum annealing protocols and approaches, such as the reverse quantum annealing considered in this section, is an active area of research.

\subsection{Portfolio Optimization on a Trapped-Ion Device}
As discussed in Section \ref{sec:quantportfolioopt}, the convex mean-variance portfolio optimization problem with linear equality constraints can be solved as a linear system of equations. The method of Lagrange multipliers turns the convex quadratic program (Equation \eqref{opt_equation}) into

\begin{align}
    \label{linear_system}
    \begin{bmatrix}
    0 & 0 & \Vec{r}^\mathsf{T}  \\
    0 & 0 & \Vec{p}^{\mathsf{T}} \\
    \Vec{r} & \Vec{p} & \Sigma
    \end{bmatrix}
    \begin{bmatrix}
    \eta \\ \theta \\ \Vec{w}
    \end{bmatrix}
    =
    \begin{bmatrix}
    \mu \\ \xi \\ \Vec{0}
    \end{bmatrix},
    \end{align}
    where $\eta, \theta$ are the dual variables.
    This lends itself to quantum linear systems algorithms (Section \ref{sec:hhlsection}). 
    
To reiterate, solving a QLSP consists of returning a quantum state, with bounded error, that is proportional to the solution of a linear system $A\vec{x} = \vec{b}$. Quantum algorithms for this problem on sparse and well-conditioned matrices potentially achieve an exponential speedup in the system size $N$. Although a QLS algorithm cannot provide classical access to amplitudes of the solution state while maintaining the exponential speedup in $N$, potentially useful financial statistics can be obtained from the state.

The QLS algorithm  discussed in this section is HHL. This algorithm is typically viewed as one that is intended for the fault-tolerant era of quantum computation. Two of the main components of the algorithm that contribute to this constraint are quantum phase estimation (Section \ref{sec:qpesection}) and the loading of the inverse of the eigenvalue estimations onto the amplitudes of an ancilla, called eigenvalue inversion \cite{harrow2009quantum}.\footnote{Hamiltonian simulation also contributes to this constraint.} Yalovetzky et al.~\cite{yalovetzky2021nisqhhl} developed an algorithm called NISQ-HHL with the goal of applying it to portfolio optimization problems on NISQ devices. While it does not retain the asymptotic speedup of HHL, the authors' approach can potentially provide benefits in practice. They applied their algorithm to small mean-variance portfolio optimization problems and executed them on the trapped-ion Quantinuum System Model H1 device \cite{pino2021demonstration}.

\begin{figure}[ht]
  \centering
  \includegraphics[width=0.37\textwidth]{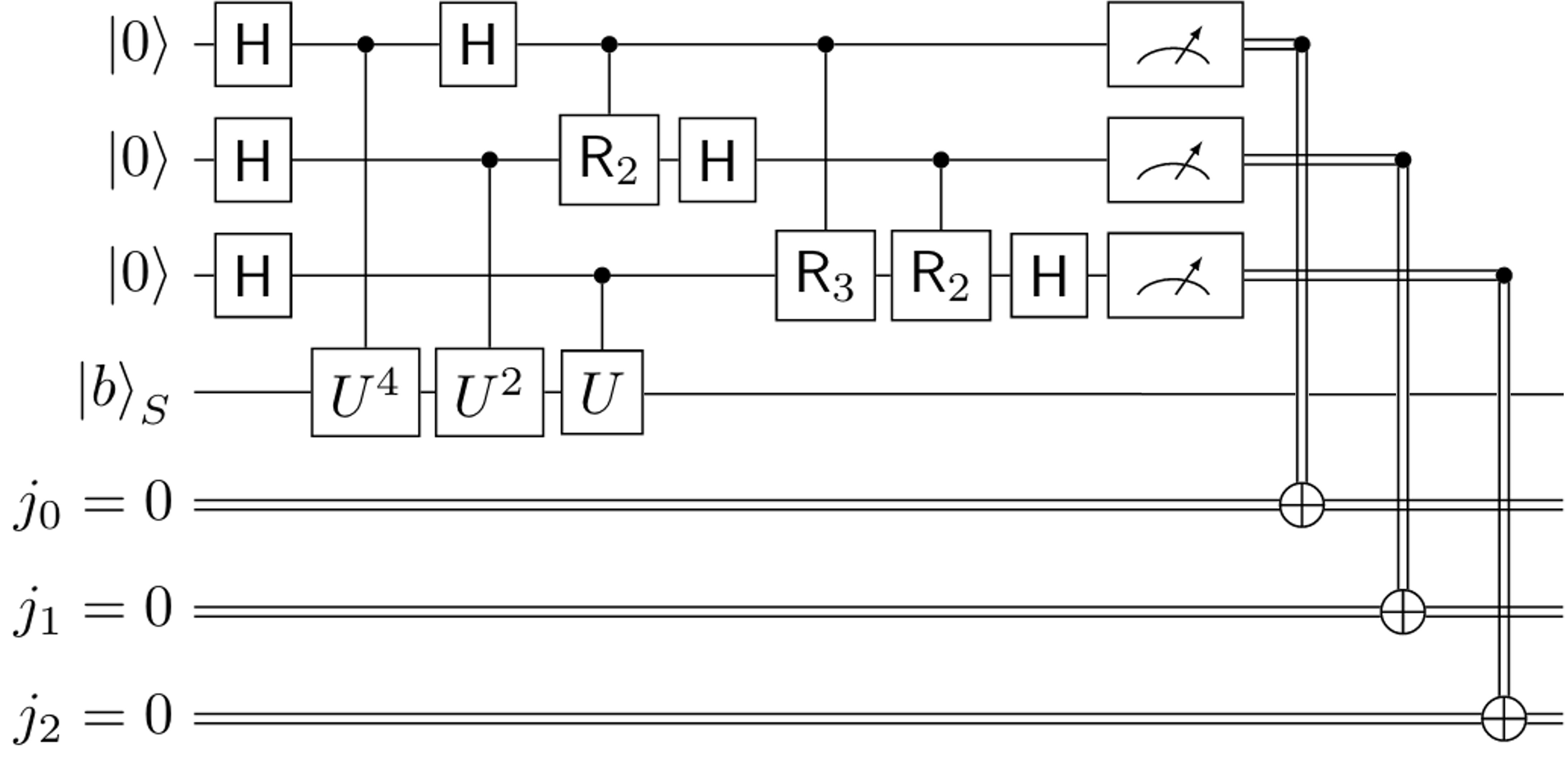}
   \includegraphics[width=0.62\textwidth]{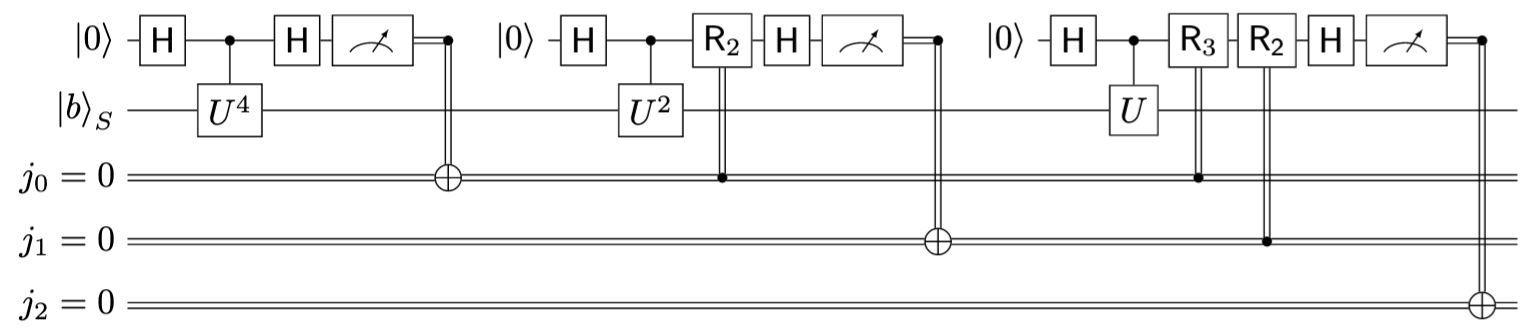}

     \caption{Circuits for estimating the eigenvalues of the unitary operator $U$ to three bits using QPE (left circuit) or QCL-QPE (right circuit). $S$ is the register that $U$ is applied to, and $j$ is a classical register; $\mathsf{H}$ refers to the Hadamard gate; and $\mathsf{R}_\mathsf{k}$,  for $\mathsf{k}=2,3$, are the standard phase gates used by the quantum Fourier transform \cite{nielsen2002quantum}. Adapted from \cite{yalovetzky2021nisqhhl}.}
  \label{diagram:qpe}
\end{figure}

QPE is typically difficult to run on near-term devices because of the large number of ancillary qubits and controlled operations required for the desired eigenvalue precision (depth scales as $O(\frac{1}{\delta})$ for precision $\delta$). An alternative realization of QPE reduces the number of ancillary qubits to one and replaces all controlled-phase gates in the quantum Fourier transform component of QPE with classically controlled single-qubit gates \cite{shors2nplus3}. This method requires mid-circuit measurements, ground-state resets, and quantum conditional logic (QCL). These features are available on the Quantinuum System Model H1 device. The paper calls this method QCL-QPE. The differences between these two QPE methods are highlighted in Figure \ref{diagram:qpe}.

\begin{figure}[ht]
    \centering
    \includegraphics[width=6cm, height=4cm]{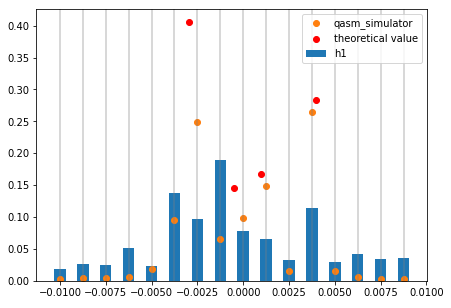}
    \includegraphics[width=6cm, height=4cm]{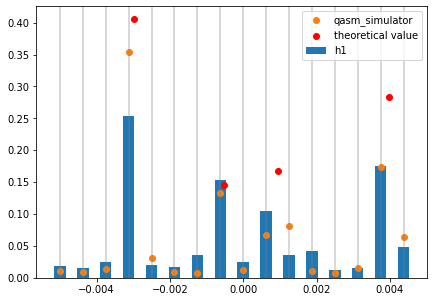}
\caption{Probability distributions over the eigenvalue estimations from the QCL-QPE run using $\gamma = 50$ (left) and $\gamma = 100$ (right). The blue bars represent the experimental results on the H1 machine---$1,000$ shots (left) and $2,000$ shots (right)---and we compare them with the theoretical eigenvalues (classically calculated using a numerical solver) represented with the red dots and the results from the Qiskit \cite{aleksandrowicz2019qiskit} QASM simulator represented with the orange dots. Adapted from \cite{yalovetzky2021nisqhhl}.} 
\label{fig:time_evolution_exp}
\end{figure}

With regard to eigenvalue inversion, implementing the controlled rotations on near-term devices typically requires a multicontrolled rotation for every possible eigenvalue that can be estimated by QPE, such as the uniformly controlled rotation gate \cite{mottonen2004transformation}. This circuit is exponentially deep in the number of qubits. Fault-tolerant implementations can take advantage of quantum arithmetic methods to implement efficient algorithms for the required $\arcsin$ computation. An alternative approach for the near term is a hybrid version of HHL \cite{lee2019hybrid}, which the developers of NISQ-HHL extended. This approach utilizes QPE to first estimate the required eigenvalues to perform the rotations for. While the number of controlled rotations is now reduced, this approach does not retain the asymptotic speedup provided by HHL because of the loss of coherence. A novel procedure was developed by the authors for scaling the matrix by a factor $\gamma$ to make it easier to distinguish the eigenvalues in the output distribution of QPE and improve performance. Figure \ref{fig:time_evolution_exp}  displays experimental results collected from the Quantinuum H1 device to show the effect of the scaling procedure.\footnote{Let $\Pi_\lambda$ denote the eigenspace projector for an eigenvalue $\lambda$ of the Hermitian matrix in Equation \eqref{linear_system}, and  let $\ket{b}$ be a quantum state proportional to the right side of Equation \eqref{linear_system}. To clarify, the y-axis value of the red dot corresponding to $\lambda$, on either of the histograms in Figure \ref{fig:time_evolution_exp}, is $\lVert \Pi_\lambda \ket{b}\rVert_2^2$.}

These methods were executed on the Quantinuum H1 device for a few small portfolios (Table \ref{table:comparision_eigenvalue inversion_inner_product}). A controlled-SWAP test \cite{patentBeusoleil} was performed between the state from HHL and the solution to the linear system obtained classically (loaded onto a quantum state) \cite{yalovetzky2021nisqhhl}. The number of H1 two-qubit \textsf{ZZMax} gates used is also  displayed in Table \ref{table:comparision_eigenvalue inversion_inner_product}.

\begin{table}[ht]
 \centering
    \begin{tabular}{lccc}
      & HHL & NISQ-HHL & Relative Change \\ \hline
      Rotations & 7  & 6 & -14.29$\%$  \\ 
      Eigenvalue Inversion Depth & 138 & 120 & -13.04$\%$\\
      Inner Product in Simulation & 0.59  & 0.83 & 40.67$\%$ \\
      Inner Product on H$1$-$1$ & 0.42 & 0.46 & 9.52$\%$
  \end{tabular}
  \caption{Comparison of the number of rotations in the eigenvalue inversion, \textsf{ZZMax} depth of the HHL plus the controlled-\textsf{SWAP} test circuits and the inner product calculated from the Qiskit state vector simulator and the Quantinuum H1 results with $3,000$ shots. Adapted from \cite{yalovetzky2021nisqhhl}.}
  \label{table:comparision_eigenvalue inversion_inner_product}
\end{table}

Statistics such as risk can be computed efficiently from the quantum state. Sampling can also be performed if the solution state is sparse enough \cite{rebentrost2018quantum}.
\section{Conclusion and Outlook}
\label{sec:conclusion}

Quantum computers are expected to surpass the computational capabilities of classical computers and achieve disruptive impact on numerous industry  sectors, such as global energy and materials, pharmaceuticals and medical products, telecommunication, travel,  logistics, and finance.  In particular, the finance sector has a history of creation and first adaptation of new technologies.  This is true also when it comes to quantum computing.  In fact, finance is  estimated to be the first industry sector to benefit from quantum computing, not only in the medium and long terms, but even in the short term, because of the large number of financial use cases that lend themselves to quantum computing and  their amenability to be solved effectively even in the presence of approximations.

A common misconception about quantum computing is that quantum computers are simply faster processors and hence speedups will be automatically achieved by porting a classical algorithm from a classical computer to a quantum one---just like moving a program to a system-upgraded hardware.
In reality, quantum computers are fundamentally different from classical computers, so much so that the algorithmic process for solving an application has to be completely redesigned based on the architecture of the underlying quantum hardware.

Most of the use cases that characterize the finance industry sector have high computational complexity, and for this reason they lend themselves to quantum computing.
However, the computational complexity of a problem per se is not a guarantee that quantum computing can make a difference.
To assess commercial viability, we must first answer a couple questions. First, is there in fact any need for improved speed of calculations or computational-accuracy for the given application? Determining whether there is a  gap in what the current technology can provide us is essential and is a question that financial players typically ask when doing proof of concept projects. Second, when adapting a classical finance application to quantum computing, can that application achieve quantum speedup and, if so, when will that  happen based on the estimated evolution of the underlying quantum hardware?

Today's quantum computers are not yet capable of solving real-life-scale problems in the industry more efficiently and more accurately than classical computers can.  Nevertheless, proofs of concept for quantum advantage and even quantum supremacy have already been provided (see Section \ref{sec:quantumconcepts}). There is hope that the situation will change in the coming years with demonstrations of quantum advantage. It is crucial for enterprises, and particularly financial institutions, to use the current time to become quantum ready in order to avoid being left behind when quantum computers become operational in a production environment.

We are in the early stages of the quantum revolution, yet we are already observing a strong potential for quantum technology to transform the financial industry. So far, the community has developed potential quantum solutions for portfolio optimization, derivatives pricing, risk modeling, and several problems in the realm of artificial intelligence and machine learning, such as fraud detection and NLP.

In this paper we have provided a comprehensive review of quantum computing for finance, specifically focusing on quantum algorithms that can solve computationally challenging financial problems. We have described the current landscape of the state of the industry. We hope this work will be used by the scientific community, both in industry and in academia, not only as a reference, but also as a source of information to identify new opportunities and advance the state of the art.

\section*{Competing Interests}
The authors declare that they have no competing interests.

\section*{Funding}
C.N.G's work was supported in part by the U.S. Department of Energy, Office of Science, Office of Workforce Development for Teachers and Scientists (WDTS) under the Science Undergraduate Laboratory Internships Program (SULI).
Y.A.'s work at Argonne National Laboratory was supported by the U.S. Department of Energy, Office of Science, under contract DE-AC02-06CH11357.

\section*{Disclaimer}
This paper was prepared for information purposes by the teams of researchers from the various institutions identified above, including the Future Lab for Applied Research and Engineering (FLARE) group of JPMorgan Chase Bank, N.A..  This paper is not a product of the Research Department of JPMorgan Chase \& Co. or its affiliates.  Neither JPMorgan Chase \& Co. nor any of its affiliates make any explicit or implied representation or warranty and none of them accept any liability in connection with this paper, including, but limited to, the completeness, accuracy, reliability of information contained herein and the potential legal, compliance, tax or accounting effects thereof.  This document is not intended as investment research or investment advice, or a recommendation, offer or solicitation for the purchase or sale of any security, financial instrument, financial product or service, or to be used in any way for evaluating the merits of participating in any transaction.

\section*{Authors' Contributions}
Dylan Herman, Cody Ning Googin, and Xiaoyuan Liu contributed equally to the manuscript. All authors contributed to the writing and reviewed and approved the final manuscript.

\section*{Acronyms}
\begin{acronym}[]  
\acro { NISQ }       Noisy-Intermediate Scale Quantum
\acro { AQC } Adiabatic Quantum Computing
\acro { qRAM } Quantum Random Access Memory
\acro { QCL } Quantum Conditional Logic
\acro { VQE }        Variational Quantum Eigensolver
\acro { QAOA }       Quantum Approximate Optimization Algorithm
\acro { QML }        Quantum Machine Learning
\acro { PCA } Principal Component Analysis
\acro { TDA } Topological Data Analysis
\acro { QNN } Quantum Neural Network
\acro { PQC } Parameterized Quantum Circuit
\acro { GAN } Generative Adversarial Network
\acro { SVM } Support Vector Machine
\acro { PH } Persistent Homology
\acro { GP } Gaussian Process
\acro { NP } Non-Deterministic Polynomial
\acro { VQA } Variational Quantum Algorithm
\acro { QPE }        Quantum Phase Estimation
\acro { NLP }        Natural Language Processing 
\acro { RL } Reinforcement Learning
\acro { QAA }        Quantum Amplitude Amplification
\acro { QAE }        Quantum Amplitude Estimation 
\acro { QLSP }       Quantum Linear Systems Problem
\acro { QLS } Quantum Linear System
\acro { QLSA } Quantum Linear Systems Algorithm
\acro { HHL }        Harrow Hassidim Lloyd
\acro { QSVT } Quantum Singular Value Transform
\acro { QSVE } Quantum Singular Value Estimation
\acro { CKS } Childs, Kothari, and Somma
\acro { QA } Quantum Annealing
\acro { IP  } Integer Programming
\acro { IQP  } Integer Quadratic Programming
\acro { QUBO } Quadratic Unconstrained Binary Optimization
\acro{ QAOA }      Quantum Approximate Optimization Algorithm
\acro { ITE } Imaginary-Time Evolution
\acro{ VarQITE }         Variational Quantum Imaginary Time Evolution
\acro{ MCI } Monte Carlo Integration
\acro{ QMCI } Quantum Monte Carlo Integration
\acro{ VaR } Value at Risk
\acro{ CVaR } Conditional Value at Risk
\acro{ CDO } Collateralized Debt Obligation
\acro { ECR } Economic Capital Requirement
\acro { CPU } Central Processing Unit
\acro { GPU } Graphics Processing Unit
\end{acronym}

\bibliography{text/ref}

\begin{thebibliography}{100}

\bibitem{mckinsey_quantum}
Alexandre M{\'e}nard, Ivan Ostojic, Mark Patel, and Daniel Volz.
\newblock
  \href{https://www.mckinsey.com/business-functions/mckinsey-digital/our-insights/a-game-plan-for-quantum-computing}{A
  game plan for quantum computing}, 2020.

\bibitem{nielsen2002quantum}
Michael~A. Nielsen and Isaac~L. Chuang.
\newblock {\em Quantum Computation and Quantum Information}.
\newblock Cambridge University Press, 2010.

\bibitem{egger2020quantum}
Daniel~J Egger, Claudio Gambella, Jakub Marecek, Scott McFaddin, Martin
  Mevissen, Rudy Raymond, Andrea Simonetto, Stefan Woerner, and Elena Yndurain.
\newblock Quantum computing for finance: State-of-the-art and future prospects.
\newblock {\em IEEE Transactions on Quantum Engineering}, 1:1--24, 2020.

\bibitem{bouland2020prospects}
Adam Bouland, Wim van Dam, Hamed Joorati, Iordanis Kerenidis, and Anupam
  Prakash.
\newblock Prospects and challenges of quantum finance.
\newblock {\em \href{https://arxiv.org/abs/2011.06492}{arXiv preprint
  arXiv:2011.06492}}, 2020.

\bibitem{orus2019quantum}
Roman Orus, Samuel Mugel, and Enrique Lizaso.
\newblock Quantum computing for finance: Overview and prospects.
\newblock {\em Reviews in Physics}, 4:100028, 2019.

\bibitem{pistoia2021quantum}
Marco Pistoia, Syed~Farhan Ahmad, Akshay Ajagekar, Alexander Buts, Shouvanik
  Chakrabarti, Dylan Herman, Shaohan Hu, Andrew Jena, Pierre Minssen, Pradeep
  Niroula, Arthur Rattew, Yue Sun, and Romina Yalovetzky.
\newblock {Quantum Machine Learning for Finance}.
\newblock {\em IEEE/ACM International Conference On Computer Aided Design
  (ICCAD)}, November 2021.
\newblock {ICCAD Special Session Paper}.

\bibitem{gomez2022survey}
Andr{\'e}s G{\'o}mez, {\'A}lvaro Leitao, Alberto Manzano, Daniele Musso,
  Mar{\'\i}a~R Nogueiras, Gustavo Ord{\'o}{\~n}ez, and Carlos V{\'a}zquez.
\newblock A survey on quantum computational finance for derivatives pricing and
  {VaR}.
\newblock {\em Archives of Computational Methods in Engineering}, pages 1--27,
  2022.

\bibitem{griffin2021quantum}
Paul Griffin and Ritesh Sampat.
\newblock Quantum computing for supply chain finance.
\newblock In {\em 2021 IEEE International Conference on Services Computing
  (SCC)}, pages 456--459. IEEE, 2021.

\bibitem{ganapathy2021quantum}
Apoorva Ganapathy.
\newblock Quantum computing in high frequency trading and fraud detection.
\newblock {\em Engineering International}, 9(2):61--72, 2021.

\bibitem{researchandmarkets}
{Research and Markets}.
\newblock
  \href{https://www.researchandmarkets.com/reports/5240250/financial-services-global-market-report-2021?utm_source=BW&utm_medium=PressRelease&utm_code=jdpplg&utm_campaign=1510809+-+Global+Financial+Services+Market+Outlook+2021-2030}{Financial
  Services Global Market Report 2021: {COVID}-19 Impact and Recovery to 2030},
  2021.

\bibitem{basel}
Basel~Committee on~Banking~Supervision.
\newblock \href{https://www.bis.org/bcbs/publ/d424_hlsummary.pdf}{High-level
  summary of {Basel III} reforms}, 2017.

\bibitem{glasserman2004monte}
Paul Glasserman.
\newblock {\em Monte Carlo methods in financial engineering}, volume~53.
\newblock Springer, 2004.

\bibitem{TheFutureofFinance}
Heath~P. Terry, Debra Schwartz, and Tina Sun.
\newblock
  \href{https://www.gspublishing.com/content/research/en/reports/2015/03/13/27d3b5ca-f425-4b11-af87-e63ea3dcc121.pdf}{The
  Future of Finance: Volume 3: The socialization of finance}, 2015.

\bibitem{Salesforce_trends}
Salesforce.
\newblock
  \href{https://www.salesforce.com/news/stories/financial-services-leaders/}{Trends
  in Financial Services}, 2020.

\bibitem{mckinsey_data}
Tucker Bailey, Soumya Banerjee, Christopher Feeney, and Heather Hogsett.
\newblock
  \href{https://www.mckinsey.com/business-functions/risk-and-resilience/our-insights/cybersecurity-emerging-challenges-and-solutions-for-the-boards-of-financial-services-companies}{Cybersecurity:
  Emerging challenges and solutions for the boards of financial-services
  companies}, 2020.

\bibitem{shor1997}
Peter~W. Shor.
\newblock Polynomial-time algorithms for prime factorization and discrete
  logarithms on a quantum computer.
\newblock {\em SIAM Journal on Computing}, 26(5):1484–1509, Oct 1997.

\bibitem{2021factoring}
Craig Gidney and Martin Ekerå.
\newblock How to factor 2048 bit {RSA} integers in 8 hours using 20 million
  noisy qubits.
\newblock {\em Quantum}, 5:433, Apr 2021.

\bibitem{roetteler2017quantum}
Martin Roetteler, Michael Naehrig, Krysta~M Svore, and Kristin Lauter.
\newblock Quantum resource estimates for computing elliptic curve discrete
  logarithms.
\newblock In {\em International Conference on the Theory and Application of
  Cryptology and Information Security}, pages 241--270, 2017.

\bibitem{childs2010quantum}
Andrew~M Childs and Wim Van~Dam.
\newblock Quantum algorithms for algebraic problems.
\newblock {\em Reviews of Modern Physics}, 82(1):1, 2010.

\bibitem{bernstein2017post}
D.~J. Bernstein and T.~Lange.
\newblock Post-quantum cryptography.
\newblock {\em Nature}, 2017.

\bibitem{gisin2002quantum}
Nicolas Gisin, Gr{\'e}goire Ribordy, Wolfgang Tittel, and Hugo Zbinden.
\newblock Quantum cryptography.
\newblock {\em Reviews of Modern Physics}, 74(1):145, 2002.

\bibitem{amy2016estimating}
Matthew Amy, Olivia Di~Matteo, Vlad Gheorghiu, Michele Mosca, Alex Parent, and
  John Schanck.
\newblock Estimating the cost of generic quantum pre-image attacks on {SHA-2
  and SHA-3}.
\newblock In {\em International Conference on Selected Areas in Cryptography},
  pages 317--337. Springer, 2016.

\bibitem{hosoyamada2021quantum}
Akinori Hosoyamada and Yu~Sasaki.
\newblock Quantum collision attacks on reduced {SHA-256 and SHA-512}.
\newblock {\em IACR Cryptol. ePrint Arch.}, 2021:292, 2021.

\bibitem{jaques2020implementing}
Samuel Jaques, Michael Naehrig, Martin Roetteler, and Fernando Virdia.
\newblock Implementing {Grover} oracles for quantum key search on {AES and
  LowMC}.
\newblock {\em Advances in Cryptology--EUROCRYPT 2020}, 12106:280, 2020.

\bibitem{Bonnetain_encryption}
Xavier Bonnetain, André Schrottenloher, and Ferdinand Sibleyras.
\newblock Beyond quadratic speedups in quantum attacks on symmetric schemes.
\newblock {\em IACR-EUROCRYPT-2022}, 10 2021.

\bibitem{wolsey1999integer}
Laurence~A. Wolsey and George~L Nemhauser.
\newblock {\em Integer and combinatorial optimization}, volume~55.
\newblock John Wiley \& Sons, 1999.

\bibitem{boyd2004convex}
Stephen Boyd and Lieven Vandenberghe.
\newblock {\em Convex Optimization}.
\newblock Cambridge University Press, 2004.

\bibitem{alexeev2021prx}
Yuri Alexeev, Dave Bacon, Kenneth~R. Brown, Robert Calderbank, Lincoln~D. Carr,
  Frederic~T. Chong, Brian DeMarco, Dirk Englund, Edward Farhi, Bill Fefferman,
  Alexey~V. Gorshkov, Andrew Houck, Jungsang Kim, Shelby Kimmel, Michael Lange,
  Seth Lloyd, Mikhail~D. Lukin, Dmitri Maslov, Peter Maunz, Christopher Monroe,
  John Preskill, Martin Roetteler, Martin~J. Savage, and Jeff Thompson.
\newblock Quantum computer systems for scientific discovery.
\newblock {\em {PRX} Quantum}, 2(1), February 2021.

\bibitem{kumar2017cmos}
Gagnesh Kumar and Sunil Agrawal.
\newblock {CMOS} limitations and futuristic carbon allotropes.
\newblock In {\em 2017 8th IEEE Annual Information Technology, Electronics and
  Mobile Communication Conference (IEMCON)}, pages 68--71. IEEE, 2017.

\bibitem{arute2019quantum}
Frank Arute, Kunal Arya, Ryan Babbush, Dave Bacon, Joseph~C. Bardin, Rami
  Barends, Rupak Biswas, Sergio Boixo, Fernando G. S.~L. Brandao, David~A.
  Buell, et~al.
\newblock Quantum supremacy using a programmable superconducting processor.
\newblock {\em Nature}, 574(7779):505--510, 2019.

\bibitem{preskill2012quantum}
John Preskill.
\newblock Quantum computing and the entanglement frontier.
\newblock {\em \href{https://arxiv.org/abs/1203.5813}{arXiv preprint
  arXiv:1203.5813}}, 2012.

\bibitem{2018preskillnisq}
John Preskill.
\newblock Quantum computing in the {NISQ} era and beyond.
\newblock {\em Quantum}, 2:79, Aug 2018.

\bibitem{shankar2012principles}
Ramamurti Shankar.
\newblock {\em Principles of quantum mechanics}.
\newblock Springer Science \& Business Media, 1994.

\bibitem{veblen1918projective}
Oswald Veblen and John~Wesley Young.
\newblock {\em Projective geometry}, volume~2.
\newblock Ginn, 1918.

\bibitem{roman2005advanced}
Steven Roman.
\newblock {\em Advanced linear algebra}.
\newblock Springer, 2008.

\bibitem{dallard1979product}
John~D. Dollard and Charles~N. Friedman.
\newblock {\em Product Integration with Application to Differential Equations}.
\newblock Encyclopedia of Mathematics and its Applications. London:
  Addison-Wesley, 1984.

\bibitem{kitaev2002classical}
A.~Yu. Kitaev, A.~H. Shen, and M.~N. Vyalyi.
\newblock {\em Classical and Quantum Computation}.
\newblock American Mathematical Society, USA, 2002.

\bibitem{2020qudits}
Yuchen Wang, Zixuan Hu, Barry~C. Sanders, and Sabre Kais.
\newblock Qudits and high-dimensional quantum computing.
\newblock {\em Frontiers in Physics}, 8, Nov 2020.

\bibitem{buck2021continuous}
Samantha Buck, Robin Coleman, and Hayk Sargsyan.
\newblock Continuous variable quantum algorithms: an introduction.
\newblock {\em \href{https://arxiv.org/abs/2107.02151}{arXiv preprint
  arXiv:2107.02151}}, 2021.

\bibitem{wang2021comparative}
Dong-Sheng Wang.
\newblock A comparative study of universal quantum computing models: towards a
  physical unification a comparative study of universal quantum computing
  models: towards a physical unification.
\newblock {\em Quantum Engineering}, page e85, 2021.

\bibitem{aharonov2008adiabatic}
Dorit Aharonov, Wim Van~Dam, Julia Kempe, Zeph Landau, Seth Lloyd, and Oded
  Regev.
\newblock Adiabatic quantum computation is equivalent to standard quantum
  computation.
\newblock {\em SIAM Review}, 50(4):755--787, 2008.

\bibitem{lidar2013quantum}
Daniel~A. Lidar and Todd~A. Brun.
\newblock {\em Quantum Error Correction}.
\newblock Cambridge University Press, 2013.

\bibitem{2012surfacecode}
A.~G. Fowler, M.~Mariantoni, J.~M. Martinis, and A.~N. Cleland.
\newblock Surface codes: Towards practical large-scale quantum computation.
\newblock {\em Physical Review A}, 86(3), Sep 2012.

\bibitem{rattew2021efficient}
Arthur~G Rattew, Yue Sun, Pierre Minssen, and Marco Pistoia.
\newblock The efficient preparation of normal distributions in quantum
  registers.
\newblock {\em Quantum}, 5:609, 2021.

\bibitem{Wu2019}
Xin-Chuan Wu, Sheng Di, Emma~Maitreyee Dasgupta, Franck Cappello, Hal Finkel,
  Yuri Alexeev, and Frederic~T. Chong.
\newblock Full-state quantum circuit simulation by using data compression.
\newblock {\em Proceedings of the International Conference for High Performance
  Computing, Networking, Storage and Analysis}, November 2019.

\bibitem{wu2019full}
Xin-Chuan Wu, Sheng Di, Emma~Maitreyee Dasgupta, Franck Cappello, Hal Finkel,
  Yuri Alexeev, and Frederic~T Chong.
\newblock Full-state quantum circuit simulation by using data compression.
\newblock In {\em Proceedings of the High Performance Computing, Networking,
  Storage and Analysis International Conference (SC19)}, Denver, CO, USA, 2019.
  IEEE Computer Society.

\bibitem{wu2018memory}
Xin-Chuan Wu, Sheng Di, Franck Cappello, Hal Finkel, Yuri Alexeev, and Frederic
  Chong.
\newblock Memory-efficient quantum circuit simulation by using lossy data
  compression.
\newblock In {\em Proceedings of the 3rd International Workshop on Post-{Moore}
  Era Supercomputing {(PMES) at SC18}}, Denver, CO, USA, 2018.

\bibitem{wu2018amplitude}
Xin-Chuan Wu, Sheng Di, Franck Cappello, Hal Finkel, Yuri Alexeev, and
  Frederic~T Chong.
\newblock Amplitude-aware lossy compression for quantum circuit simulation.
\newblock In {\em Proceedings of 4th International Workshop on Data Reduction
  for Big Scientific Data (DRBSD-4) at SC18}, 2018.

\bibitem{lykov2021large}
Danylo Lykov, Roman Schutski, Alexey Galda, Valerii Vinokur, and Yuri Alexeev.
\newblock Tensor network quantum simulator with step-dependent parallelization.
\newblock {\em arXiv preprint arXiv:2012.02430}, 2020.

\bibitem{LykovGPU}
Danylo Lykov, Angela Chen, Huaxuan Chen, Kristopher Keipert, Zheng Zhang, Tom
  Gibbs, and Yuri Alexeev.
\newblock Performance evaluation and acceleration of the {QTensor} quantum
  circuit simulator on {GPUs}.
\newblock {\em 2021 {IEEE}/{ACM} Second International Workshop on Quantum
  Computing Software ({QCS})}, nov 2021.

\bibitem{lykov_diagonal}
Danylo Lykov and Yuri Alexeev.
\newblock Importance of diagonal gates in tensor network simulations.
\newblock 2021.

\bibitem{Albash_2018}
Tameem Albash and Daniel~A. Lidar.
\newblock Adiabatic quantum computation.
\newblock {\em Reviews of Modern Physics}, 90(1), Jan 2018.

\bibitem{Roland_2002}
Jérémie Roland and Nicolas~J. Cerf.
\newblock Quantum search by local adiabatic evolution.
\newblock {\em Physical Review A}, 65(4), Mar 2002.

\bibitem{2008Biamonte}
Jacob~D. Biamonte and Peter~J. Love.
\newblock Realizable {Hamiltonians} for universal adiabatic quantum computers.
\newblock {\em Physical Review A}, 78(1), Jul 2008.

\bibitem{superconductingengineers}
P.~Krantz, M.~Kjaergaard, F.~Yan, T.~P. Orlando, S.~Gustavsson, and W.~D.
  Oliver.
\newblock A quantum engineer's guide to superconducting qubits.
\newblock {\em Applied Physics Reviews}, 6(2):021318, 2019.

\bibitem{trappedion}
Colin~D. Bruzewicz, John Chiaverini, Robert McConnell, and Jeremy~M. Sage.
\newblock Trapped-ion quantum computing: Progress and challenges.
\newblock {\em Applied Physics Reviews}, 6(2):021314, 2019.

\bibitem{vepsalainen2020impact}
Antti~P. Veps{\"a}l{\"a}inen, Amir~H. Karamlou, John~L. Orrell, Akshunna~S.
  Dogra, Ben Loer, Francisca Vasconcelos, David~K. Kim, Alexander~J. Melville,
  Bethany~M. Niedzielski, Jonilyn~L. Yoder, et~al.
\newblock Impact of ionizing radiation on superconducting qubit coherence.
\newblock {\em Nature}, 584(7822):551--556, 2020.

\bibitem{2020crosstalk}
Mohan Sarovar, Timothy Proctor, Kenneth Rudinger, Kevin Young, Erik Nielsen,
  and Robin Blume-Kohout.
\newblock Detecting crosstalk errors in quantum information processors.
\newblock {\em Quantum}, 4:321, Sep 2020.

\bibitem{2021bravyierrormitigation}
Sergey Bravyi, Sarah Sheldon, Abhinav Kandala, David~C. Mckay, and Jay~M.
  Gambetta.
\newblock Mitigating measurement errors in multiqubit experiments.
\newblock {\em Phys. Rev. A}, 103(4), 2021.

\bibitem{2010dynamicdecoupling}
Jacob~R. West, Daniel~A. Lidar, Bryan~H. Fong, and Mark~F. Gyure.
\newblock High fidelity quantum gates via dynamical decoupling.
\newblock {\em Phys. Rev. Lett.}, 105(23), Dec 2010.

\bibitem{kitaev2003fault}
A.~Yu. Kitaev.
\newblock Fault-tolerant quantum computation by anyons.
\newblock {\em Annals of Physics}, 303(1):2--30, 2003.

\bibitem{fowler2014quantifying}
Austin~G Fowler and John~M Martinis.
\newblock Quantifying the effects of local many-qubit errors and nonlocal
  two-qubit errors on the surface code.
\newblock {\em Physical Review A}, 89(3):032316, 2014.

\bibitem{pino2021demonstration}
J.~Pino, Joan Dreiling, C.~Figgatt, J.~Gaebler, S.~Moses, M.~Allman,
  C.~Baldwin, Michael Foss-Feig, D.~Hayes, K.~Mayer, C.~Ryan-Anderson, and
  Brian Neyenhuis.
\newblock Demonstration of the trapped-ion quantum {CCD} computer architecture.
\newblock {\em Nature}, 2021.

\bibitem{cormen2009introduction}
Thomas~H. Cormen, Charles~E. Leiserson, Ronald~L. Rivest, and Clifford Stein.
\newblock {\em
  \href{https://mitpress.mit.edu/books/introduction-algorithms-third-edition}{Introduction
  to algorithms}}.
\newblock {MIT} press, 2009.

\bibitem{arora2009computational}
Sanjeev Arora and Boaz Barak.
\newblock {\em Computational complexity: a modern approach}.
\newblock Cambridge University Press, 2009.

\bibitem{ambainis2017understanding}
Andris Ambainis.
\newblock Understanding quantum algorithms via query complexity.
\newblock In {\em Proceedings of the International Congress of Mathematicians:
  Rio de Janeiro 2018}, pages 3265--3285. World Scientific, 2018.

\bibitem{grover1996fast}
Lov~K Grover.
\newblock A fast quantum mechanical algorithm for database search.
\newblock In {\em Proceedings of the twenty-eighth annual ACM symposium on
  Theory of Computing}, pages 212--219, 1996.

\bibitem{1997strengths}
Charles~H. Bennett, Ethan Bernstein, Gilles Brassard, and Umesh Vazirani.
\newblock Strengths and weaknesses of quantum computing.
\newblock {\em SIAM Journal on Computing}, 26(5):1510–1523, 1997.

\bibitem{kerenidis2016quantum}
I.~Kerenidis and A.~Prakash.
\newblock Quantum recommendation systems.
\newblock {\em \href{https://arxiv.org/abs/1603.08675}{arXiv preprint
  arXiv:1603.08675}}, 2016.

\bibitem{Giovannetti_2008}
Vittorio Giovannetti, Seth Lloyd, and Lorenzo Maccone.
\newblock Quantum random access memory.
\newblock {\em Physical Review Letters}, 100(16), April 2008.

\bibitem{harrow2020small}
Aram~W Harrow.
\newblock Small quantum computers and large classical data sets.
\newblock {\em \href{https://arxiv.org/abs/2004.00026}{arXiv preprint
  arXiv:2004.00026}}, 2020.

\bibitem{ambainis2010quantum}
Andris Ambainis.
\newblock Quantum search with variable times.
\newblock {\em Theory of Computing Systems}, 47(3):786--807, 2010.

\bibitem{QAE}
Gilles Brassard, Peter Høyer, Michele Mosca, and Alain Tapp.
\newblock Quantum amplitude amplification and estimation.
\newblock {\em Quantum Computation and Information}, page 53–74, 2002.

\bibitem{berry2014exponential}
Dominic~W Berry, Andrew~M Childs, Richard Cleve, Robin Kothari, and Rolando~D
  Somma.
\newblock Exponential improvement in precision for simulating sparse
  {Hamiltonians}.
\newblock In {\em Proceedings of the forty-sixth annual ACM symposium on Theory
  of Computing}, pages 283--292, 2014.

\bibitem{1998tightbounds}
M.~Boyer, G.~Brassard, P.~H{\o}yer, and A.~Tapp.
\newblock Tight bounds on quantum searching.
\newblock {\em Fortschritte der Physik: Progress of Physics}, 46(4-5):493--505,
  1998.

\bibitem{grover2005fixed}
Lov~K. Grover.
\newblock Fixed-point quantum search.
\newblock {\em Phys. Rev. Lett.}, 2005.

\bibitem{2003searchboundederror}
Peter Høyer, Michele Mosca, and Ronald de~Wolf.
\newblock Quantum search on bounded-error inputs.
\newblock {\em Lecture Notes in Computer Science}, page 291–299, 2003.

\bibitem{ambainis2010variable}
Andris Ambainis.
\newblock Variable time amplitude amplification and quantum algorithms for
  linear algebra problems.
\newblock In {\em STACS'12 (29th Symposium on Theoretical Aspects of Computer
  Science)}, volume~14, pages 636--647. LIPIcs, 2012.

\bibitem{kitaev1995quantum}
A.~Yu. Kitaev.
\newblock Quantum measurements and the {Abelian} stabilizer problem.
\newblock {\em \href{https://arxiv.org/abs/quant-ph/9511026}{arXiv preprint
  quant-ph/9511026}}, 1995.

\bibitem{Cleve_1998}
R.~Cleve, A.~Ekert, C.~Macchiavello, and M.~Mosca.
\newblock Quantum algorithms revisited.
\newblock {\em Proceedings of the Royal Society of London. Series A:
  Mathematical, Physical and Engineering Sciences}, 454(1969):339--–354, Jan
  1998.

\bibitem{prakash2014quantum}
Anupam Prakash.
\newblock {\em Quantum algorithms for linear algebra and machine learning}.
\newblock
  \href{https://www2.eecs.berkeley.edu/Pubs/TechRpts/2014/EECS-2014-211.html}{University
  of California, Berkeley}, 2014.

\bibitem{suzuki2020amplitude}
Yohichi Suzuki, Shumpei Uno, Rudy Raymond, Tomoki Tanaka, Tamiya Onodera, and
  Naoki Yamamoto.
\newblock Amplitude estimation without phase estimation.
\newblock {\em Quantum Information Processing}, 19(2), Jan 2020.

\bibitem{Grinko_2021}
Dmitry Grinko, Julien Gacon, Christa Zoufal, and Stefan Woerner.
\newblock Iterative quantum amplitude estimation.
\newblock {\em npj Quantum Information}, 7(1), Mar 2021.

\bibitem{giurgica2020low}
Tudor Giurgica-Tiron, Iordanis Kerenidis, Farrokh Labib, Anupam Prakash, and
  William Zeng.
\newblock Low depth algorithms for quantum amplitude estimation.
\newblock {\em \href{https://arxiv.org/abs/2012.03348}{arXiv preprint
  arXiv:2012.03348}}, 2020.

\bibitem{varqae}
Kirill Plekhanov, Matthias Rosenkranz, Mattia Fiorentini, and Michael Lubasch.
\newblock Variational quantum amplitude estimation.
\newblock {\em \href{https://arxiv.org/abs/2109.03687}{arXiv preprint
  arXiv:2109.03687}}, 2021.

\bibitem{dervovic2018quantum}
Danial Dervovic, Mark Herbster, Peter Mountney, Simone Severini, Na{\"\i}ri
  Usher, and Leonard Wossnig.
\newblock Quantum linear systems algorithms: a primer.
\newblock {\em \href{https://arxiv.org/abs/1802.08227}{arXiv preprint
  arXiv:1802.08227}}, 2018.

\bibitem{Childs_2017}
Andrew~M. Childs, Robin Kothari, and Rolando~D. Somma.
\newblock Quantum algorithm for systems of linear equations with exponentially
  improved dependence on precision.
\newblock {\em SIAM Journal on Computing}, 46(6):1920–1950, Jan 2017.

\bibitem{harrow2009quantum}
Aram~W. Harrow, Avinatan Hassidim, and Seth Lloyd.
\newblock Quantum algorithm for linear systems of equations.
\newblock {\em Phys. Rev. Lett.}, 103:150502, Oct 2009.

\bibitem{aaronson2015read}
Scott Aaronson.
\newblock {Read the Fine Print}.
\newblock {\em Nature Physics}, 11(4):291--293, 04 2015.

\bibitem{Berry_2006}
Dominic~W. Berry, Graeme Ahokas, Richard Cleve, and Barry~C. Sanders.
\newblock Efficient quantum algorithms for simulating sparse {Hamiltonians}.
\newblock {\em Communications in Mathematical Physics}, 270(2):359–--371, Dec
  2006.

\bibitem{2021trotter}
Andrew~M. Childs, Yuan Su, Minh~C. Tran, Nathan Wiebe, and Shuchen Zhu.
\newblock {T}heory of {Trotter} error with commutator scaling.
\newblock {\em Physical Review X}, 11(1), Feb 2021.

\bibitem{Berry_2015}
Dominic~W. Berry, Andrew~M. Childs, and Robin Kothari.
\newblock Hamiltonian simulation with nearly optimal dependence on all
  parameters.
\newblock {\em 2015 IEEE 56th Annual Symposium on Foundations of Computer
  Science}, Oct 2015.

\bibitem{Low_2017}
Guang~Hao Low and Isaac~L. Chuang.
\newblock Optimal {Hamiltonian} simulation by quantum signal processing.
\newblock {\em Phys. Rev. Lett.}, 118(1), Jan 2017.

\bibitem{costa2021optimal}
Pedro Costa, Dong An, Yuval~R Sanders, Yuan Su, Ryan Babbush, and Dominic~W
  Berry.
\newblock Optimal scaling quantum linear systems solver via discrete adiabatic
  theorem.
\newblock {\em \href{https://arxiv.org/abs/2111.08152}{arXiv preprint
  arXiv:2111.08152}}, 2021.

\bibitem{Wossnig_2018}
Leonard Wossnig, Zhikuan Zhao, and Anupam Prakash.
\newblock Quantum linear system algorithm for dense matrices.
\newblock {\em Physical Review Letters}, 120(5), Jan 2018.

\bibitem{Kerenidis_2020LS}
Iordanis Kerenidis and Anupam Prakash.
\newblock Quantum gradient descent for linear systems and least squares.
\newblock {\em Physical Review A}, 101(2), Feb 2020.

\bibitem{chakraborty2018power}
Shantanav Chakraborty, Andr{\'a}s Gily{\'e}n, and Stacey Jeffery.
\newblock The power of block-encoded matrix powers: improved regression
  techniques via faster {Hamiltonian} simulation.
\newblock {\em \href{https://arxiv.org/abs/1804.01973}{arXiv preprint
  arXiv:1804.01973}}, 2018.

\bibitem{gribling2021improving}
Sander Gribling, Iordanis Kerenidis, and D{\'a}niel Szil{\'a}gyi.
\newblock Improving quantum linear system solvers via a gradient descent
  perspective.
\newblock {\em \href{https://arxiv.org/abs/2109.04248}{arXiv preprint
  arXiv:2109.04248}}, 2021.

\bibitem{gilyen2019}
András Gilyén, Yuan Su, Guang~Hao Low, and Nathan Wiebe.
\newblock Quantum singular value transformation and beyond: exponential
  improvements for quantum matrix arithmetics.
\newblock {\em Proceedings of the 51st Annual ACM SIGACT Symposium on Theory of
  Computing}, Junee 2019.

\bibitem{dranov1998discrete}
Alexander Dranov, Johannes Kellendonk, and Rudolf Seiler.
\newblock Discrete time adiabatic theorems for quantum mechanical systems.
\newblock {\em Journal of Mathematical Physics}, 39(3):1340--1349, 1998.

\bibitem{kerenidis2018quantum}
Iordanis Kerenidis and Anupam Prakash.
\newblock A quantum interior point method for {LP}s and {SDP}s.
\newblock {\em ACM Transactions on Quantum Computing}, 1(1), Oct 2020.

\bibitem{frieze2004fast}
Alan Frieze, Ravi Kannan, and Santosh Vempala.
\newblock Fast {Monte-Carlo} algorithms for finding low-rank approximations.
\newblock {\em Journal of the ACM (JACM)}, 51(6):1025--1041, 2004.

\bibitem{Tang_2019}
Ewin Tang.
\newblock A quantum-inspired classical algorithm for recommendation systems.
\newblock In {\em Proceedings of the 51st Annual ACM SIGACT Symposium on Theory
  of Computing}. Association for Computing Machinery, 2019.

\bibitem{Chia_2020}
Nai-Hui Chia, András Gilyén, Tongyang Li, Han-Hsuan Lin, Ewin Tang, and
  Chunhao Wang.
\newblock Sampling-based sublinear low-rank matrix arithmetic framework for
  dequantizing quantum machine learning.
\newblock {\em Proceedings of the 52nd Annual ACM SIGACT Symposium on Theory of
  Computing}, June 2020.

\bibitem{lawler2010random}
Gregory~F. Lawler and Vlada Limic.
\newblock {\em Random Walk: A Modern Introduction}.
\newblock Cambridge Studies in Advanced Mathematics. Cambridge University
  Press, 2010.

\bibitem{black2019pricing}
Fischer Black and Myron Scholes.
\newblock The pricing of options and corporate liabilities.
\newblock In {\em World Scientific Reference on Contingent Claims Analysis in
  Corporate Finance: Volume 1: Foundations of CCA and Equity Valuation}, pages
  3--21. World Scientific, 2019.

\bibitem{klebaner2012introduction}
F.~C. Klebaner.
\newblock {\em Introduction to Stochastic Calculus with Applications}.
\newblock Imperial College Press, 3rd edition, 2012.

\bibitem{lovasz1993random}
L{\'a}szl{\'o} Lov{\'a}sz.
\newblock Random walks on graphs: A survey, combinatorics, {Paul Erdos is
  Eighty}.
\newblock {\em Bolyai Soc. Math. Stud.}, 2, Jan 1993.

\bibitem{norris1998markov}
J.~R. Norris.
\newblock {\em Markov Chains}.
\newblock Cambridge Series in Statistical and Probabilistic Mathematics.
  Cambridge University Press, 1997.

\bibitem{Kempe_2003}
J.~Kempe.
\newblock Quantum random walks: An introductory overview.
\newblock {\em Contemporary Physics}, 44(4):307–327, July 2003.

\bibitem{szegedy2004quantum}
Mario Szegedy.
\newblock Quantum speed-up of {Markov} chain based algorithms.
\newblock In {\em 45th Annual IEEE symposium on foundations of computer
  science}, 2004.

\bibitem{2011Magniez}
Frédéric Magniez, Ashwin Nayak, Jérémie Roland, and Miklos Santha.
\newblock Search via quantum walk.
\newblock {\em SIAM Journal on Computing}, 40(1):14--164, Jan 2011.

\bibitem{2003walksearchgrover}
Neil Shenvi, Julia Kempe, and K.~Birgitta Whaley.
\newblock Quantum random-walk search algorithm.
\newblock {\em Physical Review A}, 67(5), May 2003.

\bibitem{ambainis2004coins}
Andris Ambainis, Julia Kempe, and Alexander Rivosh.
\newblock Coins make quantum walks faster.
\newblock In {\em
  \href{https://dl.acm.org/doi/10.5555/1070432.1070590}{Proceedings of the
  Sixteenth Annual ACM-SIAM Symposium on Discrete Algorithms}}, SODA '05, page
  1099–1108, USA, 2005. Society for Industrial and Applied Mathematics.

\bibitem{ambainis2001one}
Andris Ambainis, Eric Bach, Ashwin Nayak, Ashvin Vishwanath, and John Watrous.
\newblock One-dimensional quantum walks.
\newblock In {\em Proceedings of the thirty-third annual ACM symposium on
  Theory of Computing}, pages 37--49, 2001.

\bibitem{aharonov2001quantum}
D.~Aharonov, A.~Ambainis, J.~Kempe, and U.~Vazirani.
\newblock Quantum walks on graphs.
\newblock In {\em Proceedings of the thirty-third annual ACM symposium on
  Theory of Computing}, pages 50--59, 2001.

\bibitem{moore2002quantum}
Cristopher Moore and Alexander Russell.
\newblock Quantum walks on the hypercube.
\newblock In {\em Proceedings of the 6th International Workshop on
  Randomization and Approximation Techniques}, pages 164--–178. Springer,
  2002.

\bibitem{ambainis2007quantum}
Andris Ambainis.
\newblock Quantum walk algorithm for element distinctness.
\newblock {\em SIAM Journal on Computing}, 37(1):210--239, 2007.

\bibitem{apers2019unified}
Simon Apers, Andr{\'a}s Gily{\'e}n, and Stacey Jeffery.
\newblock A unified framework of quantum walk search.
\newblock {\em \href{https://arxiv.org/abs/1912.04233}{arXiv preprint
  arXiv:1912.04233}}, 2019.

\bibitem{somma2007quantum}
R.~D. Somma, S.~Boixo, H.~Barnum, and E.~Knill.
\newblock Quantum simulations of classical annealing processes.
\newblock {\em Phys. Rev. Lett.}, Sep 2008.

\bibitem{2005ChildsSubset}
Andrew~M. Childs and Jason~M. Eisenberg.
\newblock Quantum algorithms for subset finding.
\newblock {\em \href{https://dl.acm.org/doi/10.5555/2011656.2011663}{Quantum
  Info. Comput.}}, 5(7):593–604, Nov 2005.

\bibitem{ambainis2019quadratic}
Andris Ambainis, Andr\'{a}s Gily\'{e}n, Stacey Jeffery, and Martins Kokainis.
\newblock Quadratic speedup for finding marked vertices by quantum walks.
\newblock In {\em Proceedings of the 52nd Annual ACM SIGACT Symposium on Theory
  of Computing}, pages 412--–424, New York, NY, USA, 2020. Association for
  Computing Machinery.

\bibitem{Farhi_1998}
Edward Farhi and Sam Gutmann.
\newblock Quantum computation and decision trees.
\newblock {\em Physical Review A}, 58(2):915--928, Aug 1998.

\bibitem{Childs_2002}
Andrew~M. Childs, Edward Farhi, and Sam Gutmann.
\newblock An example of the difference between quantum and classical random
  walks.
\newblock {\em Quantum Information Processing}, 2002.

\bibitem{Childs_2003}
Andrew~M. Childs, Richard Cleve, Enrico Deotto, Edward Farhi, Sam Gutmann, and
  Daniel~A. Spielman.
\newblock Exponential algorithmic speedup by a quantum walk.
\newblock {\em Proceedings of the thirty-fifth ACM symposium on Theory of
  computing - STOC ’03}, 2003.

\bibitem{kempe2002quantum}
Julia Kempe.
\newblock Quantum random walks hit exponentially faster.
\newblock {\em \href{https://arxiv.org/abs/quant-ph/0205083}{arXiv preprint
  quant-ph/0205083}}, 2002.

\bibitem{ruan2020quantum}
Yue Ruan, Samuel Marsh, Xilin Xue, Xi~Li, Zhihao Liu, and Jingbo Wang.
\newblock Quantum approximate algorithm for {NP} optimization problems with
  constraints.
\newblock {\em \href{https://arxiv.org/abs/2002.00943}{arXiv preprint
  arXiv:2002.00943}}, 2020.

\bibitem{Childs_2009}
Andrew~M. Childs.
\newblock On the relationship between continuous- and discrete-time quantum
  walk.
\newblock {\em Communications in Mathematical Physics}, 294(2):581–603, Oct
  2009.

\bibitem{2012walkscomp}
Salvador~Elías Venegas-Andraca.
\newblock Quantum walks: a comprehensive review.
\newblock {\em Quantum Information Processing}, 11(5):101--1106, Jul 2012.

\bibitem{Cerezo_2021}
M.~Cerezo, Andrew Arrasmith, Ryan Babbush, Simon~C. Benjamin, Suguru Endo,
  Keisuke Fujii, Jarrod~R. McClean, Kosuke Mitarai, Xiao Yuan, Lukasz Cincio,
  and et~al.
\newblock Variational quantum algorithms.
\newblock {\em Nature Reviews Physics}, 3(9):625--–644, Aug 2021.

\bibitem{bittel2021training}
Lennart Bittel and Martin Kliesch.
\newblock Training variational quantum algorithms is {NP}-hard.
\newblock {\em Phys. Rev. Lett.}, 127(12):120502, 2021.

\bibitem{Peruzzo_2014}
Alberto Peruzzo, Jarrod McClean, Peter Shadbolt, Man-Hong Yung, Xiao-Qi Zhou,
  Peter~J. Love, Alán Aspuru-Guzik, and Jeremy~L. O’Brien.
\newblock A variational eigenvalue solver on a photonic quantum processor.
\newblock {\em Nat. Commun.}, 5(1), July 2014.

\bibitem{bravoprieto2020variational}
Carlos Bravo-Prieto, Ryan LaRose, Marco Cerezo, Yigit Subasi, Lukasz Cincio,
  and Patrick~J Coles.
\newblock Variational quantum linear solver.
\newblock {\em \href{https://arxiv.org/abs/1909.05820}{arXiv preprint
  arXiv:1909.05820}}, 2019.

\bibitem{huang2019nearterm}
Hsin-Yuan Huang, Kishor Bharti, and Patrick Rebentrost.
\newblock Near-term quantum algorithms for linear systems of equations with
  regression loss functions.
\newblock {\em New Journal of Physics}, 23(11):113021, Nov 2021.

\bibitem{2021vqaloss}
Patrick Huembeli and Alexandre Dauphin.
\newblock Characterizing the loss landscape of variational quantum circuits.
\newblock {\em Quantum Science and Technology}, 6(2):025011, Feb 2021.

\bibitem{farhi2014quantum}
Edward Farhi, Jeffrey Goldstone, and Sam Gutmann.
\newblock A quantum approximate optimization algorithm.
\newblock {\em \href{https://arxiv.org/abs/1411.4028}{arXiv preprint
  arXiv:1411.4028}}, 2014.

\bibitem{Hadfield_2019}
Stuart Hadfield, Zhihui Wang, Bryan O’Gorman, Eleanor Rieffel, Davide
  Venturelli, and Rupak Biswas.
\newblock From the quantum approximate optimization algorithm to a quantum
  alternating operator ansatz.
\newblock {\em Algorithms}, 12(2):34, Feb 2019.

\bibitem{Yuan_2019}
Xiao Yuan, Suguru Endo, Qi~Zhao, Ying Li, and Simon~C. Benjamin.
\newblock Theory of variational quantum simulation.
\newblock {\em Quantum}, 3:191, Oct 2019.

\bibitem{mclachlan1964variational}
A.D. McLachlan.
\newblock A variational solution of the time-dependent {S}chr\"{o}dinger
  equation.
\newblock {\em Molecular Physics}, 8(1):39--44, 1964.

\bibitem{Mitarai_2018}
K.~Mitarai, M.~Negoro, M.~Kitagawa, and K.~Fujii.
\newblock Quantum circuit learning.
\newblock {\em Physical Review A}, 98(3), Sep 2018.

\bibitem{farhi2000quantum}
Edward Farhi, Jeffrey Goldstone, Sam Gutmann, and Michael Sipser.
\newblock Quantum computation by adiabatic evolution.
\newblock {\em \href{https://arxiv.org/abs/quant-ph/0001106}{arXiv preprint
  quant-ph/0001106}}, 2000.

\bibitem{van1987simulated}
Peter~JM Van~Laarhoven and Emile~HL Aarts.
\newblock Simulated annealing.
\newblock In {\em Simulated annealing: Theory and applications}, pages 7--15.
  Springer, 1987.

\bibitem{2016narrowqa}
Kostyantyn Kechedzhi and Vadim~N. Smelyanskiy.
\newblock Open-system quantum annealing in mean-field models with exponential
  degeneracy.
\newblock {\em Physical Review X}, 6(2), May 2016.

\bibitem{Denchev_2016}
Vasil~S. Denchev, Sergio Boixo, Sergei~V. Isakov, Nan Ding, Ryan Babbush, Vadim
  Smelyanskiy, John Martinis, and Hartmut Neven.
\newblock What is the computational value of finite-range tunneling?
\newblock {\em Physical Review X}, 6(3), Aug 2016.

\bibitem{farhi2002quantum}
Edward Farhi, Jeffrey Goldstone, and Sam Gutmann.
\newblock Quantum adiabatic evolution algorithms versus simulated annealing.
\newblock {\em \href{https://arxiv.org/abs/quant-ph/0201031}{arXiv preprint
  quant-ph/0201031}}, 2002.

\bibitem{Ajagekar_2020}
Akshay Ajagekar, Travis Humble, and Fengqi You.
\newblock Quantum computing based hybrid solution strategies for large-scale
  discrete-continuous optimization problems.
\newblock {\em Computers \& Chemical Engineering}, 132:106630, Jan 2020.

\bibitem{neal1993probabilistic}
Radford~M Neal.
\newblock {\em Probabilistic inference using {Markov chain Monte Carlo
  methods}}.
\newblock
  \href{https://www.cs.toronto.edu/~radford/review.abstract.html}{Department of
  Computer Science, University of Toronto Toronto, ON, Canada}, 1993.

\bibitem{robertmontecarlo}
Christian~P Robert, George Casella, and George Casella.
\newblock {\em Monte Carlo statistical methods}, volume~2.
\newblock Springer, 2004.

\bibitem{homem2014monte}
Tito {Homem-de-Mello} and Güzin Bayraksan.
\newblock {Monte Carlo} sampling-based methods for stochastic optimization.
\newblock {\em Surveys in Operations Research and Management Science},
  19(1):56--85, 2014.

\bibitem{brassard2011optimal}
Gilles Brassard, Frederic Dupuis, Sebastien Gambs, and Alain Tapp.
\newblock An optimal quantum algorithm to approximate the mean and its
  application for approximating the median of a set of points over an arbitrary
  distance.
\newblock {\em arXiv preprint arXiv:1106.4267}, 2011.

\bibitem{heinrich2002quantum}
S.~Heinrich.
\newblock Quantum summation with an application to integration.
\newblock {\em Journal of Complexity}, 18(1):1--50, 2002.

\bibitem{Montanaro_2015}
Ashley Montanaro.
\newblock Quantum speedup of {Monte Carlo} methods.
\newblock {\em Proceedings of the Royal Society A: Mathematical, Physical and
  Engineering Sciences}, 471(2181):20150301, Sep 2015.

\bibitem{ceperley1986quantum}
D.~Ceperley and B.~Alder.
\newblock Quantum {Monte Carlo}.
\newblock {\em Science}, 231(4738):555--560, 1986.

\bibitem{Chakrabarti_2021}
Shouvanik Chakrabarti, Rajiv Krishnakumar, Guglielmo Mazzola, Nikitas
  Stamatopoulos, Stefan Woerner, and William~J. Zeng.
\newblock A threshold for quantum advantage in derivative pricing.
\newblock {\em Quantum}, 5:463, June 2021.

\bibitem{haner2018optimizing}
Thomas H{\"a}ner, Martin Roetteler, and Krysta~M Svore.
\newblock Optimizing quantum circuits for arithmetic.
\newblock {\em \href{https://arxiv.org/abs/1805.12445}{arXiv preprint
  arXiv:1805.12445}}, 2018.

\bibitem{cornelissen2021}
Arjan Cornelissen and Sofiene Jerbi.
\newblock Quantum algorithms for multivariate {Monte Carlo} estimation.
\newblock 2021.

\bibitem{grover2002creating}
Lov Grover and Terry Rudolph.
\newblock Creating superpositions that correspond to efficiently integrable
  probability distributions.
\newblock {\em \href{https://arxiv.org/abs/quant-ph/0208112}{arXiv preprint
  quant-ph/0208112}}, 2002.

\bibitem{Herbert_2021}
Steven Herbert.
\newblock No quantum speedup with grover-rudolph state preparation for quantum
  {Monte Carlo} integration.
\newblock {\em Physical Review E}, 103(6), Jun 2021.

\bibitem{Zoufal_2019}
Christa Zoufal, Aurélien Lucchi, and Stefan Woerner.
\newblock Quantum generative adversarial networks for learning and loading
  random distributions.
\newblock {\em npj Quantum Information}, 5(1), 2019.

\bibitem{Dupire94pricingwith}
Bruno Dupire, The Black–scholes~Model (see Black, and Gives Options.
\newblock Pricing with a smile.
\newblock {\em Risk Magazine}, pages 18--20, 1994.

\bibitem{maruyama1954transition}
Gisiro Maruyama.
\newblock On the transition probability functions of the {Markov} process.
\newblock {\em Nat. Sci. Rep. Ochanomizu Univ}, 5:10--20, 1954.

\bibitem{Kaneko_2020}
Kazuya Kaneko, Koichi Miyamoto, Naoyuki Takeda, and Kazuyoshi Yoshino.
\newblock Quantum pricing with a smile: Implementation of local volatility
  model on quantum computer.
\newblock 2020.

\bibitem{Miyamoto_2020}
Koichi Miyamoto and Kenji Shiohara.
\newblock Reduction of qubits in a quantum algorithm for {Monte Carlo}
  simulation by a pseudo-random-number generator.
\newblock {\em Physical Review A}, 102(2), aug 2020.

\bibitem{giles2008multilevel}
Michael~B Giles.
\newblock Multilevel {Monte Carlo} path simulation.
\newblock {\em Operations research}, 56(3):607--617, 2008.

\bibitem{An_2021}
Dong An, Noah Linden, Jin-Peng Liu, Ashley Montanaro, Changpeng Shao, and Jiasu
  Wang.
\newblock Quantum-accelerated multilevel monte carlo methods for stochastic
  differential equations in mathematical finance.
\newblock {\em Quantum}, 5:481, June 2021.

\bibitem{herbert2021quantum}
Steven Herbert.
\newblock Quantum {Monte-Carlo} integration: The full advantage in minimal
  circuit depth.
\newblock {\em \href{https://arxiv.org/abs/2105.09100}{arXiv preprint
  arXiv:2105.09100}}, 2021.

\bibitem{Babbush_2021}
Ryan Babbush, Jarrod~R. McClean, Michael Newman, Craig Gidney, Sergio Boixo,
  and Hartmut Neven.
\newblock Focus beyond quadratic speedups for error-corrected quantum
  advantage.
\newblock {\em PRX Quantum}, 2(1), March 2021.

\bibitem{gomes2021adaptive}
Niladri Gomes, Anirban Mukherjee, Feng Zhang, Thomas Iadecola, Cai-Zhuang Wang,
  Kai-Ming Ho, Peter~P Orth, and Yong-Xin Yao.
\newblock Adaptive variational quantum imaginary time evolution approach for
  ground state preparation.
\newblock {\em Advanced Quantum Technologies}, page 2100114, 2021.

\bibitem{rattewkoczor}
Arthur~G. Rattew and Bálint Koczor.
\newblock Preparing arbitrary continuous functions in quantum registers with
  logarithmic complexity.
\newblock 2022.

\bibitem{kac1949distributions}
Mark Kac.
\newblock On distributions of certain {Wiener} functionals.
\newblock {\em Transactions of the American Mathematical Society}, 65(1):1--13,
  1949.

\bibitem{feynman2005principle}
Richard~P Feynman.
\newblock The principle of least action in quantum mechanics.
\newblock In {\em Feynman's Thesis—A New Approach to Quantum Theory}, pages
  1--69. World Scientific, 2005.

\bibitem{grossmann2007numerical}
Christian Grossmann, Hans-G{\"o}rg Roos, and Martin Stynes.
\newblock {\em Numerical treatment of partial differential equations}, volume
  154.
\newblock Springer, 2007.

\bibitem{shen2011spectral}
Jie Shen, Tao Tang, and Li-Lian Wang.
\newblock {\em Spectral methods: algorithms, analysis and applications},
  volume~41.
\newblock Springer Science \& Business Media, 2011.

\bibitem{cao2013quantum}
Yudong Cao, Anargyros Papageorgiou, Iasonas Petras, Joseph Traub, and Sabre
  Kais.
\newblock Quantum algorithm and circuit design solving the {Poisson} equation.
\newblock {\em New Journal of Physics}, 15(1):013021, 2013.

\bibitem{berry2014high}
Dominic~W Berry.
\newblock High-order quantum algorithm for solving linear differential
  equations.
\newblock {\em Journal of Physics A: Mathematical and Theoretical},
  47(10):105301, 2014.

\bibitem{berry2017quantum}
Dominic~W Berry, Andrew~M Childs, Aaron Ostrander, and Guoming Wang.
\newblock Quantum algorithm for linear differential equations with
  exponentially improved dependence on precision.
\newblock {\em Communications in Mathematical Physics}, 356(3):1057--1081,
  2017.

\bibitem{childs2021high}
Andrew~M Childs, Jin-Peng Liu, and Aaron Ostrander.
\newblock High-precision quantum algorithms for partial differential equations.
\newblock {\em Quantum}, 5:574, 2021.

\bibitem{montanaro2016quantum}
Ashley Montanaro and Sam Pallister.
\newblock Quantum algorithms and the finite element method.
\newblock {\em Physical Review A}, 93(3):032324, 2016.

\bibitem{costa2019quantum}
Pedro~CS Costa, Stephen Jordan, and Aaron Ostrander.
\newblock Quantum algorithm for simulating the wave equation.
\newblock {\em Physical Review A}, 99(1):012323, 2019.

\bibitem{fillion2019simple}
Fran{\c{c}}ois Fillion-Gourdeau and Emmanuel Lorin.
\newblock Simple digital quantum algorithm for symmetric first-order linear
  hyperbolic systems.
\newblock {\em Numerical Algorithms}, 82(3):1009--1045, 2019.

\bibitem{carleman1932application}
Torsten Carleman.
\newblock Application de la th{\'e}orie des {\'e}quations int{\'e}grales
  lin{\'e}aires aux syst{\`e}mes d'{\'e}quations diff{\'e}rentielles non
  lin{\'e}aires.
\newblock {\em Acta Mathematica}, 59:63--87, 1932.

\bibitem{kowalski1991nonlinear}
Krzysztof Kowalski and W-H Steeb.
\newblock {\em Nonlinear dynamical systems and Carleman linearization}.
\newblock World Scientific, 1991.

\bibitem{liu2021efficient}
Jin-Peng Liu, Herman~{\O}ie Kolden, Hari~K Krovi, Nuno~F Loureiro, Konstantina
  Trivisa, and Andrew~M Childs.
\newblock Efficient quantum algorithm for dissipative nonlinear differential
  equations.
\newblock {\em Proceedings of the National Academy of Sciences}, 118(35), 2021.

\bibitem{leyton2008quantum}
Sarah~K Leyton and Tobias~J Osborne.
\newblock A quantum algorithm to solve nonlinear differential equations.
\newblock {\em arXiv preprint arXiv:0812.4423}, 2008.

\bibitem{lloyd2020quantum}
Seth Lloyd, Giacomo De~Palma, Can Gokler, Bobak Kiani, Zi-Wen Liu, Milad
  Marvian, Felix Tennie, and Tim Palmer.
\newblock Quantum algorithm for nonlinear differential equations.
\newblock {\em arXiv preprint arXiv:2011.06571}, 2020.

\bibitem{joseph2020koopman}
Ilon Joseph.
\newblock {Koopman--von Neumann} approach to quantum simulation of nonlinear
  classical dynamics.
\newblock {\em Physical Review Research}, 2(4):043102, 2020.

\bibitem{jin2022quantum}
Shi Jin and Nana Liu.
\newblock Quantum algorithms for computing observables of nonlinear partial
  differential equations.
\newblock {\em arXiv preprint arXiv:2202.07834}, 2022.

\bibitem{jin2003multi}
Shi Jin and Xiantao Li.
\newblock {Multi-phase computations of the semiclassical limit of the
  Schr{\"o}dinger equation and related problems: Whitham vs Wigner}.
\newblock {\em Physica D: Nonlinear Phenomena}, 182(1-2):46--85, 2003.

\bibitem{fontanela2021quantum}
Filipe Fontanela, Antoine Jacquier, and Mugad Oumgari.
\newblock A quantum algorithm for linear {PDEs} arising in finance.
\newblock {\em SIAM Journal on Financial Mathematics}, 12(4):SC98--SC114, 2021.

\bibitem{alghassi2021variational}
Hedayat Alghassi, Amol Deshmukh, Noelle Ibrahim, Nicolas Robles, Stefan
  Woerner, and Christa Zoufal.
\newblock A variational quantum algorithm for the {Feynman--Kac} formula.
\newblock {\em \href{https://arxiv.org/abs/2108.10846}{arXiv preprint
  arXiv:2108.10846}}, 2021.

\bibitem{lubasch2020variational}
Michael Lubasch, Jaewoo Joo, Pierre Moinier, Martin Kiffner, and Dieter Jaksch.
\newblock Variational quantum algorithms for nonlinear problems.
\newblock {\em Physical Review A}, 101(1):010301, 2020.

\bibitem{kyriienko2021solving}
Oleksandr Kyriienko, Annie~E Paine, and Vincent~E Elfving.
\newblock Solving nonlinear differential equations with differentiable quantum
  circuits.
\newblock {\em Physical Review A}, 103(5):052416, 2021.

\bibitem{kyriienko2022protocols}
Oleksandr Kyriienko, Annie~E Paine, and Vincent~E Elfving.
\newblock Protocols for trainable and differentiable quantum generative
  modelling.
\newblock {\em \href{https://arxiv.org/abs/2202.08253}{arXiv preprint
  arXiv:2202.08253}}, 2022.

\bibitem{Paine2021}
Annie~E. Paine, Vincent~E. Elfving, and Oleksandr Kyriienko.
\newblock Quantum quantile mechanics: Solving stochastic differential equations
  for generating time-series.
\newblock 2021.

\bibitem{steinbrecher2008quantile}
Gyorgy Steinbrecher and William Shaw.
\newblock Quantile mechanics.
\newblock {\em European Journal of Applied Mathematics - EUR J APPL MATH}, 19,
  08 2007.

\bibitem{gentle2003random}
James~E Gentle.
\newblock {\em Random number generation and Monte Carlo methods}, volume 381.
\newblock Springer, 2003.

\bibitem{Kubo_2021}
Kenji Kubo, Yuya~O. Nakagawa, Suguru Endo, and Shota Nagayama.
\newblock Variational quantum simulations of stochastic differential equations.
\newblock {\em Physical Review A}, 103(5), May 2021.

\bibitem{Boyle1986OptionVU}
Phelim~P. Boyle.
\newblock Option valuation using a three jump process.
\newblock 1986.

\bibitem{follmer2016stochastic}
Hans F{\"o}llmer and Alexander Schied.
\newblock Stochastic finance.
\newblock In {\em Stochastic Finance}. de Gruyter, 2016.

\bibitem{weinzierl2000introduction}
Stefan Weinzierl.
\newblock Introduction to {Monte Carlo} methods.
\newblock {\em \href{https://arxiv.org/abs/hep-ph/0006269}{arXiv preprint
  hep-ph/0006269}}, 2000.

\bibitem{boyle1977options}
Phelim~P Boyle.
\newblock Options: A {Monte Carlo} approach.
\newblock {\em Journal of Financial Economics}, 4(3):323--338, 1977.

\bibitem{Stamatopoulos_2020}
Nikitas Stamatopoulos, Daniel~J. Egger, Yue Sun, Christa Zoufal, Raban Iten,
  Ning Shen, and Stefan Woerner.
\newblock Option pricing using quantum computers.
\newblock {\em Quantum}, 4:291, Jul 2020.

\bibitem{Rebentrost_2018}
Patrick Rebentrost, Brajesh Gupt, and Thomas~R. Bromley.
\newblock Quantum computational finance: {Monte Carlo} pricing of financial
  derivatives.
\newblock {\em Physical Review A}, 98(2), Aug 2018.

\bibitem{Miyamoto_derivatives}
Koichi Miyamoto and Kenji Kubo.
\newblock Pricing multi-asset derivatives by finite difference method on a
  quantum computer, 2021.

\bibitem{linden2020}
Noah Linden, Ashley Montanaro, and Changpeng Shao.
\newblock Quantum vs. classical algorithms for solving the heat equation.
\newblock 2020.

\bibitem{gonzalez2021}
Javier Gonzalez-Conde, Ángel Rodríguez-Rozas, Enrique Solano, and Mikel Sanz.
\newblock Simulating option price dynamics with exponential quantum speedup.
\newblock 2021.

\bibitem{shiryaev2007optimal}
Albert~N Shiryaev.
\newblock {\em Optimal stopping rules}, volume~8.
\newblock Springer Science \& Business Media, 2007.

\bibitem{longstaff2001valuing}
Francis~A Longstaff and Eduardo~S Schwartz.
\newblock Valuing {American} options by simulation: a simple least-squares
  approach.
\newblock {\em The Review of Financial Studies}, 14(1):113--147, 2001.

\bibitem{Doriguello_2021}
João~F. Doriguello, Alessandro Luongo, Jinge Bao, Patrick Rebentrost, and
  Miklos Santha.
\newblock Quantum algorithm for stochastic optimal stopping problems with
  applications in finance, 2021.

\bibitem{Miyamoto_options}
Koichi Miyamoto.
\newblock Bermudan option pricing by quantum amplitude estimation and chebyshev
  interpolation, 2021.

\bibitem{tang2021quantum}
Hao Tang, Anurag Pal, Tian-Yu Wang, Lu-Feng Qiao, Jun Gao, and Xian-Min Jin.
\newblock Quantum computation for pricing the collateralized debt obligations.
\newblock {\em Quantum Engineering}, 3(4):e84, 2021.

\bibitem{li2000default}
David~X. Li.
\newblock On default correlation: A copula function approach.
\newblock {\em The Journal of Fixed Income}, 9(4):43--54, 2000.

\bibitem{barndorff1997normal}
Ole~E. Barndorff-Nielsen.
\newblock Normal inverse {Gaussian} distributions and stochastic volatility
  modelling.
\newblock {\em Scandinavian Journal of Statistics}, 24(1):1--13, 1997.

\bibitem{hong}
L.~Jeff Hong, Zhaolin Hu, and Guangwu Liu.
\newblock {Monte Carlo} methods for value-at-risk and conditional
  value-at-risk: A review.
\newblock {\em ACM Trans. Model. Comput. Simul.}, 24(4), Nov 2014.

\bibitem{Woerner_2019}
Stefan Woerner and Daniel~J. Egger.
\newblock Quantum risk analysis.
\newblock {\em npj Quantum Information}, 5(1), Feb 2019.

\bibitem{egger2019credit}
D.~J. Egger, R.~Garcia Gutierrez, J.~Mestre, and S.~Woerner.
\newblock Credit risk analysis using quantum computers.
\newblock {\em IEEE Transactions on Computers}, 70(12):2136--2145, Dec 2021.

\bibitem{stamatopoulos2021}
Nikitas Stamatopoulos, Guglielmo Mazzola, Stefan Woerner, and William~J. Zeng.
\newblock Towards quantum advantage in financial market risk using quantum
  gradient algorithms.
\newblock 2021.

\bibitem{jordan2005}
Stephen~P. Jordan.
\newblock Fast quantum algorithm for numerical gradient estimation.
\newblock {\em Phys. Rev. Lett.}, 95:050501, July 2005.

\bibitem{gaw}
Andr\'{a}s Gily\'{e}n, Srinivasan Arunachalam, and Nathan Wiebe.
\newblock Optimizing quantum optimization algorithms via faster quantum
  gradient computation.
\newblock In {\em Proceedings of the Thirtieth Annual ACM-SIAM Symposium on
  Discrete Algorithms}, SODA '19, page 1425–1444, USA, 2019. Society for
  Industrial and Applied Mathematics.

\bibitem{low2019hamiltonian}
Guang~Hao Low and Isaac~L Chuang.
\newblock Hamiltonian simulation by qubitization.
\newblock {\em Quantum}, 3:163, 2019.

\bibitem{li2005}
Jianping Li.
\newblock General explicit difference formulas for numerical differentiation.
\newblock {\em Journal of Computational and Applied Mathematics},
  183(1):29--52, 2005.

\bibitem{cornelissen2019quantum}
Arjan Cornelissen.
\newblock Quantum gradient estimation of gevrey functions.
\newblock {\em arXiv preprint arXiv:1909.13528}, 2019.

\bibitem{green2015xva}
Andrew Green.
\newblock {\em XVA: credit, funding and capital valuation adjustments}.
\newblock John Wiley \& Sons, 2015.

\bibitem{han2022}
Jeong~Yu Han and Patrick Rebentrost.
\newblock Quantum advantage for multi-option portfolio pricing and valuation
  adjustments.
\newblock 2022.

\bibitem{Alcazar_2022}
Javier Alcazar, Andrea Cadarso, Amara Katabarwa, Marta Mauri, Borja Peropadre,
  Guoming Wang, and Yudong Cao.
\newblock Quantum algorithm for credit valuation adjustments.
\newblock {\em New Journal of Physics}, 24(2):023036, feb 2022.

\bibitem{Wang_2021}
Guoming Wang, Dax~Enshan Koh, Peter~D. Johnson, and Yudong Cao.
\newblock Minimizing estimation runtime on noisy quantum computers.
\newblock {\em {PRX} Quantum}, 2(1), March 2021.

\bibitem{Koh_2022}
Dax~Enshan Koh, Guoming Wang, Peter~D. Johnson, and Yudong Cao.
\newblock Foundations for {Bayesian} inference with engineered likelihood
  functions for robust amplitude estimation.
\newblock {\em Journal of Mathematical Physics}, 63(5):052202, may 2022.

\bibitem{hromkovivc2013algorithmics}
J.~Hromkovi{\v{c}}.
\newblock {\em Algorithmics for hard problems: introduction to combinatorial
  optimization, randomization, approximation, and heuristics}.
\newblock Springer Science \& Business Media, 2013.

\bibitem{nesterov1994interior}
Yurii Nesterov and Arkadii Nemirovskii.
\newblock {\em Interior-point polynomial algorithms in convex programming}.
\newblock SIAM, 1994.

\bibitem{braineMBOtransacitons}
Lee Braine, Daniel~J. Egger, Jennifer Glick, and Stefan Woerner.
\newblock Quantum algorithms for mixed binary optimization applied to
  transaction settlement.
\newblock {\em IEEE Transactions on Quantum Engineering}, 2:1--8, 2021.

\bibitem{kochenberger2014unconstrained}
Gary Kochenberger, Jin-Kao Hao, Fred Glover, Mark Lewis, Zhipeng L{\"u}, Haibo
  Wang, and Yang Wang.
\newblock The unconstrained binary quadratic programming problem: a survey.
\newblock {\em Journal of combinatorial optimization}, 28(1):58--81, 2014.

\bibitem{okada2019efficient}
S.~Okada, M.~Ohzeki, and S.~Taguchi.
\newblock Efficient partition of integer optimization problems with one-hot
  encoding.
\newblock {\em Scientific Reports}, 2019.

\bibitem{IsingNP}
Andrew Lucas.
\newblock Ising formulations of many {NP} problems.
\newblock {\em Frontiers in Physics}, 2:5, 2014.

\bibitem{Kadowaki_1998}
Tadashi Kadowaki and Hidetoshi Nishimori.
\newblock Quantum annealing in the transverse {Ising} model.
\newblock {\em Physical Review E}, 58(5):5355--–5363, Nov 1998.

\bibitem{chancellor2017circuit}
Nicholas Chancellor, Stefan Zohren, and Paul~A Warburton.
\newblock Circuit design for multi-body interactions in superconducting quantum
  annealing systems with applications to a scalable architecture.
\newblock {\em npj Quantum Information}, 3(1):1--7, 2017.

\bibitem{domino2021quadratic}
Krzysztof Domino, Akash Kundu, {\"O}zlem Salehi, and Krzysztof Krawiec.
\newblock Quadratic and higher-order unconstrained binary optimization of
  railway dispatching problem for quantum computing.
\newblock {\em \href{https://arxiv.org/abs/2107.03234}{arXiv preprint
  arXiv:2107.03234}}, 2021.

\bibitem{king2019quantum}
James King, Masoud Mohseni, William Bernoudy, Alexandre Fr{\'e}chette, Hossein
  Sadeghi, Sergei~V Isakov, Hartmut Neven, and Mohammad~H Amin.
\newblock Quantum-assisted genetic algorithm.
\newblock {\em \href{https://arxiv.org/abs/1907.00707}{arXiv preprint
  arXiv:1907.00707}}, 2019.

\bibitem{Ohzeki_annealer}
Masayuki Ohzeki.
\newblock Breaking limitation of quantum annealer in solving optimization
  problems under constraints.
\newblock {\em Scientific Reports}, 10, 02 2020.

\bibitem{lobe2021minor}
Elisabeth Lobe and Annette Lutz.
\newblock Minor embedding in broken chimera and pegasus graphs is
  {NP}-complete.
\newblock {\em \href{https://arxiv.org/abs/2110.08325}{arXiv preprint
  arXiv:2110.08325}}, 2021.

\bibitem{cai2014practical}
Jun Cai, William~G Macready, and Aidan Roy.
\newblock A practical heuristic for finding graph minors.
\newblock {\em \href{https://arxiv.org/abs/1406.2741}{arXiv preprint
  arXiv:1406.2741}}, 2014.

\bibitem{goodrich2017optimizing}
Timothy Goodrich, Blair~D. Sullivan, and T.~Humble.
\newblock Optimizing adiabatic quantum program compilation using a
  graph-theoretic framework.
\newblock {\em Quantum Information Processing}, 17:1--26, 2018.

\bibitem{venturelli2015quantum}
Davide Venturelli, Salvatore Mandr\`a, Sergey Knysh, Bryan O'Gorman, Rupak
  Biswas, and Vadim Smelyanskiy.
\newblock Quantum optimization of fully connected spin glasses.
\newblock {\em Phys. Rev. X}, 5:031040, Sep 2015.

\bibitem{lechner2015quantum}
Wolfgang Lechner, Philipp Hauke, and Peter Zoller.
\newblock A quantum annealing architecture with all-to-all connectivity from
  local interactions.
\newblock {\em Science Advances}, 1(9):e1500838, 2015.

\bibitem{ender2021parity}
Kilian Ender, Roeland ter Hoeven, Benjamin~E Niehoff, Maike Drieb-Sch{\"o}n,
  and Wolfgang Lechner.
\newblock Parity quantum optimization: Compiler.
\newblock {\em \href{https://arxiv.org/abs/2105.06233}{arXiv preprint
  arXiv:2105.06233}}, 2021.

\bibitem{driebschon2021parity}
Maike Drieb-Sch{\"o}n, Younes Javanmard, Kilian Ender, and Wolfgang Lechner.
\newblock Parity quantum optimization: Encoding constraints.
\newblock {\em \href{https://arxiv.org/abs/2105.06235}{arXiv preprint
  arXiv:2105.06235}}, 2021.

\bibitem{fellner2021parity}
Michael Fellner, Kilian Ender, Roeland ter Hoeven, and Wolfgang Lechner.
\newblock Parity quantum optimization: Benchmarks.
\newblock {\em \href{https://arxiv.org/abs/2105.06240}{arXiv preprint
  arXiv:2105.06240}}, 2021.

\bibitem{silverio2021pulser}
Henrique Silv{\'e}rio, Sebasti{\'a}n Grijalva, Constantin Dalyac, Lucas
  Leclerc, Peter~J Karalekas, Nathan Shammah, Mourad Beji, Louis-Paul Henry,
  and Lo{\"\i}c Henriet.
\newblock Pulser: An open-source package for the design of pulse sequences in
  programmable neutral-atom arrays.
\newblock {\em \href{https://arxiv.org/abs/2104.15044}{arXiv preprint
  arXiv:2104.15044}}, 2021.

\bibitem{Henriet2020quantumcomputing}
Lo{\"{i}}c Henriet, Lucas Beguin, Adrien Signoles, Thierry Lahaye, Antoine
  Browaeys, Georges-Olivier Reymond, and Christophe Jurczak.
\newblock Quantum computing with neutral atoms.
\newblock {\em Quantum}, 4:327, September 2020.

\bibitem{ebadi20}
S.~Ebadi, A.~Keesling, M.~Cain, T.~T. Wang, H.~Levine, D.~Bluvstein,
  G.~Semeghini, A.~Omran, J.-G. Liu, R.~Samajdar, X.-Z. Luo, B.~Nash, X.~Gao,
  B.~Barak, E.~Farhi, S.~Sachdev, N.~Gemelke, L.~Zhou, S.~Choi, H.~Pichler,
  S.-T. Wang, M.~Greiner, V.~Vuletić, and M.~D. Lukin.
\newblock Quantum optimization of maximum independent set using rydberg atom
  arrays.
\newblock {\em Science}, 376(6598):1209--1215, 2022.

\bibitem{Shaydulin2021}
Ruslan Shaydulin, Stuart Hadfield, Tad Hogg, and Ilya Safro.
\newblock Classical symmetries and the quantum approximate optimization
  algorithm.
\newblock {\em Quantum Information Processing}, 20(11):359, Oct 2021.

\bibitem{shaydulin2019multistart}
Ruslan Shaydulin, Ilya Safro, and Jeffrey Larson.
\newblock Multistart methods for quantum approximate optimization.
\newblock In {\em 2019 IEEE High Performance Extreme Computing Conference
  (HPEC)}, pages 1--8, 2019.

\bibitem{galda2021transferability}
Alexey Galda, Xiaoyuan Liu, Danylo Lykov, Yuri Alexeev, and Ilya Safro.
\newblock Transferability of optimal {QAOA} parameters between random graphs.
\newblock In {\em 2021 IEEE International Conference on Quantum Computing and
  Engineering (QCE)}, pages 171--180. IEEE, 2021.

\bibitem{streif2020training}
Michael Streif and Martin Leib.
\newblock Training the quantum approximate optimization algorithm without
  access to a quantum processing unit.
\newblock {\em Quantum Science and Technology}, 5(3):034008, 2020.

\bibitem{khairy2019learning}
Sami Khairy, Ruslan Shaydulin, Lukasz Cincio, Yuri Alexeev, and Prasanna
  Balaprakash.
\newblock Learning to optimize variational quantum circuits to solve
  combinatorial problems.
\newblock In {\em Proceedings of the Thirty-Fourth AAAI Conference on
  Artificial Intelligence (AAAI)}, 2020.

\bibitem{shaydulin2019evaluating}
Ruslan Shaydulin and Yuri Alexeev.
\newblock Evaluating quantum approximate optimization algorithm: A case study.
\newblock In {\em Proceedings of the 2nd International Workshop on Quantum
  Computing for Sustainable Computing}, 2019.

\bibitem{wang2022quantum}
Zhen-Duo Wang, Pei-Lin Zheng, Biao Wu, and Yi~Zhang.
\newblock Quantum dropout for efficient quantum approximate optimization
  algorithm on combinatorial optimization problems.
\newblock {\em arXiv preprint arXiv:2203.10101}, 2022.

\bibitem{liu2022sparsification}
Xiaoyuan Liu, Ruslan Shaydulin, and Ilya Safro.
\newblock Quantum approximate optimization algorithm with sparsified phase
  operator, 2022.

\bibitem{gulania2022dynamics}
Sahil Gulania, Bo~Peng, Yuri Alexeev, and Niranjan Govind.
\newblock Quantum time dynamics of {1D-Heisenberg models employing the
  Yang--Baxter} equation for circuit compression.
\newblock 2021.

\bibitem{harrigan2021quantum}
Matthew~P Harrigan, Kevin~J Sung, Matthew Neeley, Kevin~J Satzinger, Frank
  Arute, Kunal Arya, Juan Atalaya, Joseph~C Bardin, Rami Barends, Sergio Boixo,
  et~al.
\newblock Quantum approximate optimization of non-planar graph problems on a
  planar superconducting processor.
\newblock {\em Nature Physics}, 17(3):332--336, 2021.

\bibitem{otterbach2017unsupervised}
Johannes~S Otterbach, Riccardo Manenti, Nasser Alidoust, A~Bestwick, M~Block,
  B~Bloom, S~Caldwell, N~Didier, E~Schuyler Fried, S~Hong, et~al.
\newblock Unsupervised machine learning on a hybrid quantum computer.
\newblock {\em arXiv preprint arXiv:1712.05771}, 2017.

\bibitem{niroula2022}
Pradeep Niroula, Ruslan Shaydulin, Romina Yalovetzky, Pierre Minssen, Dylan
  Herman, Shaohan Hu, and Marco Pistoia.
\newblock Constrained quantum optimization for extractive summarization on a
  trapped-ion quantum computer.
\newblock 2022.

\bibitem{fedorov2022vqe}
Dmitry~A Fedorov, Bo~Peng, Niranjan Govind, and Yuri Alexeev.
\newblock {VQE method: A short survey and recent developments}.
\newblock {\em Materials Theory}, 6(1):1--21, 2022.

\bibitem{evqe}
Arthur~G. Rattew, Shaohan Hu, Marco Pistoia, Richard Chen, and Steve Wood.
\newblock {A Domain-agnostic, Noise-resistant, Hardware-efficient Evolutionary
  Variational Quantum Eigensolver}.
\newblock {\em \href{https://arxiv.org/abs/1910.09694}{arXiv preprint
  arXiv:1910.09694}}, 2020.

\bibitem{filtervar}
David Amaro, Carlo Modica, Matthias Rosenkranz, Mattia Fiorentini, Marcello
  Benedetti, and Michael Lubasch.
\newblock Filtering variational quantum algorithms for combinatorial
  optimization.
\newblock {\em \href{https://arxiv.org/abs/2106.10055}{arXiv preprint
  arXiv:2106.10055}}, 2021.

\bibitem{xuchen2022}
Xuchen You, Shouvanik Chakrabarti, and Xiaodi Wu.
\newblock A convergence theory for over-parameterized variational quantum
  eigensolvers.
\newblock 2022.

\bibitem{naber2012geometry}
Gregory~L Naber.
\newblock {\em The Geometry of Minkowski Spacetime: An Introduction to the
  Mathematics of the Special Theory of Relativity}, volume~92.
\newblock Springer Science \& Business Media, 01 2012.

\bibitem{osterwalder1973axioms}
Konrad Osterwalder and Robert Schrader.
\newblock {Axioms for Euclidean Green's functions}.
\newblock {\em Communications in mathematical physics}, 31(2):83--112, 1973.

\bibitem{wick1954properties}
G.~C. Wick.
\newblock Properties of {Bethe-Salpeter} wave functions.
\newblock {\em Phys. Rev.}, 96:1124--1134, Nov 1954.

\bibitem{McArdle_2019}
Sam McArdle, Tyson Jones, Suguru Endo, Ying Li, Simon~C. Benjamin, and Xiao
  Yuan.
\newblock Variational ansatz-based quantum simulation of imaginary time
  evolution.
\newblock {\em npj Quantum Information}, 5(1), Sep 2019.

\bibitem{2019ITE}
Mario Motta, Chong Sun, Adrian T.~K. Tan, Matthew~J. O’Rourke, Erika Ye,
  Austin~J. Minnich, Fernando G. S.~L. Brandão, and Garnet Kin-Lic Chan.
\newblock Determining eigenstates and thermal states on a quantum computer
  using quantum imaginary time evolution.
\newblock {\em Nature Physics}, 16(2):205–210, Nov 2019.

\bibitem{Mitarai_2019}
Kosuke Mitarai and Keisuke Fujii.
\newblock Methodology for replacing indirect measurements with direct
  measurements.
\newblock {\em Physical Review Research}, 1(1), Aug 2019.

\bibitem{effvarte}
Marcello Benedetti, Mattia Fiorentini, and Michael Lubasch.
\newblock Hardware-efficient variational quantum algorithms for time evolution.
\newblock {\em Phys. Rev. Research}, 3:033083, Jul 2021.

\bibitem{Zoufal_2021}
Christa Zoufal, Aurélien Lucchi, and Stefan Woerner.
\newblock Variational quantum {Boltzmann} machines.
\newblock {\em Quantum Machine Intelligence}, 3(1), Feb 2021.

\bibitem{Chen_2020}
Ming-Cheng Chen, Ming Gong, Xiaosi Xu, Xiao Yuan, Jian-Wen Wang, Can Wang,
  Chong Ying, Jin Lin, Yu~Xu, Yulin Wu, and et~al.
\newblock Demonstration of adiabatic variational quantum computing with a
  superconducting quantum coprocessor.
\newblock {\em Physical Review Letters}, 125(18), Oct 2020.

\bibitem{durrhoyer1996GroverMin}
Christoph D\"{u}rr and Peter H\o{}yer.
\newblock A quantum algorithm for finding the minimum.
\newblock {\em \href{https://arxiv.org/abs/quant-ph/9607014}{arXiv preprint
  quant-ph/9607014}}, 1996.

\bibitem{baritompa2005grover}
William~P Baritompa, David~W Bulger, and Graham~R Wood.
\newblock Grover's quantum algorithm applied to global optimization.
\newblock {\em SIAM Journal on Optimization}, 15(4):1170--1184, 2005.

\bibitem{bulger2003implementing}
David Bulger, William~P. Baritompa, and Graham~R. Wood.
\newblock Implementing pure adaptive search with {Grover's} quantum algorithm.
\newblock {\em Journal of optimization theory and applications},
  116(3):517--529, 2003.

\bibitem{Gilliam2021groveradaptive}
Austin Gilliam, Stefan Woerner, and Constantin Gonciulea.
\newblock Grover adaptive search for constrained polynomial binary
  optimization.
\newblock {\em {Quantum}}, 5:428, 2021.

\bibitem{bertsimas1997introduction}
Dimitris Bertsimas and John~N Tsitsiklis.
\newblock {\em Introduction to linear optimization}, volume~6.
\newblock Athena Scientific Belmont, MA, 1997.

\bibitem{nannicini2021fast}
Giacomo Nannicini.
\newblock Fast quantum subroutines for the simplex method.
\newblock In {\em Integer Programming and Combinatorial Optimization - 22nd
  International Conference, {IPCO} 2021, Atlanta, GA, USA, May 19-21, 2021,
  Proceedings}, volume 12707 of {\em Lecture Notes in Computer Science}, pages
  311--325. Springer, 2021.

\bibitem{Kerenidis_2021}
Iordanis Kerenidis, Anupam Prakash, and Dániel Szilágyi.
\newblock Quantum algorithms for second-order cone programming and support
  vector machines.
\newblock {\em Quantum}, 5:427, April 2021.

\bibitem{monteiro2000polynomial}
Renato Monteiro and Takashi Tsuchiya.
\newblock Polynomial convergence of primal-dual algorithms for the second-order
  cone program based on the {MZ}-family of directions.
\newblock {\em Mathematical programming}, 88(1):61--83, 2000.

\bibitem{brandao2017quantum}
Fernando~G.S.L. Brandao and Krysta~M. Svore.
\newblock Quantum speed-ups for solving semidefinite programs.
\newblock In {\em 2017 IEEE 58th Annual Symposium on Foundations of Computer
  Science (FOCS)}, pages 415--426, 2017.

\bibitem{brandao2019quantum}
Fernando G. S.~L. Brand{\~a}o, Amir Kalev, Tongyang Li, Cedric Yen-Yu Lin,
  Krysta~M. Svore, and Xiaodi Wu.
\newblock Quantum {SDP} solvers: Large speed-ups, optimality, and applications
  to quantum learning.
\newblock In {\em 46th International Colloquium on Automata, Languages, and
  Programming (ICALP 2019)}, volume 132 of {\em Leibniz International
  Proceedings in Informatics (LIPIcs)}, pages 27:1--27:14. Schloss
  Dagstuhl--Leibniz-Zentrum fuer Informatik, 2019.

\bibitem{2020sdpvanApeldoorn}
Joran van Apeldoorn, András Gilyén, Sander Gribling, and Ronald de~Wolf.
\newblock Quantum {SDP}-solvers: Better upper and lower bounds.
\newblock {\em Quantum}, 4:230, Feb 2020.

\bibitem{arora2007combinatorial}
Sanjeev Arora and Satyen Kale.
\newblock A combinatorial, primal-dual approach to semidefinite programs.
\newblock {\em J. ACM}, 63(2), may 2016.

\bibitem{shaydulin2019hybrid}
Ruslan Shaydulin, Hayato Ushijima-Mwesigwa, Christian~FA Negre, Ilya Safro,
  Susan~M Mniszewski, and Yuri Alexeev.
\newblock A hybrid approach for solving optimization problems on small quantum
  computers.
\newblock {\em Computer}, 52(6):18--26, 2019.

\bibitem{shaydulin2019network}
Ruslan Shaydulin, Hayato Ushijima-Mwesigwa, Ilya Safro, Susan Mniszewski, and
  Yuri Alexeev.
\newblock Network community detection on small quantum computers.
\newblock {\em Advanced Quantum Technologies}, 2(9):1900029, 2019.

\bibitem{ushijima2021multilevel}
Hayato Ushijima-Mwesigwa, Ruslan Shaydulin, Christian F.~A. Negre, Susan~M.
  Mniszewski, Yuri Alexeev, and Ilya Safro.
\newblock Multilevel combinatorial optimization across quantum architectures.
\newblock {\em ACM Transactions on Quantum Computing}, 2(1), feb 2021.

\bibitem{chapuis2019finding}
Guillaume Chapuis, Hristo Djidjev, Georg Hahn, and Guillaume Rizk.
\newblock Finding maximum cliques on the {D-Wave} quantum annealer.
\newblock {\em Journal of Signal Processing Systems}, 91(3):363--377, 2019.

\bibitem{markowitzmpt1952}
Harry Markowitz.
\newblock Portfolio selection.
\newblock {\em The Journal of Finance}, 7(1):77--91, 1952.

\bibitem{benati2007mixed}
S.~Benati and R.~Rizzi.
\newblock A mixed integer linear programming formulation of the optimal
  mean/value-at-risk portfolio problem.
\newblock {\em European Journal of Operational Research}, 2007.

\bibitem{mansini2015linear}
Renata Mansini, W\l{}odzimierz Ogryczak, and M.~Grazia Speranza.
\newblock {\em Linear and mixed integer programming for portfolio
  optimization}.
\newblock Springer, 2015.

\bibitem{cornuejols1977exceptional}
Gerard Cornuejols, Marshall~L Fisher, and George~L Nemhauser.
\newblock Exceptional paper—location of bank accounts to optimize float: An
  analytic study of exact and approximate algorithms.
\newblock {\em Management Science}, 23(8):789--810, 1977.

\bibitem{Portfolio_asset}
Angad Kalra, Faisal Qureshi, and Michael Tisi.
\newblock Portfolio asset identification using graph algorithms on a quantum
  annealer.
\newblock {\em Available at SSRN 3333537}, 2018.

\bibitem{Rounds_risk}
Gili Rosenberg and Maxwell Rounds.
\newblock Long-short minimum risk parity optimization using a quantum or
  digital annealer.
\newblock {\em
  \href{https://1qbit.com/whitepaper/long-short-minimum-risk-parity-optimization-quantum-or-digital-annealer/}{White
  Paper 1Qbit}}, 2018.

\bibitem{hodson2019portfolio}
Mark Hodson, Brendan Ruck, Hugh Ong, David Garvin, and Stefan Dulman.
\newblock Portfolio rebalancing experiments using the quantum alternating
  operator ansatz.
\newblock {\em \href{https://arxiv.org/abs/1911.05296}{arXiv preprint
  arXiv:1911.0529}6}, 2019.

\bibitem{slate2020quantum}
N.~Slate, E.~Matwiejew, S.~Marsh, and J.~B. Wang.
\newblock Quantum walk-based portfolio optimisation.
\newblock {\em {Quantum}}, 5:513, July 2021.

\bibitem{Rosenberg_2016}
Gili Rosenberg, Poya Haghnegahdar, Phil Goddard, Peter Carr, Kesheng Wu, and
  Marcos~Lopez de~Prado.
\newblock Solving the optimal trading trajectory problem using a quantum
  annealer.
\newblock {\em IEEE Journal of Selected Topics in Signal Processing},
  10(6):1053–1060, Sep 2016.

\bibitem{mugel2020dynamic}
Samuel Mugel, Carlos Kuchkovsky, Escol\'astico S\'anchez, Samuel
  Fern\'andez-Lorenzo, Jorge Luis-Hita, Enrique Lizaso, and Rom\'an Or\'us.
\newblock Dynamic portfolio optimization with real datasets using quantum
  processors and quantum-inspired tensor networks.
\newblock {\em Phys. Rev. Research}, 4:013006, Jan 2022.

\bibitem{Grant_2021}
Erica Grant, Travis~S. Humble, and Benjamin Stump.
\newblock Benchmarking quantum annealing controls with portfolio optimization.
\newblock {\em Physical Review Applied}, 15(1), Jan 2021.

\bibitem{kerenidis2019quantumPortOpt}
Iordanis Kerenidis, Anupam Prakash, and D\'{a}niel Szil\'{a}gyi.
\newblock Quantum algorithms for portfolio optimization.
\newblock In {\em Proceedings of the 1st ACM Conference on Advances in
  Financial Technologies}, AFT '19, pages 147–--155, New York, NY, USA, 2019.
  Association for Computing Machinery.

\bibitem{rebentrost2018quantum}
Patrick Rebentrost and Seth Lloyd.
\newblock Quantum computational finance: quantum algorithm for portfolio
  optimization.
\newblock {\em \href{https://arxiv.org/abs/1811.03975}{arXiv preprint
  arXiv:1811.03975}}, 2018.

\bibitem{gilyen_stoch_regr2018}
András Gilyén, Seth Lloyd, and Ewin Tang.
\newblock Quantum-inspired low-rank stochastic regression with logarithmic
  dependence on the dimension.
\newblock 2018.

\bibitem{Arrazola_2020}
Juan~Miguel Arrazola, Alain Delgado, Bhaskar~Roy Bardhan, and Seth Lloyd.
\newblock Quantum-inspired algorithms in practice.
\newblock {\em Quantum}, 4:307, aug 2020.

\bibitem{Swap_netting}
G.~Rosenberg, C.~Adolphs, A.~Milne, and A.~Lee.
\newblock Swap netting using a quantum annealer.
\newblock {\em
  \href{https://1qbit.com/whitepaper/swap-netting-using-a-quantum-annealer/}{White
  Paper 1Qbit}}, 2016.

\bibitem{stulz2010credit}
Ren{\'e}~M Stulz.
\newblock Credit default swaps and the credit crisis.
\newblock {\em Journal of Economic Perspectives}, 24(1):73--92, 2010.

\bibitem{hu1989swaps}
Henry~TC Hu.
\newblock Swaps, the modern process of financial innovation and the
  vulnerability of a regulatory paradigm.
\newblock {\em U. Pa. L. Rev.}, 138:333, 1989.

\bibitem{bicksler1986economic}
James Bicksler and Andrew~H. Chen.
\newblock An economic analysis of interest rate swaps.
\newblock {\em The Journal of Finance}, 41(3):645--655, 1986.

\bibitem{duffie1996swap}
Darrell Duffie and Ming Huang.
\newblock Swap rates and credit quality.
\newblock {\em The Journal of Finance}, 51(3):921--949, 1996.

\bibitem{shleifer1997limits}
Andrei Shleifer and Robert~W. Vishny.
\newblock The limits of arbitrage.
\newblock {\em The Journal of Finance}, 52(1):35--55, 1997.

\bibitem{delbaen2006mathematics}
Freddy Delbaen and Walter Schachermayer.
\newblock {\em The mathematics of arbitrage}.
\newblock Springer Science \& Business Media, 2006.

\bibitem{Rosenberg_arbitrage}
Gilli Rosenberg.
\newblock Finding optimal arbitrage opportunities using a quantum annealer.
\newblock {\em \href{https://1qbit.com/whitepaper/arbitrage/}{White Paper
  1Qbit}}, 2016.

\bibitem{cherkassky1999negative}
Boris~V Cherkassky and Andrew~V Goldberg.
\newblock Negative-cycle detection algorithms.
\newblock {\em Mathematical Programming}, 85(2), 1999.

\bibitem{soon2011currency}
Wanmei Soon and Heng-Qing Ye.
\newblock Currency arbitrage detection using a binary integer programming
  model.
\newblock {\em International Journal of Mathematical Education in Science and
  Technology}, 42(3):369--376, 2011.

\bibitem{Optimal_Feature}
Andrew Milne, Maxwell Rounds, and Phil Goddard.
\newblock Optimal feature selection in credit scoring and classification using
  a quantum annealer.
\newblock {\em
  \href{https://1qbit.com/whitepaper/optimal-feature-selection-in-credit-scoring-classification-using-quantum-annealer/}{White
  Paper 1Qbit}}, 2017.

\bibitem{financial_model}
Matthew Elliott, Benjamin Golub, and Matthew~O. Jackson.
\newblock Financial networks and contagion.
\newblock {\em American Economic Review}, 104(10):3115--53, October 2014.

\bibitem{hemenway2016sensitivity}
Brett Hemenway and Sanjeev Khanna.
\newblock Sensitivity and computational complexity in financial networks.
\newblock {\em Algorithmic Finance}, 5(3-4):95--110, 2016.

\bibitem{grabel2003predicting}
Ilene Grabel.
\newblock Predicting financial crisis in developing economies: astronomy or
  astrology?
\newblock {\em \href{https://www.jstor.org/stable/40325412 }{Eastern Economic
  Journal}}, 29(2):243--258, 2003.

\bibitem{Classical_crash}
Arturo Estrella and Frederic~S. Mishkin.
\newblock {Predicting U.S.} recessions: Financial variables as leading
  indicators.
\newblock {\em \href{http://www.jstor.org/stable/2646728}{The Review of
  Economics and Statistics}}, 1998.

\bibitem{Financial_crash}
Román Orús, Samuel Mugel, and Enrique Lizaso.
\newblock Forecasting financial crashes with quantum computing.
\newblock {\em Phys. Rev. A}, 99(6), Jun 2019.

\bibitem{plesch2011quantum}
Martin Plesch and \ifmmode \check{C}\else~\v{C}\fi{}aslav Brukner.
\newblock Quantum-state preparation with universal gate decompositions.
\newblock {\em Phys. Rev. A}, 83:032302, Mar 2011.

\bibitem{mcclean2016theory}
Jarrod~R McClean, Jonathan Romero, Ryan Babbush, and Al{\'{a}}n Aspuru-Guzik.
\newblock The theory of variational hybrid quantum-classical algorithms.
\newblock {\em New Journal of Physics}, 18(2):023023, Feb 2016.

\bibitem{huang2021power}
Hsin-Yuan Huang, Michael Broughton, Masoud Mohseni, Ryan Babbush, Sergio Boixo,
  Hartmut Neven, and Jarrod~R McClean.
\newblock Power of data in quantum machine learning.
\newblock {\em Nature Communications}, 12(1):1--9, 2021.

\bibitem{wiebe2012quantum}
Nathan Wiebe, Daniel Braun, and Seth Lloyd.
\newblock Quantum algorithm for data fitting.
\newblock {\em Phys. Rev. Lett.}, 109:050505, Aug 2012.

\bibitem{wang2017quantum}
Guoming Wang.
\newblock Quantum algorithm for linear regression.
\newblock {\em Phys. Rev. A}, 96:012335, Jul 2017.

\bibitem{date2020adiabatic}
Prasanna Date and Thomas Potok.
\newblock Adiabatic quantum linear regression.
\newblock {\em Scientific reports}, 11(1):21905, 2021.

\bibitem{zhao2019quantum}
Zhikuan Zhao, Jack~K. Fitzsimons, and Joseph~F. Fitzsimons.
\newblock Quantum-assisted {Gaussian} process regression.
\newblock {\em Physical Review A}, 99(5):052331, 2019.

\bibitem{PhysRevA.98.032309}
K.~Mitarai, M.~Negoro, M.~Kitagawa, and K.~Fujii.
\newblock Quantum circuit learning.
\newblock {\em Phys. Rev. A}, 98:032309, Sep 2018.

\bibitem{muller2001introduction}
K.-R. Muller, S.~Mika, G.~Ratsch, K.~Tsuda, and B.~Scholkopf.
\newblock An introduction to kernel-based learning algorithms.
\newblock {\em IEEE Transactions on Neural Networks}, 12(2):181--201, 2001.

\bibitem{rebentrost2014quantum}
Patrick Rebentrost, Masoud Mohseni, and Seth Lloyd.
\newblock Quantum support vector machine for big data classification.
\newblock {\em Phys. Rev. Lett.}, 2014.

\bibitem{lloyd2013quantum}
Seth Lloyd, Masoud Mohseni, and Patrick Rebentrost.
\newblock Quantum algorithms for supervised and unsupervised machine learning.
\newblock {\em \href{https://arxiv.org/abs/1307.0411}{arXiv preprint
  arXiv:1307.0411}}, 2013.

\bibitem{Havl_ek_2019}
Vojtěch Havlíček, Antonio~D. Córcoles, Kristan Temme, Aram~W. Harrow,
  Abhinav Kandala, Jerry~M. Chow, and Jay~M. Gambetta.
\newblock Supervised learning with quantum-enhanced feature spaces.
\newblock {\em Nature}, 567(7747):209–212, March 2019.

\bibitem{shaydulin2021bandwidth}
Ruslan Shaydulin and Stefan~M. Wild.
\newblock Importance of kernel bandwidth in quantum machine learning.
\newblock 2021.

\bibitem{canatar2022}
Abdulkadir Canatar, Evan Peters, Cengiz Pehlevan, Stefan~M. Wild, and Ruslan
  Shaydulin.
\newblock Bandwidth enables generalization in quantum kernel models.
\newblock 2022.

\bibitem{Kublerkernels2021}
Jonas~M. Kübler, Simon Buchholz, and Bernhard Schölkopf.
\newblock The inductive bias of quantum kernels.
\newblock 2021.

\bibitem{wiebe2014quantum}
N.~Wiebe, A.~Kapoor, and K.~Svore.
\newblock Quantum algorithms for nearest-neighbor methods for supervised and
  unsupervised learning.
\newblock {\em \href{https://dl.acm.org/doi/10.5555/2871393.2871400}{Quantum
  Information \& Computation}}, 15, 2015.

\bibitem{ruan2017quantum}
Yue Ruan, Xiling Xue, Heng Liu, Jianing Tan, and Xi~Li.
\newblock Quantum algorithm for k-nearest neighbors classification based on the
  metric of {Hamming} distance.
\newblock {\em International Journal of Theoretical Physics},
  56(11):3496--3507, 2017.

\bibitem{basheer2021quantum}
Afrad Basheer, A~Afham, and Sandeep~K Goyal.
\newblock Quantum $k$-nearest neighbors algorithm.
\newblock {\em \href{https://arxiv.org/abs/2003.09187}{arXiv preprint
  arXiv:2003.09187}}, 2020.

\bibitem{macqueen1967some}
James MacQueen.
\newblock Some methods for classification and analysis of multivariate
  observations.
\newblock In {\em
  \href{https://projecteuclid.org/proceedings/berkeley-symposium-on-mathematical-statistics-and-probability/proceedings-of-the-fifth-berkeley-symposium-on-mathematical-statistics-and-probability-volume-1-statistics/toc/bsmsp/1200512974}{Proceedings
  of the fifth Berkeley symposium on mathematical statistics and probability}},
  volume~1, pages 281--297. Oakland, CA, USA, 1967.

\bibitem{kerenidis2018qmeans}
Iordanis Kerenidis, Jonas Landman, Alessandro Luongo, and Anupam Prakash.
\newblock q-means: A quantum algorithm for unsupervised machine learning.
\newblock {\em \href{https://arxiv.org/abs/1812.03584}{arXiv preprint
  arXiv:1812.03584}}, 2018.

\bibitem{khan2019kmeans}
Sumsam~Ullah Khan, Ahsan~Javed Awan, and Gemma Vall-Llosera.
\newblock K-means clustering on noisy intermediate scale quantum computers.
\newblock {\em \href{https://arxiv.org/abs/1909.12183}{arXiv preprint
  arXiv:1909.12183}}, 2019.

\bibitem{PhysRevA.101.012326}
Hideyuki Miyahara, Kazuyuki Aihara, and Wolfgang Lechner.
\newblock Quantum expectation-maximization algorithm.
\newblock {\em Phys. Rev. A}, 101:012326, Jan 2020.

\bibitem{ng2002spectral}
Andrew~Y. Ng, Michael~I. Jordan, and Yair Weiss.
\newblock On spectral clustering: Analysis and an algorithm.
\newblock In {\em
  \href{https://papers.nips.cc/paper/2001/hash/801272ee79cfde7fa5960571fee36b9b-Abstract.html}{Proceedings
  of the 14th International Conference on Neural Information Processing
  Systems: Natural and Synthetic}}, pages 849–--856, 2001.

\bibitem{daskin2017quantum}
Ammar Daskin.
\newblock Quantum spectral clustering through a biased phase estimation
  algorithm.
\newblock {\em
  \href{https://dergipark.org.tr/en/pub/twmsjaem/issue/55715/761759}{TWMS
  Journal of Applied and Engineering Mathematics}}, 10(1):24--33, 2017.

\bibitem{apers2020quantum}
Simon Apers and Ronald de~Wolf.
\newblock Quantum speedup for graph sparsification, cut approximation and
  {L}aplacian solving.
\newblock In {\em 2020 IEEE 61st Annual Symposium on Foundations of Computer
  Science (FOCS)}, pages 637--648, 2020.

\bibitem{kerenidis2021quantum}
I.~Kerenidis and J.~Landman.
\newblock Quantum spectral clustering.
\newblock {\em Physical Review A}, 103(4):042415, 2021.

\bibitem{Otterbach_ML}
J.~S. Otterbach, R.~Manenti, N.~Alidoust, A.~Bestwick, M.~Block, B.~Bloom,
  S.~Caldwell, N.~Didier, E.~Schuyler Fried, S.~Hong, P.~Karalekas, C.~B.
  Osborn, A.~Papageorge, E.~C. Peterson, G.~Prawiroatmodjo, N.~Rubin, Colm~A.
  Ryan, D.~Scarabelli, M.~Scheer, E.~A. Sete, P.~Sivarajah, Robert~S. Smith,
  A.~Staley, N.~Tezak, W.~J. Zeng, A.~Hudson, Blake~R. Johnson, M.~Reagor,
  M.~P. da~Silva, and C.~Rigetti.
\newblock Unsupervised machine learning on a hybrid quantum computer, 2017.

\bibitem{Aimeur_Clustering}
E.~Aïmeur, G.~Brassard, and S.~Gambs.
\newblock Quantum clustering algorithms.

\bibitem{Aimeur_ML}
E.~Aïmeur, G.~Brassard, and S.~Gambs.
\newblock Quantum speed-up for unsupervised learning.
\newblock {\em Mach Learn}, 20:261--287, 2013.

\bibitem{Kerenidis}
Iordanis Kerenidis, Alessandro Luongo, and Anupam Prakash.
\newblock Quantum expectation-maximization for {G}aussian mixture models.
\newblock In Hal~Daumé III and Aarti Singh, editors, {\em Proceedings of the
  37th International Conference on Machine Learning}, volume 119 of {\em
  Proceedings of Machine Learning Research}, pages 5187--5197. PMLR, 13--18 Jul
  2020.

\bibitem{Kumar_2018}
Vaibhaw Kumar, Gideon Bass, Casey Tomlin, and Joseph Dulny.
\newblock Quantum annealing for combinatorial clustering.
\newblock {\em Quantum Information Processing}, 17(2), jan 2018.

\bibitem{lloyd2014quantum}
Seth Lloyd, Masoud Mohseni, and Patrick Rebentrost.
\newblock Quantum principal component analysis.
\newblock {\em Nature Physics}, 10(9):631--–633, July 2014.

\bibitem{He_2020}
Chen He, Jiazhen Li, Weiqi Liu, and Z.~Jane Wang.
\newblock A low complexity quantum principal component analysis algorithm,
  2020.

\bibitem{Yu_2019}
CH. Yu, F.~Gao, and S.~et~al Lin.
\newblock Quantum data compression by principal component analysis.
\newblock {\em Quantum Inf Process}, 18:249, 2019.

\bibitem{lin2019improved}
Jie Lin, Wan-Su Bao, Shuo Zhang, Tan Li, and Xiang Wang.
\newblock An improved quantum principal component analysis algorithm based on
  the quantum singular threshold method.
\newblock {\em Physics Letters A}, 383(24):2862--2868, 2019.

\bibitem{Bellante}
Armando Bellante, Alessandro Luongo, and Stefano Zanero.
\newblock Quantum algorithms for data representation and analysis, 2021.

\bibitem{Li_2020}
YaoChong Li, Ri-Gui Zhou, RuiQing Xu, WenWen Hu, and Ping Fan.
\newblock Quantum algorithm for the nonlinear dimensionality reduction with
  arbitrary kernel.
\newblock {\em Quantum Science and Technology}, 6(1):014001, Nov 2020.

\bibitem{Cong_2016}
Iris Cong and Luming Duan.
\newblock Quantum discriminant analysis for dimensionality reduction and
  classification.
\newblock {\em New Journal of Physics}, 18(7):073011, July 2016.

\bibitem{Kerendis_2020}
Iordanis Kerenidis and Alessandro Luongo.
\newblock Classification of the {MNIST} data set with quantum slow feature
  analysis.
\newblock {\em Phys. Rev. A}, 101:062327, Jun 2020.

\bibitem{lloyd2016quantum}
Seth Lloyd, Silvano Garnerone, and Paolo Zanardi.
\newblock Quantum algorithms for topological and geometric analysis of data.
\newblock {\em Nat. Commun.}, 7(1):1--7, 2016.

\bibitem{gyurik2020quantum}
Casper Gyurik, Chris Cade, and Vedran Dunjko.
\newblock Towards quantum advantage via topological data analysis.
\newblock {\em \href{https://arxiv.org/abs/2005.02607}{arXiv preprint
  arXiv:2005.02607}}, 2020.

\bibitem{ubaru2021quantum}
Shashanka Ubaru, Ismail~Yunus Akhalwaya, Mark~S Squillante, Kenneth~L Clarkson,
  and Lior Horesh.
\newblock Quantum topological data analysis with linear depth and exponential
  speedup.
\newblock {\em \href{https://arxiv.org/abs/2108.02811}{arXiv preprint
  arXiv:2108.02811}}, 2021.

\bibitem{kerendis2022}
Iordanis Kerenidis and Anupam Prakash.
\newblock Quantum machine learning with subspace states.
\newblock 2022.

\bibitem{Marcello2019}
Marcello Benedetti, Delfina Garcia-Pintos, Oscar Perdomo, Vicente
  Leyton-Ortega, Yunseong Nam, and Alejandro Perdomo-Ortiz.
\newblock A generative modeling approach for benchmarking and training shallow
  quantum circuits.
\newblock {\em npj Quantum Information}, 5(1), May 2019.

\bibitem{liu2018differentiable}
Jin-Guo Liu and Lei Wang.
\newblock Differentiable learning of quantum circuit {Born} machines.
\newblock {\em Physical Review A}, 98(6), Dec 2018.

\bibitem{coyle2021quantum}
Brian Coyle, Maxwell Henderson, Justin Chan~Jin Le, Niraj Kumar, Marco Paini,
  and Elham Kashefi.
\newblock Quantum versus classical generative modelling in finance.
\newblock {\em {Quantum Science and Technology}}, 6(2):024013, 2021.

\bibitem{zhu2021generative}
Elton~Yechao Zhu, Sonika Johri, Dave Bacon, Mert Esencan, Jungsang Kim, Mark
  Muir, Nikhil Murgai, Jason Nguyen, Neal Pisenti, Adam Schouela, et~al.
\newblock Generative quantum learning of joint probability distribution
  functions.
\newblock {\em \href{https://arxiv.org/abs/2109.06315}{arXiv preprint
  arXiv:2109.06315}}, 2021.

\bibitem{koller2009probabilistic}
Daphne Koller and Nir Friedman.
\newblock {\em
  \href{https://mitpress.mit.edu/books/probabilistic-graphical-models}{Probabilistic
  Graphical Models: Principles and Techniques}}.
\newblock The MIT Press, 2009.

\bibitem{Low_2014}
Guang~Hao LLow, Theodore~James Yoder, and Isaac~L Chuang.
\newblock Quantum inference on {Bayesian} networks.
\newblock {\em Phys. Rev. A}, 89, June 2014.

\bibitem{tucci1995quantum}
Robert~R Tucci.
\newblock Quantum {B}ayesian nets.
\newblock {\em International Journal of Modern Physics B}, 09(03):295--337,
  1995.

\bibitem{borujeni2021quantum}
S.~E. Borujeni, S.~Nannapaneni, N.~H. Nguyen, E.~C. Behrman, and J.~E. Steck.
\newblock Quantum circuit representation of {Bayesian} networks.
\newblock {\em Expert Systems with Applications}, 2021.

\bibitem{borujeni2020experimental}
Sima~E Borujeni, Nam~H Nguyen, Saideep Nannapaneni, Elizabeth~C Behrman, and
  James~E Steck.
\newblock Experimental evaluation of quantum bayesian networks on {IBM QX}
  hardware.
\newblock In {\em 2020 IEEE International Conference on Quantum Computing and
  Engineering (QCE)}, pages 372--378. IEEE, 2020.

\bibitem{KLEPAC2017391}
G.~Klepac.
\newblock Chapter 12 -- {T}he {Schrödinger} equation as inspiration for a
  client portfolio simulation hybrid system based on dynamic {Bayesian}
  networks and the {REFII} model.
\newblock In {\em Quantum Inspired Computational Intelligence}, pages 391--416.
  Morgan Kaufmann, Boston, 2017.

\bibitem{Moreira_2016}
Catarina Moreira and Andreas Wichert.
\newblock Quantum-like {Bayesian} networks for modeling decision making.
\newblock {\em Frontiers in Psychology}, 7:11, 2016.

\bibitem{goodfellow2016deep}
Ian Goodfellow, Yoshua Bengio, and Aaron Courville.
\newblock {\em \href{https://www.deeplearningbook.org/}{Deep learning}}.
\newblock MIT Press, 2016.

\bibitem{hinton2002training}
Geoffrey~E Hinton.
\newblock Training products of experts by minimizing contrastive divergence.
\newblock {\em Neural computation}, 14(8):1771--1800, 2002.

\bibitem{salakhutdinov2009deep}
Ruslan Salakhutdinov and Geoffrey Hinton.
\newblock Deep {B}oltzmann machines.
\newblock In {\em \href{http://proceedings.mlr.press/v5/}{Proceedings of the
  Twelth International Conference on Artificial Intelligence and Statistics}},
  volume~5, pages 448--455. PMLR, 2009.

\bibitem{benedetti2016estimation}
Marcello Benedetti, John Realpe-G{\'o}mez, Rupak Biswas, and Alejandro
  Perdomo-Ortiz.
\newblock Estimation of effective temperatures in quantum annealers for
  sampling applications: A case study with possible applications in deep
  learning.
\newblock {\em Physical Review A}, 94(2):022308, 2016.

\bibitem{dixit2021training}
Vivek Dixit, Raja Selvarajan, Muhammad~A Alam, Travis~S Humble, and Sabre Kais.
\newblock Training restricted {B}oltzmann machines with a d-wave quantum
  annealer.
\newblock {\em Front. Phys.}, 9:589626, 2021.

\bibitem{amin2018quantum}
Mohammad~H. Amin, Evgeny Andriyash, Jason Rolfe, Bohdan Kulchytskyy, and Roger
  Melko.
\newblock Quantum {B}oltzmann machine.
\newblock {\em Phys. Rev. X}, 8:021050, May 2018.

\bibitem{lloyd2018quantum}
Seth Lloyd and Christian Weedbrook.
\newblock Quantum generative adversarial learning.
\newblock {\em Phys. Rev. Lett.}, 121:040502, July 2018.

\bibitem{Beer_2020}
Kerstin Beer, Dmytro Bondarenko, Terry Farrelly, Tobias~J. Osborne, Robert
  Salzmann, Daniel Scheiermann, and Ramona Wolf.
\newblock Training deep quantum neural networks.
\newblock {\em Nature Communications}, 11(1), Feb 2020.

\bibitem{cao2017quantum}
Yudong Cao, Gian~Giacomo Guerreschi, and Al{\'a}n Aspuru-Guzik.
\newblock Quantum neuron: an elementary building block for machine learning on
  quantum computers.
\newblock {\em \href{https://arxiv.org/abs/1711.11240}{arXiv preprint
  arXiv:1711.11240}}, 2017.

\bibitem{jia2019orthogonal}
Shuai Li, Kui Jia, Yuxin Wen, Tongliang Liu, and Dacheng Tao.
\newblock Orthogonal deep neural networks.
\newblock {\em IEEE transactions on pattern analysis and machine intelligence},
  43(4):1352--—1368, April 2021.

\bibitem{kerenidis2021classical}
Iordanis Kerenidis, Jonas Landman, and Natansh Mathur.
\newblock Classical and quantum algorithms for orthogonal neural networks.
\newblock {\em \href{https://arxiv.org/abs/2106.07198}{arXiv preprint
  arXiv:2106.07198}}, 2021.

\bibitem{allcock2020quantum}
Jonathan Allcock, Chang-Yu Hsieh, Iordanis Kerenidis, and Shengyu Zhang.
\newblock Quantum algorithms for feedforward neural networks.
\newblock {\em ACM Transactions on Quantum Computing}, 1(1):1--24, 2020.

\bibitem{farhi2018classification}
Edward Farhi and Hartmut Neven.
\newblock Classification with quantum neural networks on near term processors.
\newblock {\em \href{https://arxiv.org/abs/1802.06002}{arXiv preprint
  arXiv:1802.06002}}, 2018.

\bibitem{henderson2020quanvolutional}
Maxwell Henderson, Samriddhi Shakya, Shashindra Pradhan, and Tristan Cook.
\newblock Quanvolutional neural networks: powering image recognition with
  quantum circuits.
\newblock {\em Quantum Machine Intelligence}, 2(1):1--9, 2020.

\bibitem{killoran2019continuous}
Nathan Killoran, Thomas~R. Bromley, Juan~Miguel Arrazola, Maria Schuld,
  Nicol\'as Quesada, and Seth Lloyd.
\newblock Continuous-variable quantum neural networks.
\newblock {\em Phys. Rev. Research}, 1:033063, Oct 2019.

\bibitem{liu2022embeddig}
Henry Liu, Junyu Liu, Rui Liu, Henry Makhanov, Danylo Lykov, Anuj Apte, and
  Yuri Alexeev.
\newblock Embedding learning in hybrid quantum-classical neural networks, 2022.

\bibitem{schuld2021effect}
Maria Schuld, Ryan Sweke, and Johannes~Jakob Meyer.
\newblock Effect of data encoding on the expressive power of variational
  quantum-machine-learning models.
\newblock {\em Physical Review A}, 103(3):032430, 2021.

\bibitem{abbas2021power}
A.~Abbas, D.~Sutter, C.~Zoufal, A.~Lucchi, A.~Figalli, and S.~Woerner.
\newblock {The Power of Quantum Neural Networks}.
\newblock {\em Nature Computational Science}, 1(6):403--409, 2021.

\bibitem{schuld2021supervised}
Maria Schuld.
\newblock Supervised quantum machine learning models are kernel methods.
\newblock {\em \href{https://arxiv.org/abs/2101.11020}{arXiv preprint
  arXiv:2101.11020}}, 2021.

\bibitem{herman2022}
Dylan Herman, Rudy Raymond, Muyuan Li, Nicolas Robles, Antonio Mezzacapo, and
  Marco Pistoia.
\newblock Expressivity of variational quantum machine learning on the {Boolean}
  cube.
\newblock 2022.

\bibitem{jerbibeyond2022}
Sofiene Jerbi, Lukas~J. Fiderer, Hendrik~Poulsen Nautrup, Jonas~M. Kübler,
  Hans~J. Briegel, and Vedran Dunjko.
\newblock Quantum machine learning beyond kernel methods.
\newblock 2021.

\bibitem{jacot2020neural}
Arthur Jacot, Franck Gabriel, and Cl{\'e}ment Hongler.
\newblock Neural tangent kernel: Convergence and generalization in neural
  networks.
\newblock {\em \href{https://arxiv.org/abs/1806.07572}{arXiv preprint
  arXiv:1806.07572}}, 2018.

\bibitem{nakaji2021quantumenhanced}
Kouhei Nakaji, Hiroyuki Tezuka, and Naoki Yamamoto.
\newblock Quantum-enhanced neural networks in the neural tangent kernel
  framework.
\newblock {\em \href{https://arxiv.org/abs/2109.03786}{arXiv preprint
  arXiv:2109.03786}}, 2021.

\bibitem{lecun1998gradient}
Y.~Lecun, L.~Bottou, Y.~Bengio, and P.~Haffner.
\newblock Gradient-based learning applied to document recognition.
\newblock {\em Proceedings of the IEEE}, 86(11):2278--2324, 1998.

\bibitem{kerenidis2019quantum}
Iordanis Kerenidis, Jonas Landman, and Anupam Prakash.
\newblock Quantum algorithms for deep convolutional neural networks.
\newblock {\em \href{https://arxiv.org/abs/1911.01117}{arXiv preprint
  arXiv:1911.01117}}, 2019.

\bibitem{shen2021qfcnn}
Feihong Shen and Jun Liu.
\newblock Qfcnn: Quantum fourier convolutional neural network.
\newblock {\em \href{https://arxiv.org/abs/2106.10421}{arXiv preprint
  arXiv:2106.10421}}, 2021.

\bibitem{cong2019quantum}
Iris Cong, Soonwon Choi, and Mikhail~D. Lukin.
\newblock Quantum convolutional neural networks.
\newblock {\em Nature Physics}, 15(12):1273--1278, 2019.

\bibitem{verdon2019quantum}
Guillaume Verdon, Trevor McCourt, Enxhell Luzhnica, Vikash Singh, Stefan
  Leichenauer, and Jack Hidary.
\newblock Quantum graph neural networks.
\newblock {\em \href{https://arxiv.org/abs/1909.12264}{arXiv preprint
  arXiv:1909.12264}}, 2019.

\bibitem{quatgraphrecugnn}
Jaeho Choi, Seunghyeok Oh, and Joongheon Kim.
\newblock A tutorial on quantum graph recurrent neural network ({QGRNN}).
\newblock In {\em 2021 International Conference on Information Networking
  (ICOIN)}, pages 46--49, 2021.

\bibitem{dong2008quantum}
Daoyi Dong, Chunlin Chen, Hanxiong Li, and Tzyh-Jong Tarn.
\newblock Quantum reinforcement learning.
\newblock {\em IEEE Transactions on Systems, Man, and Cybernetics, Part B
  (Cybernetics)}, 38(5):1207--1220, 2008.

\bibitem{paparo2014quantum}
G.~D. Paparo, V.~Dunjko, A.~Makmal, M.~A. Martin-Delgado, and H.~J. Briegel.
\newblock Quantum speedup for active learning agents.
\newblock {\em Phys. Rev. X}, 2014.

\bibitem{chen2020variational}
Samuel Yen-Chi Chen, Chao-Han~Huck Yang, Jun Qi, Pin-Yu Chen, Xiaoli Ma, and
  Hsi-Sheng Goan.
\newblock Variational quantum circuits for deep reinforcement learning.
\newblock {\em IEEE Access}, 8:141007--141024, 2020.

\bibitem{chen2021variational}
Samuel Yen-Chi Chen, Chih-Min Huang, Chia-Wei Hsing, Hsi-Sheng Goan, and
  Ying-Jer Kao.
\newblock Variational quantum reinforcement learning via evolutionary
  optimization.
\newblock {\em Machine Learning: Science and Technology}, 2021.

\bibitem{lockwood2020reinforcement}
Owen Lockwood and Mei Si.
\newblock Reinforcement learning with quantum variational circuits.
\newblock In {\em
  \href{https://ojs.aaai.org//index.php/AIIDE/article/view/7437}{Proceedings of
  the AAAI Conference on Artificial Intelligence and Interactive Digital
  Entertainment}}, volume~16, pages 245--251. AAAI Press, 2020.

\bibitem{jerbi2021quantum}
Sofiene Jerbi, Lea~M. Trenkwalder, Hendrik Poulsen~Nautrup, Hans~J. Briegel,
  and Vedran Dunjko.
\newblock Quantum enhancements for deep reinforcement learning in large spaces.
\newblock {\em PRX Quantum}, 2:010328, Feb 2021.

\bibitem{crawford2016reinforcement}
Daniel Crawford, Anna Levit, Navid Ghadermarzy, Jaspreet~S Oberoi, and Pooya
  Ronagh.
\newblock Reinforcement learning using quantum {Boltzmann} machines.
\newblock {\em \href{https://arxiv.org/abs/1612.05695}{arXiv preprint
  arXiv:1612.05695}}, 2016.

\bibitem{Cherrat_2022}
El~Amine Cherrat, Iordanis Kerenidis, and Anupam Prakash.
\newblock Quantum reinforcement learning via policy iteration, 2022.

\bibitem{Cornelissen_2018}
Arjan Cornelissen.
\newblock Quantum gradient estimation and its application to quantum
  reinforcement learning.
\newblock Master's thesis, Delft, Netherlands, 2018.

\bibitem{otternlp}
Daniel~W. Otter, Julian~R. Medina, and Jugal~K. Kalita.
\newblock A survey of the usages of deep learning for natural language
  processing.
\newblock {\em IEEE Transactions on Neural Networks and Learning Systems},
  32(2):604--624, 2021.

\bibitem{naseem2021comprehensive}
Usman Naseem, Imran Razzak, Shah~Khalid Khan, and Mukesh Prasad.
\newblock A comprehensive survey on word representation models: From classical
  to state-of-the-art word representation language models.
\newblock {\em ACM Trans. Asian Low-Resour. Lang. Inf. Process.}, 20(5), Jun
  2021.

\bibitem{vaswani2017attention}
Ashish Vaswani, Noam Shazeer, Niki Parmar, Jakob Uszkoreit, Llion Jones,
  Aidan~N. Gomez, \L{}ukasz Kaiser, and Illia Polosukhin.
\newblock Attention is all you need.
\newblock In {\em
  \href{https://dl.acm.org/doi/10.5555/3295222.3295349}{Proceedings of the 31st
  International Conference on Neural Information Processing Systems}}, NIPS'17,
  2017.

\bibitem{devlin2019bert}
Jacob Devlin, Ming-Wei Chang, Kenton Lee, and Kristina Toutanova.
\newblock {BERT}: Pre-training of deep bidirectional transformers for language
  understanding.
\newblock In {\em Proceedings of the 2019 Conference of the North {A}merican
  Chapter of the Association for Computational Linguistics: Human Language
  Technologies, Volume 1 (Long and Short Papers)}. Association for
  Computational Linguistics, June 2019.

\bibitem{sybrandt2021cbag}
Justin Sybrandt and Ilya Safro.
\newblock {CBAG: Conditional biomedical abstract generation}.
\newblock {\em Plos one}, 16(7):e0253905, 2021.

\bibitem{floridi}
Luciano Floridi and Massimo Chiriatti.
\newblock {GPT}-3: Its nature, scope, limits, and consequences.
\newblock {\em Minds and Machines}, 30:1--14, 2020.

\bibitem{blodgettlanguage}
Su~Lin Blodgett, Solon Barocas, Hal Daum{\'e}~III, and Hanna Wallach.
\newblock Language (technology) is power: A critical survey of {``}bias{''} in
  {NLP}.
\newblock In {\em Proceedings of the 58th Annual Meeting of the Association for
  Computational Linguistics}. Association for Computational Linguistics, July
  2020.

\bibitem{sybrandt2020agatha}
Justin Sybrandt, Ilya Tyagin, Michael Shtutman, and Ilya Safro.
\newblock Agatha: Automatic graph mining and transformer based hypothesis
  generation approach.
\newblock In {\em Proceedings of the 29th ACM International Conference on
  Information \& Knowledge Management}, pages 2757--2764, 2020.

\bibitem{dhar2020carbon}
Payal Dhar.
\newblock The carbon impact of artificial intelligence.
\newblock {\em Nature Machine Intelligence}, 2(8):423--425, 2020.

\bibitem{coecke2010mathematical}
Bob Coecke, Mehrnoosh Sadrzadeh, and Stephen Clark.
\newblock Mathematical foundations for a compositional distributional model of
  meaning.
\newblock {\em \href{https://arxiv.org/abs/1003.4394}{arXiv preprint
  arXiv:1003.4394}}, 2010.

\bibitem{Lambek1968}
Joachim Lambek.
\newblock The mathematics of sentence structure.
\newblock {\em Journal of Symbolic Logic}, 33(4):627--628, 1968.

\bibitem{mikolov2013efficient}
Tomas Mikolov, Kai Chen, Greg Corrado, and Jeffrey Dean.
\newblock Efficient estimation of word representations in vector space.
\newblock In {\em
  \href{https://iclr.cc/archive/2013/conference-proceedings.html}{ICLR:
  Proceeding of the International Conference on Learning Representations
  Workshop}}, 2013.

\bibitem{coecke2009categories}
Bob Coecke and Eric~Oliver Paquette.
\newblock Categories for the practising physicist.
\newblock In {\em New structures for physics}, pages 173--286. Springer, 2010.

\bibitem{coecke2020foundations}
Bob Coecke, Giovanni de~Felice, Konstantinos Meichanetzidis, and Alexis Toumi.
\newblock Foundations for near-term quantum natural language processing.
\newblock {\em \href{https://arxiv.org/abs/2012.03755}{arXiv preprint
  arXiv:2012.03755}}, 2020.

\bibitem{abramsky2008categorical}
Samson Abramsky and Bob Coecke.
\newblock Categorical quantum mechanics.
\newblock {\em \href{https://arxiv.org/abs/0808.1023}{Handbook of quantum logic
  and quantum structures}}, 2:261--325, 2009.

\bibitem{Coecke_2011}
Bob Coecke and Ross Duncan.
\newblock Interacting quantum observables: categorical algebra and
  diagrammatics.
\newblock {\em New Journal of Physics}, 13(4):043016, Apr 2011.

\bibitem{de_Felice_2021}
Giovanni de~Felice, Alexis Toumi, and Bob Coecke.
\newblock {DisCoPy}: Monoidal categories in {Python}.
\newblock {\em Electronic Proceedings in Theoretical Computer Science},
  333:183--197, Feb 2021.

\bibitem{lorenz2021qnlp}
R.~Lorenz, A.~Pearson, K.~Meichanetzidis, D.~Kartsaklis, and B.~Coecke.
\newblock {QNLP} in practice: Running compositional models of meaning on a
  quantum computer.
\newblock {\em \href{https://arxiv.org/abs/2102.12846}{arXiv preprint
  arXiv:2102.12846}}, 2021.

\bibitem{meichanetzidis2020grammaraware}
Konstantinos Meichanetzidis, Alexis Toumi, Giovanni de~Felice, and Bob Coecke.
\newblock Grammar-aware question-answering on quantum computers.
\newblock {\em \href{https://arxiv.org/abs/2012.03756}{arXiv preprint
  arXiv:2012.03756}}, 2020.

\bibitem{bausch2020recurrent}
Johannes Bausch.
\newblock Recurrent quantum neural networks.
\newblock In H.~Larochelle, M.~Ranzato, R.~Hadsell, M.~F. Balcan, and H.~Lin,
  editors, {\em
  \href{https://proceedings.neurips.cc/paper/2020/hash/0ec96be397dd6d3cf2fecb4a2d627c1c-Abstract.html}{Advances
  in Neural Information Processing Systems}}, volume~33, pages 1368--1379.
  Curran Associates, Inc., 2020.

\bibitem{chen2020quantum}
Samuel Yen-Chi Chen, Shinjae Yoo, and Yao-Lung~L Fang.
\newblock Quantum long short-term memory.
\newblock {\em \href{https://arxiv.org/abs/2009.01783}{arXiv preprint
  arXiv:2009.01783}}, 2020.

\bibitem{Takaki_2021}
Yuto Takaki, Kosuke Mitarai, Makoto Negoro, Keisuke Fujii, and Masahiro
  Kitagawa.
\newblock Learning temporal data with a variational quantum recurrent neural
  network.
\newblock {\em Physical Review A}, 103(5), May 2021.

\bibitem{AHMED201619}
Mohiuddin Ahmed, Abdun {Naser Mahmood}, and Jiankun Hu.
\newblock A survey of network anomaly detection techniques.
\newblock {\em Journal of Network and Computer Applications}, 60:19--31, 2016.

\bibitem{fraudDetectionWest}
Jarrod West and Maumita Bhattacharya.
\newblock Intelligent financial fraud detection: A comprehensive review.
\newblock {\em Computers \& Security}, 57:47--66, 2016.

\bibitem{pang}
Guansong Pang, Chunhua Shen, Longbing Cao, and Anton Van~Den Hengel.
\newblock Deep learning for anomaly detection: A review.
\newblock {\em ACM Comput. Surv.}, 54(2), March 2021.

\bibitem{Herr_2021}
Daniel Herr, Benjamin Obert, and Matthias Rosenkranz.
\newblock Anomaly detection with variational quantum generative adversarial
  networks.
\newblock {\em Quantum Science and Technology}, 6(4):045004, Jul 2021.

\bibitem{schlegl2017unsupervised}
T.~Schlegl, P.~Seeböck, S.~M. Waldstein, U.~Schmidt-Erfurth, and G.~Langs.
\newblock Unsupervised anomaly detection with generative adversarial networks
  to guide marker discovery.
\newblock In {\em Information Processing in Medical Imaging}. Springer, 2017.

\bibitem{guo2021quantum}
Ming-Chao Guo, Hai-Ling Liu, Yong-Mei Li, Wen-Min Li, Su-Juan Qin, Qiao-Yan
  Wen, and Fei Gao.
\newblock Quantum algorithms for anomaly detection using amplitude estimation.
\newblock {\em \href{https://arxiv.org/abs/2109.13820}{arXiv preprint
  arXiv:2109.13820}}, 2021.

\bibitem{xu2015comprehensive}
Dongkuan Xu and Yingjie Tian.
\newblock A comprehensive survey of clustering algorithms.
\newblock {\em Annals of Data Science}, 2(2):165--193, 2015.

\bibitem{gu2020empirical}
Shihao Gu, Bryan Kelly, and Dacheng Xiu.
\newblock Empirical asset pricing via machine learning.
\newblock {\em {The Review of Financial Studies}}, 33(5):2223--2273, 2020.

\bibitem{chen2020deep}
L.~Chen, M.~Pelger, and J.~Zhu.
\newblock Deep learning in asset pricing.
\newblock {\em SSRN}, 2020.

\bibitem{Nagel_2021}
Stefan Nagel.
\newblock {\em Machine Learning in Asset Pricing}.
\newblock Princeton University Press, 2021.

\bibitem{sakuma2020application}
T.~Sakuma.
\newblock Application of deep quantum neural networks to finance.
\newblock {\em \href{https://arxiv.org/abs/2011.07319}{arXiv preprint
  arXiv:2011.07319}}, 2020.

\bibitem{cuccaro2004new}
Steven~A Cuccaro, Thomas~G Draper, Samuel~A Kutin, and David~Petrie Moulton.
\newblock A new quantum ripple-carry addition circuit.
\newblock {\em \href{https://arxiv.org/abs/quant-ph/0410184}{arXiv preprint
  quant-ph/0410184}}, 2004.

\bibitem{Venturelli_2019}
Davide Venturelli and Alexei Kondratyev.
\newblock Reverse quantum annealing approach to portfolio optimization
  problems.
\newblock {\em Quantum Machine Intelligence}, 1(1-2):17--–30, Apr 2019.

\bibitem{boothby2016fast}
Tomas Boothby, Andrew~D King, and Aidan Roy.
\newblock Fast clique minor generation in {Chimera} qubit connectivity graphs.
\newblock {\em Quantum Information Processing}, 15(1):495--508, 2016.

\bibitem{yalovetzky2021nisqhhl}
Romina Yalovetzky, Pierre Minssen, Dylan Herman, and Marco Pistoia.
\newblock {NISQ-HHL}: Portfolio optimization for near-term quantum hardware.
\newblock {\em \href{https://arxiv.org/abs/2110.15958}{arXiv preprint
  arXiv:2110.15958}}, 2021.

\bibitem{shors2nplus3}
Stephane Beauregard.
\newblock Circuit for {Shor's} algorithm using 2n+3 qubits.
\newblock {\em Quantum Info. Comput.}, 3(2):175–--185, Machr 2003.

\bibitem{mottonen2004transformation}
Mikko M\"{o}tt\"{o}nen, Juha~J. Vartiainen, Ville Bergholm, and Martti~M.
  Salomaa.
\newblock Transformation of quantum states using uniformly controlled
  rotations.
\newblock {\em Quantum Info. Comput.}, 5(6):467--–473, Sep 2005.

\bibitem{lee2019hybrid}
Yonghae Lee, Jaewoo Joo, and Soojoon Lee.
\newblock Hybrid quantum linear equation algorithm and its experimental test on
  {IBM} quantum experience.
\newblock {\em Scientific Reports}, 9, 03 2019.

\bibitem{aleksandrowicz2019qiskit}
Gadi Aleksandrowicz, Thomas Alexander, Panagiotis Barkoutsos, Luciano Bello,
  Yael Ben-Haim, David Bucher, Francisco~Jose Cabrera-Hern{\'a}ndez, Jorge
  Carballo-Franquis, Adrian Chen, Chun-Fu Chen, et~al.
\newblock Qiskit: An open-source framework for quantum computing.
\newblock 2019.

\bibitem{patentBeusoleil}
R.~G. Beausoleil, W.~J. Munro, T.~P. Spiller, and W.~K. van Dam.
\newblock Tests of quantum information.
\newblock {\em US Patent 7,559,101 B2}, 2008.

\end{thebibliography}

\end{document}